\documentclass[review,authoryear,3p,times,10pt]{elsarticle}

\usepackage{multirow,setspace,amssymb,amsmath,graphicx,color,url,bm,pdflscape}
\usepackage{lineno}
\usepackage{natbib}
\usepackage{booktabs}
\usepackage{longtable}
\usepackage{lscape}  
\usepackage[normalem]{ulem}  
\usepackage{threeparttable}
\usepackage[table]{xcolor}
\usepackage{tabularx}
\usepackage{epstopdf}
\usepackage{mathrsfs}
\usepackage{makecell}
\usepackage[bookmarks=true,colorlinks,linkcolor=blue,anchorcolor=blue,citecolor=blue,unicode]{hyperref}
\usepackage{bookmark}
\usepackage{todonotes}
\usepackage{ragged2e}
\usepackage{dcolumn}
\usepackage{dsfont}
\usepackage{listings}
\usepackage{relsize}
\usepackage{tikz}
\usepackage{subcaption}  
\usepackage[font=small]{caption}

\newcommand{\code}[1]{\texttt{#1}}

\graphicspath{{Figures/}}
\makeatletter
\def\ps@pprintTitle{%
	\let\@oddhead\@empty 
	\let\@evenhead\@empty
}
\newcolumntype{d}[1]{D{.}{.}{#1}} 
\newcommand{\MVP}{\textup{MVP}}
\newcommand{\MCP}{\textup{MCP}}
\newcommand{\MCoP}{\textup{MCoP}}
\begin{document}

\begin{frontmatter}

\title{Russia-Ukraine conflict and the quantile return connectedness of grain futures in the BRICS and international markets}

\author[SILC]{Yan-Hong Yang}

\author[SUIBE]{Ying-Hui Shao \corref{cor1}}
\ead{yinghuishao@126.com}

\author[SB,RCE,SM]{Wei-Xing Zhou \corref{cor1}}
\ead{wxzhou@ecust.edu.cn}
\cortext[cor1]{Corresponding authors.}

\address[SILC]{SILC Business School, Shanghai University, Shanghai 201899, China}
\address[SUIBE]{School of Statistics and Information, Shanghai University of International Business and Economics, Shanghai 201620, China}
\address[SB]{School of Business, East China University of Science and Technology, Shanghai 200237, China}
\address[RCE]{Research Center for Econophysics, East China University of Science and Technology, Shanghai 200237, China}
\address[SM]{School of Mathematics, East China University of Science and Technology, Shanghai 200237, China}

\begin{abstract}

This study investigates quantile-based connectedness among BRICS and international grain futures around the Russia-Ukraine conflict and milestones of the Black Sea Grain Initiative. Using a dynamic quantile VAR combined with a frequency-domain decomposition, we trace spillovers across market states and horizons. Spillovers are heterogeneous across quantiles, as the time-varying total connectedness index hovers near 95\% in the tails, remains well above the median, and is higher before the outbreak than after. Furthermore, grain type and regional proximity strengthen pairwise connectedness. South African grain futures are persistent net receivers, whereas Argentine grain futures, U.S. soybean, and Ukrainian wheat are key transmitters. In the frequency domain, short-term components dominate total spillovers. In portfolio applications, the minimum connectedness portfolio delivers a positive Sharpe ratio under both normal and lower tail conditions. Overall, the results inform asset allocation and risk management in grain futures markets under geopolitical instability and support policy formulation.

\end{abstract}
\end{frontmatter}

\section{Introduction}
\label{S1:Introduction}

Food security is a fundamental basis for the healthy and stable development of socio-economic systems. In recent years, international grain prices have experienced several sharp fluctuations due to a confluence of factors including the COVID-19 pandemic, extreme weather events, economic shocks, trade frictions, and geopolitical conflicts \citep{zhou2024impact}. These fluctuations have dramatically increased the number of people suffering from poverty and hunger, placing unprecedented pressure on the global food supply chain. According to \textsl{The State of Food Security and Nutrition in the World 2025}, severe food insecurity affected about 840 million people in 2022, 841 million in 2023, and 828 million in 2024. Including moderate cases, roughly 2.3 billion people were food insecure in 2024.\footnote{\url{https://www.fao.org/publications/home/fao-flagship-publications}}

Specifically, the Russia-Ukraine conflict, now in its fourth year, has had profound effects on the global food system \citep{naeem2024tail,zhou2024does,jiang2024impact,cui2024unveiling,gao2022global}, leaving roughly 23 million people facing hunger. Russia and Ukraine are major producers of food, energy, and fertilizers and leading exporters of wheat, barley, maize, and sunflower oil. In 2021 they accounted for 14.3\% of global wheat production, 19.0\% of barley, and 4.5\% of maize.
According to the Food and Agriculture Organization of the United Nations (FAO), by the end of 2021, the export shares of these two countries in the global market were notably high, with sunflower oil at 78.00\% (Russia 28.00\%, Ukraine 50.00\%), wheat at 32.53\% (Russia 21.99\%, Ukraine 10.54\%), barley at 29.60\% (Russia 12.93\%, Ukraine 16.67\%), and maize at 16.25\% (Russia 2.25\%, Ukraine 14.00\%). The conflict has directly disrupted critical supply chains and export routes in these countries, leading to tensions and price fluctuations in the global food market, severely affecting the food security and economic stability of countries reliant on these imports \citep{behnassi2022implications}. Furthermore, the conflict has exacerbated trade protectionism and export restrictions, driving up global food prices and causing serious concerns about food security in at least 50 countries and regions.

In July 2022, the Black Sea Grain Initiative (BSGI) was established to ease tensions by creating safe humanitarian corridors for the export of food and fertilizers.\footnote{For details, see the United Nations page on the BSGI: \url{https://www.un.org/en/black-sea-grain-initiative}.} However, amid disagreements, the initiative was suspended on 17 July 2023, again disrupting global food supplies. Over the longer run, the Russia-Ukraine conflict may reorganize global agricultural trade as countries seek greater self-sufficiency and diversify supply sources to reduce regional dependence and strengthen resilience to future shocks \citep{neik2023diversifying}. The conflict exposes the high sensitivity and vulnerability of global food security to geopolitical events and compels the international community to reconsider and reinforce food security strategies in response to the ongoing and intensifying effects of the crisis. This backdrop underscores the need to examine the dynamics of international grain markets, with a focus on the interconnections among major producers and consumers and on how conflicts such as the Russia-Ukraine war disrupt them.

The BRICS countries play a crucial role in stabilizing global grain supplies and prices, and the recent expansion of BRICS has further amplified the group's influence in global grain markets. For instance, Brazil is a leading exporter of soybeans and corn, while India and China are major producers and consumers of several grains. Focusing on BRICS helps clarify how changes in agricultural output, trade policy, and economic conditions in these economies affect global grain markets and grain futures. Moreover, existing research on grain futures largely emphasizes developed economies or adopts a global perspective \citep{jiang2024impact,zhu2024uncovering,cui2024unveiling,naeem2024tail,zhou2024does}, with relatively limited attention to BRICS as a group. As BRICS has expanded from five to ten members, assessing the role of agricultural futures within this enlarged framework has become increasingly important for both theory and practice.

Grain futures markets are key price discovery mechanisms that reflect expectations about future supply and demand, so futures prices serve as critical indicators and reliable benchmarks for spot pricing \citep{zhou2024impact}. Against this backdrop, the Russia-Ukraine conflict has materially disrupted global grain supply and increased futures volatility.
This study examines quantile connectedness and spillovers among BRICS grain futures and key international benchmarks around the Russia-Ukraine conflict and the BSGI. We map static and dynamic connectedness in the time and frequency domains using a dynamic quantile VAR and a frequency-domain decomposition \citep{barunik2018measuring,ando2022quantile}. The analysis spans the major contracts in soybeans, maize, wheat, and rice across BRICS exchanges and the United States, Argentina, Ukraine, and the Black Sea region, which enables a unified view of cross-country and cross-grain transmission under normal and tail conditions. We quantify total connectedness, directional spillovers, net effects, and net pairwise connectedness and assess quantile sensitivity across horizons. Building on these estimates, we evaluate portfolio strategies that incorporate spillovers by constructing minimum variance, minimum correlation, and minimum connectedness portfolios. We benchmark performance adjusted for risk using the Sharpe ratio, hedging effectiveness, and measures based on VaR and CVaR. This integrated design links tail- and horizon-varying spillovers to asset allocation choices and offers practical guidance for risk management in geopolitical instability.

This study contributes to the literature in three ways. First, we provide novel tail- and horizon-specific evidence on return connectedness between BRICS and international grain futures in a unified framework that integrates a dynamic quantile VAR and a frequency-domain decomposition. We characterize static and dynamic spillovers across normal and tail states and across short-, medium-, and long-term horizons, quantify a diverse set of connectedness measures, and show that grain type and regional proximity strengthen pairwise connectedness. Second, we document that geopolitical shocks reshape market structure, with tail connectedness hovering near 95\% and generally higher in the pre-outbreak than in the post-outbreak period, alongside shifts in transmitter and receiver roles around the Russia-Ukraine conflict and the BSGI. Third, we evaluate grain portfolios that embed connectedness information across conditional quantiles and find that the minimum connectedness portfolio attains the highest Sharpe ratio under normal and lower tail conditions, providing an integrated view of connectedness for practical portfolio design and risk control.

The rest of the paper is organized as follows. Section~\ref{S2:Literature} reviews the related literature. Section~\ref{S3:Data description} describes the data and summary statistics. Section~\ref{S4:Methodology} outlines the QVAR methodology. Section~\ref{S5:Spillover analysis} presents the spillover results. Section~\ref{S6:Portofolio} evaluates portfolio performance. Section~\ref{S7:conclusion} concludes with key findings and implications.

\section{Literature review}
\label{S2:Literature}

Connectedness and spillovers among commodity markets have received sustained attention, especially with the continuous financialization of commodities \citep{gong2023research,zhou2024impact,wei2023alarming,tiwari2022quantile,le2023price,zhang2023contemporaneous, jiang2024impact,zhu2024uncovering,cui2024unveiling,naeem2024tail,zhou2024does}. Our study examines how connectedness in international grain futures evolves and how the Russia-Ukraine conflict affects these dynamics. Accordingly, we review the literature in three parts, covering connectedness within grain futures markets, linkages between grain futures and other commodity futures, and the effects of the Russia-Ukraine conflict on grain futures markets.

\subsection{Connectedness within grain futures markets}

Grain commodity financialization underscores the importance of analyzing interdependencies among grain futures through the lenses of connectedness and risk spillovers \citep{ouyang2020financialization,ait2019index,just2022dynamic,hamadi2017news,wang2024dynamic,zhu2024uncovering}. This perspective helps explain market responses across economic regimes, particularly during global turbulence. Using a CoVaR approach, \cite{ke2019risk} study risk transmission between U.S. and Chinese soybean, corn, and sugar futures, and find that soybean futures transmit more risk than corn and sugar futures. Likewise, drawing on the CoVaR family, \cite{hu2024extreme} document significant and asymmetric tail-risk spillovers between the U.S. and Chinese futures for soybean and wheat, and they show that corn uniquely exhibits no downside risk spillovers. They also find that dependence structures shift between stable and crisis periods and that tail spillovers intensify around the Global Financial Crisis, the COVID-19 pandemic, and the Russia-Ukraine conflict. \cite{vzivkov2020bayesian} use an MS-GARCH model and Bayesian quantile regression to examine volatility spillovers among corn, wheat, soybean, and rice futures and report that soybean and wheat experience stronger volatility shocks from other markets, making them weaker primary portfolio assets, whereas rice receives the lowest shocks and appears most stable. \cite{li2017price} investigate lead-lag relations among U.S., Brazilian, and Chinese soybean futures and find that U.S. contracts lead long-run price changes in Brazil and China, while short-run interactions occur mainly between U.S. overnight returns and China's No. 1 soybean futures. In contrast, \cite{han2013cross} employ a structural VAR and a VECM to show that DCE contributes meaningfully to global soybean futures price discovery, with bidirectional information transmission between DCE and CBOT, challenging the view of DCE as a satellite market. Further, \cite{jiang2017dynamics} identify sizable bidirectional volatility spillovers between U.S. and Chinese agricultural futures, especially for soybean, wheat, and corn, with U.S. short-term volatility affecting China and evidence of rising Chinese pricing power. Employing a quantilogram, \cite{jiang2016spillovers} reveal bidirectional return dependence between the two markets, particularly in the tails, with a moderately stronger influence from the U.S. to China. Moreover, \cite{chen2018information} adopt a VAR-BEKK-MGARCH model that accounts for skewness to examine mean and volatility spillovers in U.S. and Chinese corn, wheat, and soybean futures and find that the U.S. plays a major role in information transmission, while China's volatility spillovers to the U.S. increase in more marketized commodities and after changes in trading structure.

\subsection{Connectedness between grain and other commodity futures markets}

Compared with the relatively limited work on connectedness within grain markets, there is extensive research on linkages between grain futures and other commodity futures \citep{cui2024unveiling,tiwari2022quantile,kumar2021time,rezitis2024assessing,kang2017dynamic,zhang2023contemporaneous,xiao2020estimating,yang2024contemporaneous}. While connectedness between grain and energy dominates, related work extends to grain linkages with precious metals and beyond.

Using a rolling window quantile VAR, \citet{tiwari2022quantile} examine volatility spillovers between energy and agricultural markets and find that corn, wheat, and soybean play important transmitter roles, with more than 30\% of their volatility attributed to network linkages with other markets. They show that agricultural commodities are more closely tied to energy than to natural gas, ethanol, or coffee, and that spillovers intensify under extreme market conditions, while the COVID-19 pandemic further amplifies both volatility and interconnectedness. \citet{kumar2021time} study the energy-food nexus with a dependence-switching copula model for corn, oats, soybeans, wheat, and WTI oil across four market states. They find that oil and grain crashes tend to coincide in crises but not in normal times, implying that investors cannot earn excess profits in both markets simultaneously. They also report pervasive return chasing and significant risk spillovers from oil to grains, particularly during the global financial crisis. \citet{rezitis2024assessing} assess volatility linkages between energy and grain futures and show asymmetric effects of external shocks, with higher cross correlations in high volatility regimes. Using Markov-switching regressions and quadrivariate VAR-DCC-GARCH and VAR-BEKK-GARCH models, they find that comovement dominates substitution among WTI crude oil, natural gas, corn, and soybeans and confirm bidirectional price volatility spillovers during turbulent periods.

\citet{kang2017dynamic} analyze spillovers among gold, silver, WTI crude oil, corn, wheat, and rice with a DECO-GARCH model and a spillover index. Findings reveal that return correlations rise during the global financial crisis and the European sovereign debt crisis, reducing diversification benefits,and that bidirectional return and volatility spillovers strengthen after the crisis. Gold and silver act as information transmitters, whereas the other commodities tend to be receivers during financial stress.
\cite{zhang2023contemporaneous} introduce a network topology approach to discern both contemporaneous and noncontemporaneous idiosyncratic risk spillovers in commodity futures markets, demonstrating that contemporaneous information is more critical in constructing networks, particularly for higher-moment risks. Key findings indicate that gold, silver, and wheat are major transmitters of volatility and kurtosis risks, while corn and silver significantly transmit skewness risks.
\cite{xiao2020estimating} find that metal futures act as net transmitters of shocks, while grain futures such as wheat and corn are particularly susceptible to these shocks. 
They further discover that nearly two thirds of volatility uncertainty in commodity futures arises from interconnected shocks and that connectedness increases notably during market turmoil.

\subsection{Impact of the Russia-Ukraine conflict on grain futures markets}

The Russia-Ukraine conflict has attracted substantial scholarly attention, particularly regarding spillovers and connectedness in global grain futures markets \citep{zhou2024impact,just2022dynamic,wu2023time,zhang2024covid,zhou2024does}. Prior research examines how this geopolitical shock increases volatility, disrupts supply chains, and heightens uncertainty, with implications for global food security and economic stability \citep{behnassi2022implications,jiang2024impact}.

\citet{zhou2024impact} propose a tail dependence framework that integrates Copula-CoVaR with an ARMA-GARCH-skewed Student-\emph{t} specification to study extreme risk spillovers between agricultural futures and spot markets before and after the conflict. They find that tail dependence in soybean, maize, wheat, and rice responds to the conflict, that risks increase especially for wheat, and that downside and upside spillovers to spot markets are significant and asymmetric across directions and periods. \citet{just2022dynamic} employ dynamic DY and BK spillover indices and show stronger integration from the post-COVID-19 recovery into the Russia-Ukraine conflict, with wheat, maize, and barley acting as main transmitters of returns and rice remaining relatively isolated. \citet{wu2023time} use a TVP-VAR-based frequency connectedness approach and document that the pandemic raises connectedness more strongly than the conflict through high- and medium-frequency channels. Initially, fossil energy transmits risk to energy crops such as corn and soybean, whereas during the conflict agricultural commodities, especially wheat and corn, become the primary transmitters. \citet{zhang2024covid} analyze U.S. and Chinese agricultural futures with wavelet coherence and a TVP-VAR and find altered correlations and lead-lag comovement for soybean and corn, limited effects for cotton, greater volatility transmission from U.S. to China, and a stronger spillover impact of the war than the pandemic, particularly for soybean and corn. \citet{zhou2024does} examine time-varying linkages between energy and agricultural commodities and document asymmetric volatility spillovers that are stronger across quantiles than at the mean. Crude oil is the main transmitter before the conflict and, together with natural gas, continues to transmit volatility thereafter, while the conflict shifts many agricultural commodities from transmitters to receivers and increases dependence on crude oil for key exports such as wheat.

Existing empirical work offers mixed evidence on the nexus between grain and other commodity futures, but risk spillovers within BRICS markets and between BRICS and international grain markets during major geopolitical events remain underexplored. To address this gap, the present study integrates time- and frequency-domain perspectives to analyze the dynamic evolution of connectedness and risk transmission between BRICS and international grain futures under varying market conditions, with particular attention to the Russia-Ukraine conflict.

\section{ Data and preliminary analysis}
\label{S3:Data description}
\subsection{Data description}

We study grain futures from BRICS exchanges, focusing on corn (maize), soybeans, wheat, and rice. Because CBOT prices are key international references for agricultural commodities, we also include a set of CBOT contracts to benchmark global pricing. We use futures rather than spot prices since futures embed market expectations and offer superior price discovery and liquidity \citep{shao2019time,yang2020time}. Unlike spot quotations, which reflect immediate delivery, futures are standardized contracts for delivery at specified maturities.
Our dataset consists of daily closes for continuous contract series from April 12, 2021 through August 11, 2023, for a total of 610 trading days. We construct the continuous series from exchange listings and handle missing observations by forward filling, and when a series starts with missing values we backfill at the beginning. The data come from Wind and Bloomberg. To study the Russia-Ukraine conflict, we split the sample into a pre-outbreak period from April 12, 2021 to February 23, 2022 and a post-outbreak period from February 24, 2022 to August 11, 2023.

Due to data limitations, we do not include Russian domestic grain futures. Instead, we proxy regional exposure with CME Group's financially settled Black Sea Wheat (Platts) futures listed on CBOT. Ukraine is a major wheat exporter and a principal party to the conflict, so we also include Ukrainian Wheat (Platts) futures listed on CBOT. Following the suspension of the BSGI on July 17, 2023, CME Group announced the cessation of trading and clearing for Ukrainian Wheat (Platts), Black Sea Wheat (Platts), and Black Sea Corn (Platts) contracts effective August 16, 2023 (SER-9233).\footnote{For detailed information, readers can refer to \url{https://www.cmegroup.com/notices/ser/2023/08/SER-9233.html}} We therefore set the sample end date to August 11, 2023. In addition, Chinese wheat futures are excluded because quotes exhibit limited variation over the employed period.
Some of the new BRICS members do not have well-established grain futures markets, so we restrict attention to the original BRICS. Although Argentina declined an invitation to join BRICS in December 2023, it is a major global exporter of corn and soybeans, and we therefore include contracts from Argentina's Matba Rofex, the largest agricultural futures and options exchange in South America. In total, we analyze 23 grain futures contracts from BRICS exchanges, Matba Rofex, and CBOT. 
Contract codes use ISO-2 prefixes to indicate the origin of the underlying crop, as detailed in the notes to Table~\ref{Tb:Statistics:Full}.

\begin{table}[htp]
	\centering
	\setlength{\abovecaptionskip}{0pt}
	\setlength{\belowcaptionskip}{10pt}
	\caption{Summary statistics of the whole sample.}
	\label{Tb:Statistics:Full}
\resizebox{\textwidth}{!}{%
\begin{tabular}{l l l l l*{4}{d{1.3}} r r @{}}            
		\toprule
		Abbrev. Code & Full Name & Exchange & Country/Region & ISO‑2 &
		\multicolumn{1}{c}{Mean ($\times 10^{-3}$)} &
		\multicolumn{1}{c}{S.D.} &
		\multicolumn{1}{c}{Skew} &
		\multicolumn{1}{c}{Ex.Kurt} &
		\multicolumn{1}{c}{JB} &
		\multicolumn{1}{c}{ADF} \\ \midrule
		BRms & Mini soybean‑CME & BMF & Brazil & BR & 0.086 & 0.014 & -0.530 & 3.925 & 419.470*** & -8.177*** \\ 
		BRs & Santos soybeans & CBOT & Brazil & BR & 0.078 & 0.013 & -0.627 & 5.139 & 709.929*** & -8.091*** \\ 
		CNs & Soybean & DCE & China & CN & -0.169 & 0.010 & 0.201 & 3.233 & 269.347*** & -7.219*** \\ 
		ZAs & Soybean & SAFEX & South Africa & ZA & 0.726 & 0.013 & -0.006 & 0.072 & 0.136 & -7.728*** \\ 
		ZAcs & Soybean‑CBOT & SAFEX & South Africa & ZA & 0.980 & 0.016 & 0.192 & 3.957 & 401.078*** & -9.029*** \\ 
		ARs & Soybean & Matba Rofex & Argentina & AR & 0.283 & 0.013 & -0.310 & 3.037 & 243.852*** & -7.494*** \\ 
		USs & Soybean & CBOT & United States & US & 0.028 & 0.016 & -1.143 & 7.723 & 1645.946*** & -7.494*** \\ 
		BRc & Cash‑settled corn & BMF & Brazil & BR & -0.665 & 0.013 & 0.184 & 2.213 & 127.755*** & -7.787*** \\ 
		INm & Maize & NCDEX & India & IN & 0.961 & 0.016 & 0.059 & 3.144 & 251.244*** & -7.212*** \\ 
		CNc & Corn & DCE & China & CN & 0.024 & 0.008 & -0.471 & 1.652 & 91.758*** & -7.837*** \\ 
		ZAwm & White maize & SAFEX & South Africa & ZA & 0.450 & 0.016 & 0.101 & 0.190 & 1.949 & -8.379*** \\ 
		ZAym & Yellow maize & SAFEX & South Africa & ZA & 0.400 & 0.014 & 0.111 & 0.232 & 2.628 & -8.402*** \\ 
		ZAc & Corn‑CBOT & SAFEX & South Africa & ZA & 0.561 & 0.020 & 0.298 & 3.104 & 253.584*** & -8.623*** \\ 
		ARm & Maize & Matba Rofex & Argentina & AR & 0.375 & 0.017 & -0.111 & 2.191 & 123.096*** & -7.398*** \\ 
		USc & Corn & CBOT & United States & US & -0.291 & 0.025 & -1.936 & 18.121 & 8713.067*** & -7.839*** \\ 
		ZAw & Wheat & SAFEX & South Africa & ZA & 0.392 & 0.009 & -0.033 & 1.547 & 60.843*** & -8.320*** \\ 
		ARw & Wheat & Matba Rofex & Argentina & AR & 0.911 & 0.019 & 0.142 & 3.856 & 379.299*** & -7.873*** \\ 
		BSw & Black Sea wheat & CBOT & Black Sea & BS & 0.341 & 0.012 & 0.498 & 5.492 & 790.550*** & -6.539*** \\ 
		USw & Wheat & CBOT & United States & US & -0.002 & 0.026 & -0.403 & 11.126 & 3157.874*** & -7.909*** \\ 
		UAw & Ukrainian wheat & CBOT & Ukraine & UA & 0.130 & 0.013 & 0.428 & 3.912 & 406.963*** & -6.603*** \\ 
		CNr & Rice & DCE & China & CN & -0.051 & 0.004 & 2.666 & 21.211 & 12137.676*** & -8.684*** \\ 
		USr & Rice & CBOT & United States & US & 0.328 & 0.015 & -1.569 & 16.669 & 7300.795*** & -9.339*** \\ 
		INb & Bajra & NCDEX & India & IN & 0.562 & 0.014 & 0.037 & 2.621 & 174.509*** & -6.511*** \\
		\bottomrule
\end{tabular}}
\begin{flushleft}
\footnotesize
\justifying Notes: 
1. Variable abbreviations follow the pattern  ``XXy'': the ISO-2 prefix \code{XX} gives the origin country/region and the suffix \code{y} differentiates contracts.  
2. The Black Sea region has no official ISO‑2 code; for consistency we use \code{BS}, which is formally assigned to the Bahamas.   
3. ``Country/Region'' denotes the \emph{origin} of the underlying commodity, whereas ``Exchange'' is the marketplace where the futures contract is traded. Thus \code{BRs} is listed on CBOT but references Brazilian soybeans shipped via the port of Santos.  
4. \code{ZAc} (corn‑CBOT), \code{ZAcs} (soybean‑CBOT) and \code{BRms} (mini soybean‑CME) are traded on SAFEX or BMF but are contractually linked to CBOT/CME prices. Asterisks (*, **, ***) denote significance at the 10\%, 5\%, and 1\% levels, respectively.
\end{flushleft} 
\end{table}

\subsection{ Preliminary analysis}

Table~\ref{Tb:Statistics:Full} reports summary statistics and diagnostic tests for the returns of 23 grain futures over the full sample. Returns are log differences, defined as $\ln(p_{i,t}/p_{i,t-1})$, where $p_{i,t}$ is the daily closing price of grain future $i$ on day $t$. As shown in Table~\ref{Tb:Statistics:Full}, USw has the highest standard deviation, whereas CNr has the lowest. Skewness and excess kurtosis are pronounced, and the Jarque-Bera test rejects normality at the 1\% level for all series except ZAs, ZAwm, and ZAym, which supports our use of a quantile VAR connectedness framework. The ADF unit-root test indicates that all return series are stationary.

\begin{figure}[!ht]
	\centering
	\includegraphics[width=0.7\linewidth]{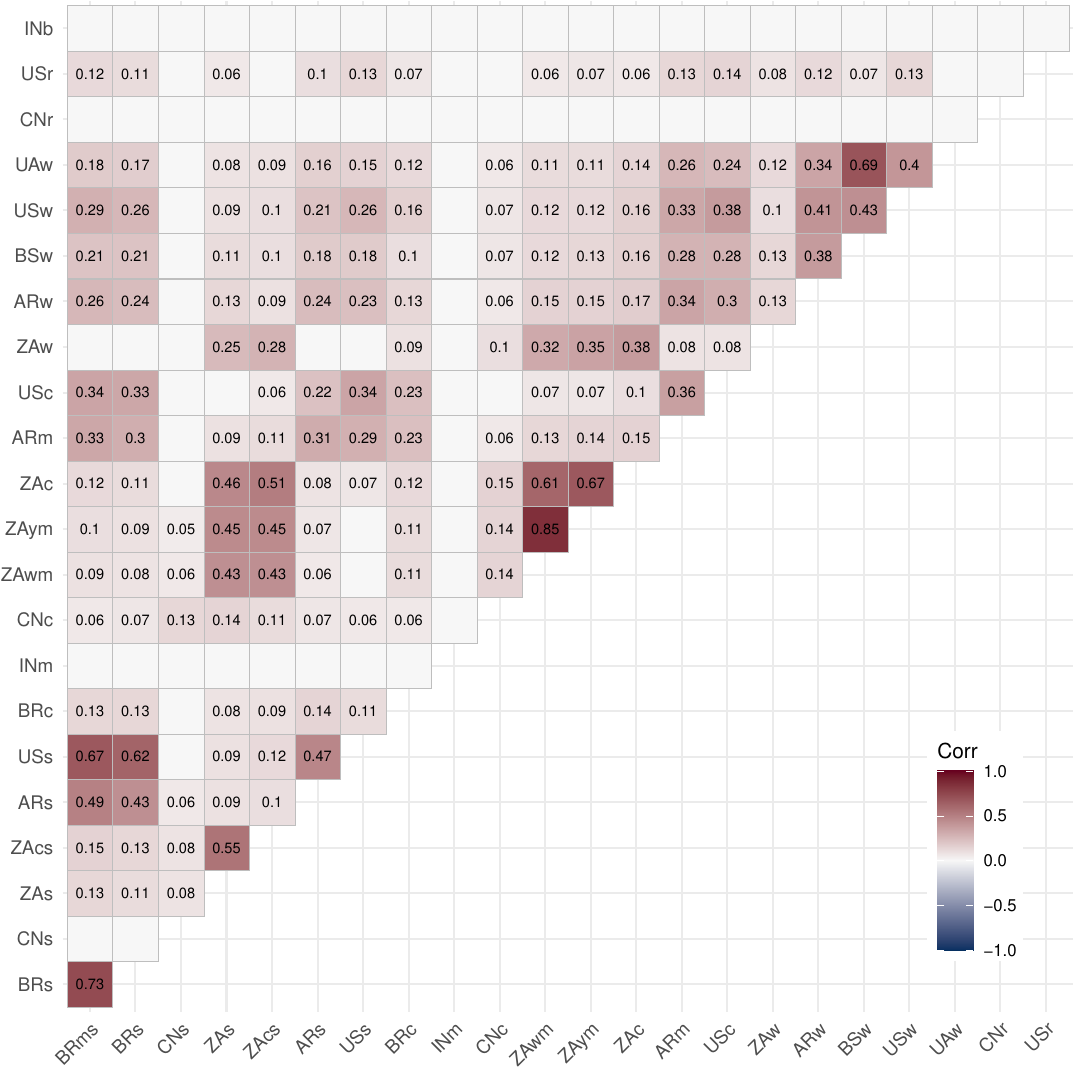}
	\caption{Heatmap of pairwise Kendall's rank correlations. Blank cells indicate correlations not significant at the 5\% level.}
	\label{Fig:correlation}
\end{figure}

Fig.~\ref{Fig:correlation} shows that nonparametric Kendall correlations are positive for all statistically significant pairs among the 23 contracts, indicating a shared direction of movement. Only INm, INb, and CNr lack significant ties with other grains. Comparing the pre- and post-conflict subsamples, 109 pairs remain significant in both windows, yet 83 of them (76.15\%) display stronger absolute correlations in the pre-conflict period.\footnote{We computed these correlations separately for the pre- and post-conflict subsamples and can provide the full set of coefficients on request.} This weakening of linear dependence is consistent with \cite{zhou2024impact}, who report a post-conflict decline in comovement between grain futures and spot returns.

\section{Methodology}
\label{S4:Methodology}

We utilize the quantile connectedness method proposed by \cite{ando2022quantile} and further elaborated upon by \cite{chatziantoniou2021interest}, to investigate the return risk spillover across quantiles in grain futures markets. As described by \cite{chatziantoniou2021interest}, the quantile-based vector autoregression  model of lag order $p$ and dimension $n$ (QVAR($p$)) is estimated in the following manner:
\begin{equation}
	\bm{y}_t = \bm{\mu}_t(\tau) + \bm{\phi}_1(\tau)\bm{y}_{t-1} + \bm{\phi}_2(\tau)\bm{y}_{t-2} + \cdots + \bm{\phi}_p(\tau)\bm{y}_{t-p} + \bm{u}_t(\tau)
\end{equation}
where $\bm{y}_t$ and $\bm{y}_{t-i}$, $i = 1,\ldots,p$ are $n \times 1$ dimensional endogenous variable vectors, $\tau$ signifies the quantile level and ranges between 0 and 1, $\bm{\mu}_t(\tau)$ is an $n \times 1$ dimensional conditional mean vector, $\bm{\phi}_i (\tau)$ is an $n \times n$ dimensional QVAR coefficient matrix, and $\bm{u}_t(\tau)$ represents an $n \times 1$ dimensional error vector associated with an $n \times n$ dimensional variance-covariance matrix, $\bm{\Sigma}(\tau)$. Utilizing Wold's theorem, the quantile vector moving average representation of infinite order, QVMA($\infty$), for the QVAR($p$)  is given by:
\begin{equation}
	\bm{y}_t = \bm{\mu}(\tau) + \sum_{j=1}^p \bm{\phi}_j (\tau)\bm{y}_{t-j} + \bm{u}_t(\tau) = \bm{\mu}(\tau) + \sum_{i=0}^{\infty} \bm{\psi}_i(\tau)\bm{u}_{t-i}.
\end{equation}

Then the $H$-step generalized forecast error variance decomposition (GFEVD) is conducted, which serves as a cornerstone of the connectedness analysis. The GFEVD reflects the proportion of the forecast error variance of a series $i$ that can be attributed to shocks in series $j$, formulated as follows:
\begin{equation}
	\theta_{ij}(H)=\frac{(\Sigma(\tau))_{jj}^{-1}\sum_{h=0}^{H}((\bm{\psi}_{h}(\tau)\Sigma(\tau))_{ij})^{2}}{\sum_{h=0}^{H}(\bm{\psi}_{h}(\tau)\Sigma(\tau)\bm{\psi}_{h}^{\prime}(\tau))_{ii}}
	\end{equation}
For comparability purposes, the decomposition values $\tilde{\theta}_{ij}(H)$ undergo normalization by their respective row sums:
\begin{equation}
	\tilde{\theta}_{ij}(H) = \frac{\theta_{ij}(H)}{\sum_{j=1}^{n} \theta_{ij}(H)}
\end{equation}
This adjustment ensures that the sum of influences for each series equals one, thus indicating the proportion of shocks in series $i$ affecting all other series $j$:
\begin{equation}
	\sum_{i=1}^{n} \tilde{\theta}_{ij}(H) = 1 \quad \text{and} \quad \sum_{j=1}^{n}\sum_{i=1}^{n} \tilde{\theta}_{ij}(H) = n
\end{equation}
Through this normalized measure, the extent to which a shock in any series impacts the entire network is encapsulated, indicating the aggregate connectedness within the system.
In a words, $\tilde{\theta}_{ij}(H)$ measures the spillover level of series $j$ to series $i$ in the quantile time domain.

Building on the frameworks established by \cite{diebold2014network} and \cite{chatziantoniou2021interest}, we proceed to compute all  quantile-based connectedness measures as follows:  
\begin{align}
	NPDC_{ij,t}(H) &= \tilde{\theta}_{ij,t}(H) - \tilde{\theta}_{ji,t}(H), \\
	TO_{i,t}(H) &= \sum_{\substack{j=1,j\neq i}}^{n} \tilde{\theta}_{ji,t}(H), \\
	FROM_{i,t}(H) &= \sum_{\substack{j=1,j\neq i}}^{n} \tilde{\theta}_{ij,t}(H), \\
	NET_{i,t}(H) &= TO_{i,t}(H) - FROM_{i,t}(H), \\
	TCI_{t}(H) &= n^{-1}\sum_{i=1}^{n} TO_{i,t}(H)
	= n^{-1}\sum_{i=1}^{n} FROM_{i,t}(H).
\end{align}
Here $NPDC_{ij,t}$ represents the net pairwise directional connectedness between series $i $ and $j$ at time $t$, indicating the difference in influence exerted by series $j$ on series $i$ and vice versa, with positive values suggesting a dominant directional influence of series $j$ over series $i$. $TO_{i,t}(H)$ is the total directional connectedness to others, quantifying the aggregate influence transmitted by series \( i \) to all other series in the network at time $t$. Conversely, $FROM_{i,t}(H)$ measures the total directional connectedness from others, encapsulating the cumulative influence received by series $i$ from all other series. $NET_{i,t}(H)$ delineates the net total directional connectedness, essentially the balance of influences, with positive values denoting a net transmitter of shocks or influence and negative values signifying a net receiver in the network structure. Lastly, $TCI_{t}(H)$ denotes the total connectedness index at time $t$, reflecting the average degree of interconnectivity and the systemic risk within the network, with higher values indicating greater overall market risk and interdependence.

We have previously concentrated on the time domain for analyzing connectedness and now move to comparable studies in the frequency domain. Employing the spectral representation of variance decompositions, as proposed by \cite{barunik2018measuring}, we delve into the frequency-based relationships in connectedness. We start with the frequency response function $\bm{\psi}(e^{-i\omega})$, and from there, define the spectral density of series $\bm{y}_t$ at frequency $\omega$. This is captured by a Fourier transform of the QVMA($\infty$) representation:
\begin{equation}
	\bm{S}_{\bm{y}}(\omega) = \sum_{h=-\infty}^{\infty} E(\bm{y}_t \bm{y}'_{t-h})e^{-i\omega h} = \boldsymbol{\psi}(e^{-i\omega h})\bm{\Sigma}_t \boldsymbol{\psi}'(e^{i\omega h}).
\end{equation}

Significantly, the frequency-based generalized forecast error variance decomposition integrates spectral density with GFEVD. Like in the time domain, it's essential to normalize the frequency GFEVD. The normalization procedure is outlined as follows:
\begin{align}
	\theta_{ij}(\omega)
	&= \frac{(\boldsymbol{\Sigma}(\tau))_{jj}^{-1}
		\left|\sum_{h=0}^{\infty}\big(\boldsymbol{\psi}(\tau)(e^{-i\omega h})\boldsymbol{\Sigma}(\tau)\big)_{ij}\right|^{2}}
	{\sum_{h=0}^{\infty}\big(\boldsymbol{\psi}(e^{-i\omega h})\boldsymbol{\Sigma}(\tau)\boldsymbol{\psi}(\tau)(e^{i\omega h})\big)_{ii}}, \\
	\tilde{\theta}_{ij}(\omega)
	&= \frac{\theta_{ij}(\omega)}{\sum_{j=1}^{n}\theta_{ij}(\omega)} .
\end{align}
Here $\tilde\theta_{ij}(\omega)$ captures the influence of a shock in the $j$th series on the spectrum of the $i$th series at frequency $\omega$, serving as a within-frequency marker. 
To gauge connectedness across various timeframes, we integrate over the frequency spectrum, $\tilde{\theta}_{ij}(d) = \int_a^b \tilde{\theta}_{ij}(\omega)d\omega$, within a certain frequency ranges defined as $d = (a, b)$: $a, b \in (-\pi, \pi)$, $a < b$, allowing us to compute connectedness measures the same as \cite{diebold2014network}’s metrics:
\begin{align}
	NPDC_{ij,t}(d) &= \tilde{\theta}_{ij,t}(d) - \tilde{\theta}_{ji,t}(d), \\
	TO_{i,t}(d) &= \sum_{\substack{j=1, j \neq i}}^{n} \tilde{\theta}_{ji,t}(d), \\
	FROM_{i,t}(d) &= \sum_{\substack{j=1, j \neq i}}^{n} \tilde{\theta}_{ij,t}(d), \\
	NET_{i,t}(d) &= TO_{i,t}(d) - FROM_{i,t}(d), \\
	TCI_{t}(d) &= n^{-1}\sum_{i=1}^{n} TO_{i,t}(d)
	= n^{-1}\sum_{i=1}^{n} FROM_{i,t}(d).
\end{align}
These measures provide a detailed assessment of directional and total connectedness within a frequency band $d=(a,b)$, thereby improving understanding of interactions in grain futures markets. Following \citet{gong2023research}, we consider three frequency bands that correspond to distinct investment horizons: the high-frequency band $d_1=(\pi/5,\pi)$ for short horizons of 1--5 days, the mid-frequency band $d_2=(\pi/20,\pi/5)$ for medium horizons of 5--20 days, and the low-frequency band $d_3=(0,\pi/20)$ for long horizons exceeding 20 days.

Correspondingly, the metrics $NPDC_{ij}(d_1)$, $TO_i(d_1)$, $FROM_i(d_1)$, $NET_i(d_1)$ and $TCI(d_1)$ reflect short-term connectedness, $NPDC_{ij}(d_2)$, $TO_i(d_2)$, $FROM_i(d_2)$, $NET_i(d_2)$ and $TCI(d_2)$ capture medium-term connectedness, and $NPDC_{ij}(d_3)$, $TO_i(d_3)$, $FROM_i(d_3)$, $NET_i(d_3)$ and $TCI(d_3)$ describe long-term connectedness across series over the corresponding horizons. In addition, the frequency-domain measures of \cite{barunik2018measuring} are additive to the time-domain measures of \cite{diebold2014network}, that is, $NPDC_{ij}(H)=\sum_d NPDC_{ij}(d)$, $TO_i(H)=\sum_d TO_i(d)$, $FROM_i(H)=\sum_d FROM_i(d)$, $NET_i(H)=\sum_d NET_i(d)$, and $TCI(H)=\sum_d TCI(d)$, where the sum is taken over $d\in\{d_1,d_2,d_3\}$. Following \cite{chatziantoniou2021interest}, we employ a QVAR with a 200-day rolling window, a lag length of one selected by BIC, and a 20-step-ahead forecast horizon.

\section{Spillover analysis}
\label{S5:Spillover analysis}

In this section, we use time domain and frequency domain connectedness methods to analyze variation in short- and long-run connectedness across conditional quantiles, from both static and dynamic perspectives.

\begin{table}[!ht]
	\centering
	\setlength{\abovecaptionskip}{0pt}
	\setlength{\belowcaptionskip}{10pt}
	\caption{Average time-domain dynamic connectedness at three conditional quantiles ($\tau=0.05, 0.50, 0.95$) over the full sample.}
	\label{Tb:Average:Dynamic:Connectedness:Full:Time}
	\resizebox{\textwidth}{!}{
		\begin{tabular*}{\textwidth}
			{@{\extracolsep{\fill}} l *{9}{d{3.2}} @{}}
			\toprule
			& \multicolumn{3}{c}{FROM}
			& \multicolumn{3}{c}{TO}
			& \multicolumn{3}{c}{NET} \\
			\cmidrule(lr){2-4}\cmidrule(lr){5-7}\cmidrule(lr){8-10}
			& \multicolumn{1}{c}{$\tau=0.05$}
			& \multicolumn{1}{c}{$\tau=0.50$}
			& \multicolumn{1}{c}{$\tau=0.95$}
			& \multicolumn{1}{c}{$\tau=0.05$}
			& \multicolumn{1}{c}{$\tau=0.50$}
			& \multicolumn{1}{c}{$\tau=0.95$}
			& \multicolumn{1}{c}{$\tau=0.05$}
			& \multicolumn{1}{c}{$\tau=0.50$}
			& \multicolumn{1}{c}{$\tau=0.95$}\\
			\midrule
			BRms & 94.14 & 71.51 & 95.45 & 108.68 & 111.87 &  92.77 &  14.54 &  40.36 &  -2.68 \\
			BRs  & 94.48 & 66.98 & 95.46 & 102.92 &  98.06 &  88.07 &   8.45 &  31.07 &  -7.39 \\
			CNs  & 94.67 & 30.64 & 94.94 &  91.09 &   9.00 &  97.73 &  -3.58 & -21.64 &   2.78 \\
			ZAs  & 95.04 & 77.40 & 95.55 &  91.18 &  39.72 &  90.60 &  -3.86 & -37.69 &  -4.94 \\
			ZAcs & 95.28 & 77.44 & 95.96 &  88.90 &  47.92 &  78.96 &  -6.38 & -29.51 & -16.99 \\
			ARs  & 94.44 & 65.28 & 94.95 &  98.29 &  82.65 & 101.56 &   3.85 &  17.37 &   6.61 \\
			USs  & 94.54 & 67.36 & 94.97 & 102.32 &  92.24 &  98.25 &   7.78 &  24.88 &   3.27 \\
			BRc  & 94.33 & 48.45 & 95.16 & 103.06 &  40.60 &  95.07 &   8.73 &  -7.85 &  -0.10 \\
			INm  & 94.04 & 15.45 & 95.60 &  97.25 &  13.89 &  83.70 &   3.21 &  -1.56 & -11.91 \\
			CNc  & 94.46 & 43.00 & 95.59 &  94.91 &  19.09 &  89.07 &   0.46 & -23.91 &  -6.52 \\
			ZAwm & 95.12 & 75.49 & 95.45 &  93.92 &  61.86 &  93.98 &  -1.21 & -13.64 &  -1.47 \\
			ZAym & 95.17 & 77.50 & 95.46 &  92.21 &  67.43 &  93.77 &  -2.96 & -10.07 &  -1.69 \\
			ZAc  & 95.24 & 77.31 & 95.73 &  89.28 &  70.52 &  87.41 &  -5.95 &  -6.79 &  -8.32 \\
			ARm  & 94.32 & 67.38 & 95.10 & 103.02 &  92.83 &  99.56 &   8.70 &  25.44 &   4.46 \\
			USc  & 95.00 & 55.86 & 95.61 &  87.11 &  70.06 &  87.45 &  -7.89 &  14.20 &  -8.16 \\
			ZAw  & 94.96 & 64.21 & 95.46 &  91.30 &  36.72 &  89.61 &  -3.65 & -27.49 &  -5.85 \\
			ARw  & 94.65 & 63.73 & 94.98 &  99.11 &  79.20 & 101.88 &   4.47 &  15.47 &   6.90 \\
			BSw  & 94.90 & 64.11 & 94.50 &  91.82 &  77.21 & 111.76 &  -3.08 &  13.09 &  17.26 \\
			USw  & 95.21 & 64.09 & 94.90 &  86.86 &  77.16 & 101.30 &  -8.35 &  13.07 &   6.40 \\
			UAw  & 94.66 & 61.39 & 94.69 &  97.71 &  72.33 & 109.18 &   3.05 &  10.94 &  14.49 \\
			CNr  & 94.42 & 15.41 & 94.51 &  81.88 &  10.49 & 108.35 & -12.53 &  -4.91 &  13.84 \\
			USr  & 94.91 & 35.67 & 95.02 &  86.70 &  23.80 & 100.05 &  -8.21 & -11.87 &   5.04 \\
			INb  & 93.99 & 18.70 & 95.38 &  98.41 &   9.73 &  90.34 &   4.42 &  -8.97 &  -5.04 \\
			\bottomrule
	\end{tabular*}}
	\begin{flushleft}
		\footnotesize
		\justifying Notes: Estimates are based on a quantile VAR with lag length 1 selected by BIC and a 20-step-ahead forecast horizon. \texttt{FROM} denotes the total directional return connectedness received from other grain futures. \texttt{TO} denotes the total directional return connectedness transmitted to other grain futures. \texttt{NET} is the difference (\texttt{TO} $-$ \texttt{FROM}), i.e., the net total directional return connectedness. 
	\end{flushleft} 
\end{table}

\subsection{Static spillover effects of returns in grain futures markets}
\subsubsection{Average dynamic connectedness measures in the time domain}

Table~\ref{Tb:Average:Dynamic:Connectedness:Full:Time} reports average dynamic connectedness in the time domain at three conditional quantiles over the full sample. Under both normal ($\tau=0.50$) and tail ($\tau=0.05, 0.95$) market states, each grain explains a larger share of its own forecast error variance than that attributed to others.\footnote{For compactness, Table~\ref{Tb:Average:Dynamic:Connectedness:Full:Time} omits the pairwise connectedness between any two grains and the total connectedness index, though the underlying calculations include them.} The gap between own and cross market contributions narrows in the tails, indicating stronger transmission under stress. 
We also document stronger system-level connectedness under extreme market conditions. The TCI is 56.71\% at $\tau=0.50$ and rises to 94.69\% and 95.24\% at $\tau=0.05$ and $\tau=0.95$, respectively. In both the extreme bearish state ($\tau=0.05$) and the extreme bullish state ($\tau=0.95$), FROM and TO directional return connectedness at the system level are higher than under normal conditions for the vast majority of grains. In particular, FROM for most grains clusters around 95\%, and most TO values exceed 90\%, with ZAcs the lowest TO in the right tail ($\tau=0.95$) at 78.96\%. 
At the median quantile ($\tau=0.50$), the five smallest TO spillovers are CNs, INb, CNr, INm, and CNc, at 9.00\%, 9.73\%, 10.49\%, 13.89\%, and 19.09\%, respectively. The five smallest FROM spillovers are CNr, INm, INb, CNs, and USr, at 15.41\%, 15.45\%, 18.70\%, 30.64\%, and 35.67\%, respectively. These patterns indicate that, under normal conditions, Chinese and Indian grain futures are relatively weak transmitters and receivers in the information network, consistent with the correlation patterns for INm, INb, and CNr in Fig.~\ref{Fig:correlation}.

Across market states, the identities of net receivers and transmitters are broadly consistent while magnitudes vary. Under extreme bearish conditions the main net receivers are CNr at -12.53\%, USw at -8.35\%, and USr at -8.21\%, whereas BRms at 14.54\%, BRc at 8.73\%, and ARm at 8.70\% are the leading net transmitters. Under extreme bullish conditions ZAcs at -16.99\%, INm at -11.91\%, and ZAc at -8.32\% absorb shocks, while BSw at 17.26\% and UAw at 14.49\% distribute them. Under normal conditions South African futures dominate the receiver side, with ZAs at -37.69\%, ZAcs at -29.51\%, and ZAw at -27.49\%, and the principal transmitters are BRms at 40.36\%, BRs at 31.07\%, ARm at 25.44\%, and USs at 24.88\%. Taken together, South African grain futures are predominantly net risk takers across quantiles, whereas Argentine grain futures, U.S. soybean, and Ukrainian wheat are persistent sources of risk spillovers. Contracts linked to the same grain type display stronger pairwise connectedness, most clearly within wheat, and spatial proximity also reinforces ties, with Black Sea and Ukrainian wheat showing markedly higher pairwise connectedness than most cross grain links.

Table~\ref{Tb:Average:Dynamic:Connectedness:Time:pre:post} provides average time-domain dynamic connectedness at several conditional quantiles for the periods before and after the conflict between Russia and Ukraine. Under normal market conditions, the TCI declines from 61.96\% before the conflict to 54.49\% after the conflict, indicating weaker interdependence across grain futures, whereas the left tail and right tail TCIs remain broadly comparable to their pre-conflict values.

Additionally, under extreme bearish conditions, the pre-outbreak pattern shows INm at -18.89\%, INb at -18.46\%, and ZAc at -16.54\% as representative net receivers, while ARw at 13.17\%, BRms at 10.70\%, and UAw at 9.78\% are representative net transmitters. After the outbreak, the receiver side shifts toward USc at -11.92\%, USr at -11.36\%, and CNr at -11.34\%, whereas transmission strengthens for BRms at 24.64\%, ARm at 13.67\%, and BRc at 9.41\%. 
The overlap of USc, USr, USw, BSw, CNs, ZAym, and ZAc across periods highlights persistent vulnerability as net receivers, whereas BRms, ARm, BRc, ARs, BRs, USs, ZAs, and UAw persist as net transmitters. From before to after the outbreak, shocks received by U.S. corn, wheat, and rice increased, the Indian market shifted from net receiver to net transmitter, Argentine wheat became a net receiver, Brazilian soybean and corn remained the dominant transmitters with strengthening intensity, Black Sea wheat remained a net receiver with larger magnitudes, and Ukrainian wheat remained a net transmitter but its transmission capacity declined from  9.78\% to 0.18\%.

Under extreme bullish market conditions, the composition of net receivers and net transmitters changes across periods but the broad pattern is clear. Before the outbreak, representative net receivers include INm at -30.41\%, ZAw at -22.08\%, ZAcs at -18.20\%, USc at -12.92\%, ZAc at -12.88\%, and CNs at -10.23\%, while the main net transmitters include UAw at 49.92\%, ARs at 16.37\%, ARm at 12.73\%, CNr at 8.61\%, and BRc at 7.41\%. After the outbreak, the receiver side is led by ZAcs at -16.77\%, INb at -13.53\%, BRs at -13.05\%, USc at -11.10\%, BRms at -9.72\%, and INm at -9.19\%, whereas the transmitter side is led by CNr at 35.36\%, USr at 18.61\%, ZAwm at 18.21\%, BSw at 13.95\%, and ZAym at 7.47\%. Several grain futures persist as receivers across periods, including INm, ZAw, ZAcs, USc, ZAc, CNs, BRms, and BRs, while ARs, BRc, CNr, USr, ZAwm, USs, and ZAym consistently act as transmitters. Unlike the left tail, Brazilian soybean appears as a receiver in the pre-outbreak right tail. From the pre- to post-outbreak period, the receiver role weakens for Indian maize, South African wheat, South African soybeans, U.S. corn, South African maize, and Chinese soybeans. U.S., Argentine, and Black Sea wheat shift from receivers to transmitters after the outbreak, with Black Sea wheat moving from -0.49\% to 13.95\%, which underscores the impact of the Russia and Ukraine conflict. By contrast, South African soybeans and Chinese corn move from transmitters before the conflict to receivers after the conflict. It is especially notable that Ukrainian wheat, a direct bellwether of the conflict, switches from the most dominant transmitter at 49.92\% before the outbreak to a receiver at -0.67\% afterward.

Regarding the median quantile reported in Table~\ref{Tb:Average:Dynamic:Connectedness:Time:pre:post}, the information transmission mechanism differs from that observed at the lower and upper quantiles across the periods before and after the outbreak. Before the outbreak, the main net receivers of shocks were ZAcs at -42.16\%, ZAs at -41.92\%, ZAc at -35.85\%, and ZAw at -34.37\%, while the leading net transmitters were ARm at 48.76\%, USs at 47.67\%, and BRms at 46.91\%. 
These patterns indicate that South African grain futures were the predominant risk absorbers, followed by China's corn, soybean, and rice futures, Brazil's corn futures, and U.S. rice futures, with India's grain futures absorbing the least. Conversely, all Argentine grain futures were chief net transmitters of shocks, with Argentine maize the most dominant and U.S. soybean futures next, while Black Sea wheat, Ukrainian wheat, Santos soybean futures, and U.S. corn and wheat futures also served as net transmitters.  After the outbreak, the receiver side was led by ZAs at -33.26\%, ZAcs at -30.51\%, ZAw at -30.09\%, and CNc at -26.81\%, and the transmitter side by BRs at 48.07\%, BRms at 42.04\%, and ARm at 24.33\%. Hence, under normal market conditions, net transmitters and net receivers are largely consistent across the two periods, except that U.S. corn shifts from a net transmitter at 31.79\% before the conflict to a net receiver at -0.71\% after the conflict. Net spillover intensities also differ between periods. From pre- to post-outbreak, transmission strength declines for all transmitters except BRs, with especially large reductions for Black Sea wheat from 28.32\% to 4.24\% and for Ukrainian wheat from 20.86\% to 2.74\%, which makes them among the weakest transmitters after the conflict.

\begin{figure}[!ht]
	\centering
	\includegraphics[width=14cm]{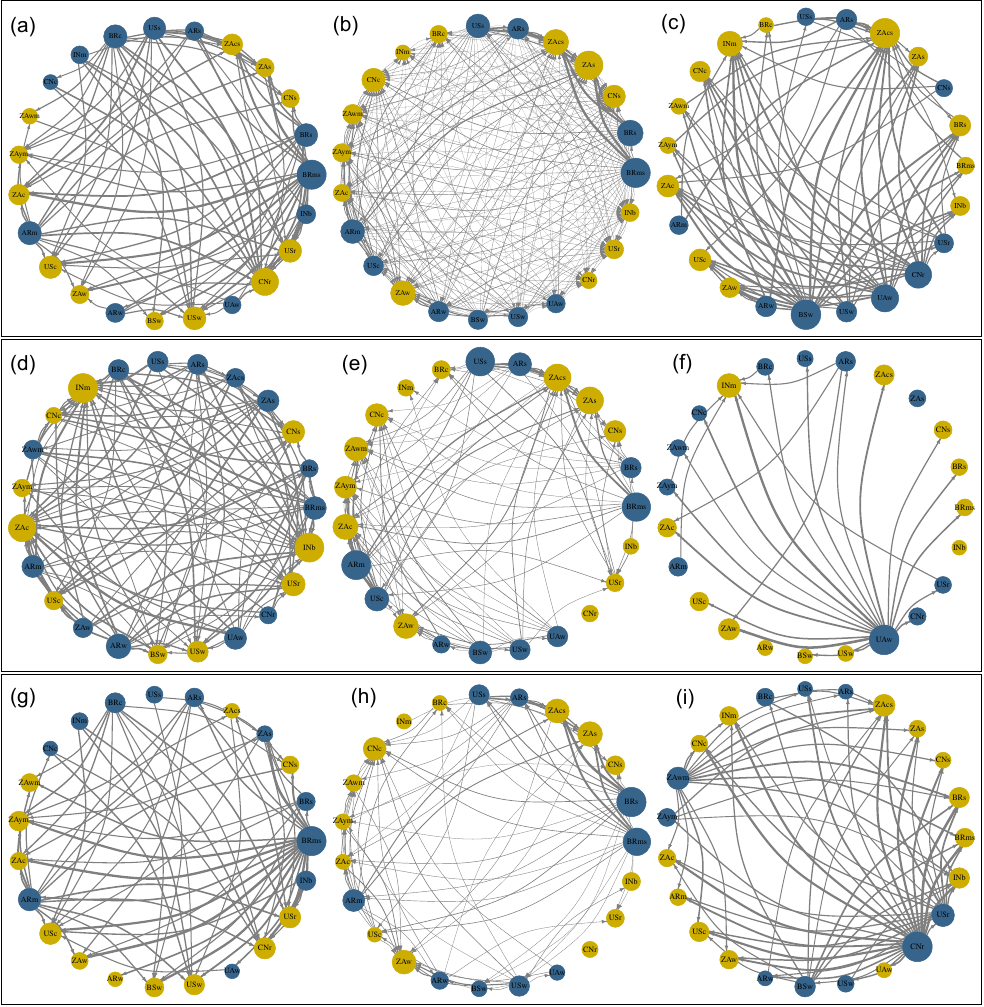}
	\caption{Network connectedness in the time domain. The upper, middle, and lower panels correspond to the full sample, the pre-outbreak subsample, and the post-outbreak subsample, respectively. Subfigures (a), (d), and (g) represent the lower tail ($\tau=0.05$). Subfigures (b), (e), and (h) depict the median ($\tau=0.50$). Subfigures (c), (f), and (i) depict the upper tail ($\tau=0.95$). Node size reflects the magnitude of spillovers transmitted to others. The direction of arrows indicates the flow of spillovers from transmitters to receivers. Arrow width, from thin to thick, indicates spillover intensity between grain futures.}
	\label{Fig:Network:time:domain}
\end{figure}

To visualize the sources, directions, and magnitudes of spillovers, Fig.~\ref{Fig:Network:time:domain} presents the network of net pairwise directional connectedness for the grain futures system at the median ($\tau=0.50$) and the lower and upper tails ($\tau=0.05,\ 0.95$). Blue nodes denote net transmitters of return shocks, whereas yellow nodes denote net receivers. Node size reflects the magnitude of net total directional connectedness, and edge thickness reflects the strength of pairwise spillovers.
Consistent with \citet{ando2022quantile} and \citet{lei2024nexus}, pairwise linkages among grain futures tighten in the tails relative to normal conditions, which aligns with stronger information transmission and more homogeneous investor behavior under stress \citep{ding2023time}. Fig.~\ref{Fig:Network:time:domain} also highlights role switching between net transmitters and receivers across market states and across the pre- and post-conflict subsamples. For instance, at the upper tail ($\tau=0.95$), Ukrainian wheat shifts from the dominant transmitter in the pre-outbreak period to a net receiver in the post-outbreak period.

\begin{figure}[!ht]
	\centering
	\includegraphics[width=12cm]{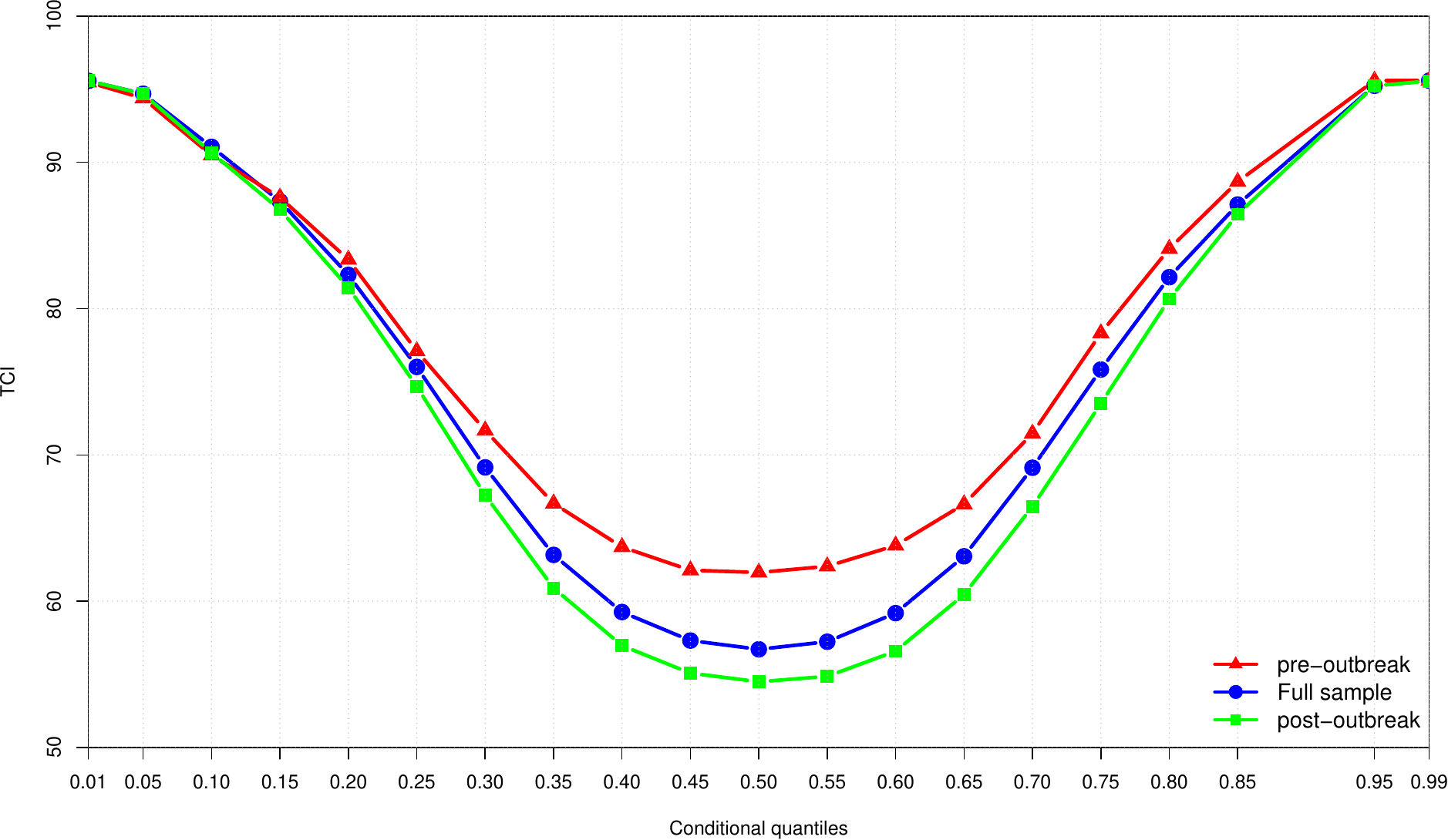}
	\caption{TCI across quantiles for the full sample and the pre- and post-outbreak subsamples. Note: Estimates are based on a 200-day rolling-window QVAR with a lag length of one selected by BIC and a 20-step-ahead generalized FEVD.}
	\label{Fig:VaryingTCI:quantiles}
\end{figure}

To examine how quantile selection affects spillovers, Fig.~\ref{Fig:VaryingTCI:quantiles} plots the TCI at several conditional quantiles for the full sample and for the pre- and post-outbreak subsamples. The TCI exhibits a distinct U-shape across quantiles, with values lowest around the median and approaching 95\% in the lower and upper tails. At every quantile the pre-outbreak TCI exceeds the post-outbreak TCI, and the gap widens as quantiles move from the tails toward the median. These patterns indicate that spillovers are most pronounced in the tails and support the use of a quantile VAR framework.

\begin{table}[!ht]
	\centering
	\setlength{\abovecaptionskip}{0pt}
	\setlength{\belowcaptionskip}{10pt}
	\caption{Average frequency-domain connectedness at $\tau=0.05,0.50,0.95$ for the full sample, with parenthesized values showing medium‐/long‐term spillovers and the rest short-term.}
	\label{Tb:Average:Dynamic:Connectedness:Full:Frequency}
	{
		\setlength{\tabcolsep}{5pt} 
		\resizebox{\textwidth}{!}{
			\begin{tabular}{l*{9}{r}}
				\toprule
				& \multicolumn{3}{c}{FROM} & \multicolumn{3}{c}{TO} & \multicolumn{3}{c}{NET}\\
				\cmidrule(lr){2-4}\cmidrule(lr){5-7}\cmidrule(lr){8-10}
				& $\tau=0.05$ & $\tau=0.50$ & $\tau=0.95$
				& $\tau=0.05$ & $\tau=0.50$ & $\tau=0.95$
				& $\tau=0.05$ & $\tau=0.50$ & $\tau=0.95$\\
				\midrule
				BRms & 73.04 & 59.81 & 69.37 & 80.74 & 92.66 & 68.03 & 7.70 & 32.85 & -1.34 \\
				& (9.76, 11.36) & (7.78, 3.92) & (13.84, 12.29) & (12.82,  15.52) & (12.82,  6.39) & (12.96,  11.54) & (3.06,  4.16) & (5.04,  2.46) & (-0.88, -0.75) \\
				BRs & 72.34 & 55.25 & 70.53 & 79.67 & 81.51 & 64.45 & 7.33 & 26.26 & -6.09 \\
				& (9.95, 12.22) & (7.80, 3.94) & (12.73, 12.26) & (10.80,  12.68) & (11.08,  5.47) & (12.38,  11.11) & (0.85,  0.46) & (3.28,  1.53) & (-0.35, -1.15) \\
				CNs & 77.54 & 25.93 & 69.48 & 75.00 & 7.65 & 70.87 & -2.53 & -18.28 & 1.39 \\
				& (8.03,  9.14) & (3.18, 1.54) & (13.52, 11.95) & (8.06,   8.16) & (0.90,  0.45) & (13.86,  13.42) & (0.03, -0.98) & (-2.27, -1.09) & (0.34,  1.47) \\
				ZAs & 67.80 & 60.09 & 63.21 & 69.51 & 32.88 & 63.20 & 1.71 & -27.21 & -0.01 \\
				& (12.39, 14.90) & (11.44, 5.87) & (16.13, 16.21) & (9.74,  11.78) & (4.52,  2.31) & (14.01,  13.30) & (-2.65, -3.12) & (-6.92, -3.56) & (-2.12, -2.91) \\
				ZAcs & 77.16 & 61.20 & 69.89 & 71.69 & 41.54 & 56.91 & -5.48 & -19.65 & -12.98 \\
				& (9.19,  8.97) & (10.75, 5.49) & (14.59, 11.54) & (8.46,   8.79) & (4.28,  2.10) & (11.22,  10.33) & (-0.73, -0.17) & (-6.47, -3.39) & (-3.37, -1.20) \\
				ARs & 66.11 & 54.49 & 69.21 & 73.37 & 67.65 & 74.57 & 7.26 & 13.16 & 5.36 \\
				& (14.67, 13.72) & (7.15, 3.64) & (13.51, 12.29) & (11.29,  13.59) & (9.97,  5.03) & (14.10,  12.76) & (-3.37, -0.13) & (2.82,  1.38) & (0.59,  0.47) \\
				USs & 75.07 & 58.51 & 71.19 & 78.56 & 76.69 & 70.37 & 3.49 & 18.18 & -0.81 \\
				& (9.12, 10.38) & (5.94, 2.91) & (12.81, 10.96) & (10.61,  13.10) & (10.36,  5.20) & (14.94,  12.89) & (1.49,  2.72) & (4.42,  2.28) & (2.14,  1.92) \\
				BRc & 71.07 & 37.32 & 66.88 & 80.17 & 32.00 & 70.23 & 9.10 & -5.32 & 3.36 \\
				& (10.66, 12.62) & (7.30, 3.82) & (15.00, 13.28) & (10.77,  12.26) & (5.70,  2.90) & (13.01,  11.96) & (0.11, -0.37) & (-1.61, -0.92) & (-1.99, -1.32) \\
				INm & 82.12 & 13.58 & 68.72 & 82.54 & 11.85 & 58.89 & 0.42 & -1.73 & -9.83 \\
				& (6.11,  5.88) & (1.25, 0.62) & (13.60, 13.30) & (7.90,   7.56) & (1.35,  0.68) & (12.78,  11.94) & (1.78,  1.69) & (0.10,  0.06) & (-0.82, -1.36) \\
				CNc & 75.94 & 33.45 & 68.06 & 77.86 & 15.26 & 67.08 & 1.92 & -18.19 & -0.99 \\
				& (9.36,  9.17) & (6.36, 3.18) & (14.53, 12.98) & (8.51,   8.85) & (2.54,  1.28) & (11.56,  10.39) & (-0.85, -0.32) & (-3.82, -1.90) & (-2.97, -2.59) \\
				ZAwm & 73.77 & 62.15 & 70.34 & 76.20 & 51.02 & 66.21 & 2.43 & -11.13 & -4.14 \\
				& (11.67,  9.70) & (8.87, 4.48) & (13.16, 11.95) & (8.80,   8.93) & (7.19,  3.65) & (13.94,  13.57) & (-2.87, -0.77) & (-1.68, -0.82) & (0.78,  1.62) \\
				ZAym & 69.82 & 63.28 & 70.49 & 74.18 & 55.91 & 62.83 & 4.37 & -7.37 & -7.65 \\
				& (12.03, 13.36) & (9.45, 4.76) & (13.43, 11.58) & (8.74,   9.10) & (7.66,  3.86) & (15.62,  14.88) & (-3.29, -4.25) & (-1.79, -0.91) & (2.19,  3.29) \\
				ZAc & 73.95 & 61.13 & 67.66 & 71.75 & 58.67 & 62.62 & -2.20 & -2.47 & -5.04 \\
				& (9.43, 11.88) & (10.70, 5.47) & (15.01, 13.07) & (8.46,   9.00) & (7.89,  3.96) & (12.72,  12.01) & (-0.96, -2.88) & (-2.81, -1.51) & (-2.30, -1.06) \\
				ARm & 72.16 & 57.51 & 68.25 & 79.46 & 75.41 & 72.74 & 7.30 & 17.90 & 4.48 \\
				& (9.85, 12.32) & (6.63, 3.24) & (14.58, 12.29) & (10.97,  12.63) & (11.58,  5.83) & (14.25,  12.46) & (1.12,  0.31) & (4.95,  2.59) & (-0.32,  0.17) \\
				USc & 79.08 & 48.56 & 69.93 & 69.78 & 59.69 & 60.96 & -9.30 & 11.13 & -8.97 \\
				& (7.93,  8.06) & (4.89, 2.41) & (13.18, 12.52) & (8.46,   8.75) & (6.91,  3.46) & (13.65,  12.65) & (0.53,  0.69) & (2.02,  1.04) & (0.47,  0.13) \\
				ZAw & 68.56 & 49.22 & 65.89 & 73.49 & 30.20 & 65.29 & 4.93 & -19.02 & -0.60 \\
				& (11.43, 15.00) & (9.91, 5.08) & (14.73, 14.80) & (8.56,   9.05) & (4.32,  2.20) & (12.62,  11.68) & (-2.87, -5.95) & (-5.59, -2.88) & (-2.11, -3.12) \\
				ARw & 77.16 & 52.19 & 67.81 & 77.96 & 65.37 & 72.28 & 0.80 & 13.19 & 4.47 \\
				& (8.33,  9.17) & (7.70, 3.84) & (13.80, 13.41) & (10.17,  10.99) & (9.19,  4.64) & (15.49,  13.83) & (1.83,  1.83) & (1.49,  0.80) & (1.69,  0.42) \\
				BSw & 77.47 & 55.86 & 69.40 & 72.09 & 63.21 & 80.02 & -5.38 & 7.35 & 10.63 \\
				& (7.63,  9.82) & (5.54, 2.71) & (13.09, 12.01) & (9.16,  10.28) & (9.29,  4.70) & (17.20,  14.89) & (1.53,  0.46) & (3.75,  1.99) & (4.11,  2.88) \\
				USw & 78.90 & 53.92 & 65.82 & 68.27 & 62.97 & 73.20 & -10.62 & 9.05 & 7.38 \\
				& (7.94,  8.44) & (6.77, 3.39) & (15.00, 14.11) & (8.73,   9.48) & (9.45,  4.74) & (14.88,  13.35) & (0.79,  1.04) & (2.67,  1.35) & (-0.12, -0.76) \\
				UAw & 79.91 & 54.07 & 70.25 & 78.38 & 58.74 & 80.50 & -1.52 & 4.67 & 10.26 \\
				& (6.86,  7.94) & (4.93, 2.39) & (13.13, 11.33) & (9.06,  10.21) & (9.03,  4.56) & (15.32,  14.00) & (2.20,  2.27) & (4.10,  2.17) & (2.19,  2.67) \\
				CNr & 81.89 & 13.23 & 66.44 & 66.62 & 9.14 & 79.20 & -15.27 & -4.09 & 12.76 \\
				& (6.27,  6.33) & (1.45, 0.73) & (13.75, 14.34) & (7.61,   7.63) & (0.90,  0.45) & (15.29,  14.57) & (1.34,  1.30) & (-0.55, -0.27) & (1.54,  0.23) \\
				USr & 75.25 & 30.67 & 66.99 & 68.97 & 19.13 & 71.30 & -6.29 & -11.54 & 4.32 \\
				& (9.89,  9.82) & (3.42, 1.58) & (14.08, 13.85) & (8.14,   9.46) & (3.10,  1.58) & (14.68,  14.65) & (-1.76, -0.36) & (-0.32,  0.00) & (0.60,  0.79) \\
				INb & 82.52 & 16.12 & 70.17 & 82.36 & 8.37 & 64.21 & -0.16 & -7.75 & -5.97 \\
				& (5.47,  6.05) & (1.72, 0.86) & (12.67, 12.61) & (8.17,   8.43) & (0.90,  0.46) & (13.37,  12.77) & (2.69,  2.38) & (-0.82, -0.41) & (0.70,  0.15) \\
				\bottomrule
	\end{tabular}}}
\end{table}

\subsubsection{Average dynamic connectedness measures in the frequency domain}

After examining static quantile connectedness in the time domain, we next analyze its frequency-domain counterpart for BRICS and international grain futures, with results in Table~\ref{Tb:Average:Dynamic:Connectedness:Full:Frequency}. Short-term TCI values at the lower, median, and upper quantiles are 75.16\%, 46.85\%, and 68.52\%, respectively, indicating a high degree of connectedness. By contrast, medium-term TCI values are 9.30\%, 6.56\%, and 13.91\%, and long-term values are 10.27\%, 3.30\%, and 12.82\%. Spillover strength therefore declines sharply from the short to the medium and long horizons. This pattern is consistent with prior studies and is often attributed to rapid information transmission that renders cross-asset spillovers short-lived \citep{le2023price,lei2024nexus}. Under normal and extreme bullish conditions, the TCI declines monotonically from short to medium to long horizons, whereas under extreme bearish conditions the long-term TCI exceeds the medium-term TCI. Pairwise connectedness follows the same ordering. We further find that, in the pre-outbreak period, the TCI and pairwise connectedness in the left tail are higher at the medium horizon than at the long horizon. In the post-outbreak period, the pattern reverses, and in both tails TCI and pairwise connectedness are lower at the medium horizon than at the long horizon. Overall, spillovers among grains are predominantly concentrated in the short term.

With respect to NET spillovers in the left tail, Table~\ref{Tb:Average:Dynamic:Connectedness:Full:Frequency} indicates that BRms, BRs, USs, ARm, ARw, and INm are net transmitters at short-, medium-, and long-term horizons, whereas ZAcs, ZAc, and USr are net receivers at all horizons. USc, BSw, USw, UAw, CNr, and INb are receivers in the short term but become transmitters at the medium and long horizons. Conversely, ZAs, ARs, CNc, ZAwm, ZAym, and ZAw are transmitters in the short term but become receivers at the medium and long horizons. Among these, only INm, INb, and ARw exhibit markedly lower net spillover intensity in the short term than at the medium and long horizons.

As for the right tail, BRms, BRs, ZAs, ZAcs, INm, CNc, ZAc, and ZAw are consistently net receivers at all horizons, whereas CNs, ARs, ARw, BSw, UAw, CNr, and USr are net transmitters. At the median quantile ($\tau=0.50$), BRms, BRs, ARs, USs, ARm, USc, ARw, BSw, USw, and UAw are net transmitters across short-, medium-, and long-term horizons, while CNs, ZAs, ZAcs, BRc, CNc, ZAwm, ZAym, ZAc, ZAw, CNr, and INb are net receivers. Apart from INm, grains maintain the same NET sign across horizons at the median. By contrast, in the tails ($\tau=0.05$ and $\tau=0.95$) many grains switch between receiver and transmitter across horizons. These patterns point to heterogeneity in spillovers across horizons and asymmetry between the left and right tails.

Additionally, under extreme market conditions ($\tau = 0.05$ and $0.95$), short-term TCI values are higher in the post-outbreak period than in the pre-outbreak period, whereas medium-term and long-term TCI values are higher in the pre-outbreak period. Under normal market conditions, this pattern reverses across all three horizons.
Examining Table~\ref{Tb:Average:Dynamic:Connectedness:frequency:pre:post}, we find the following regularities. In the left tail ($\tau=0.05$), USs, BRc, CNc, ZAw, UAw, and USr retain the same NET sign across the short-, medium-, and long-term frequency bands in both the pre- and post-outbreak periods. In the right tail ($\tau=0.95$), BRms, BRs, CNs, ZAcs, ZAym, USc, and ARw show the same stability. At the median quantile ($\tau=0.50$), all grains except INm, ZAc, USc, and UAw maintain a constant NET sign across frequencies in both periods. Overall, the conflict between Russia and Ukraine affects tail spillovers by changing not only the sign but also the magnitude across frequencies.
Furthermore, the frequency-domain connectedness patterns reported in Tables~\ref{Tb:Average:Dynamic:Connectedness:Full:Frequency} and \ref{Tb:Average:Dynamic:Connectedness:frequency:pre:post} are visualized in the networks shown in Fig.~\ref{Fig:Network:fre}, which illustrate the differential impact of the Russia-Ukraine conflict across quantile levels and frequency horizons.

\subsection{Dynamic spillover effects of returns in grain futures markets}
\subsubsection{Dynamic total connectedness measures in the time domain}

\begin{figure}[!ht]
	\centering
	\includegraphics[width=8cm]{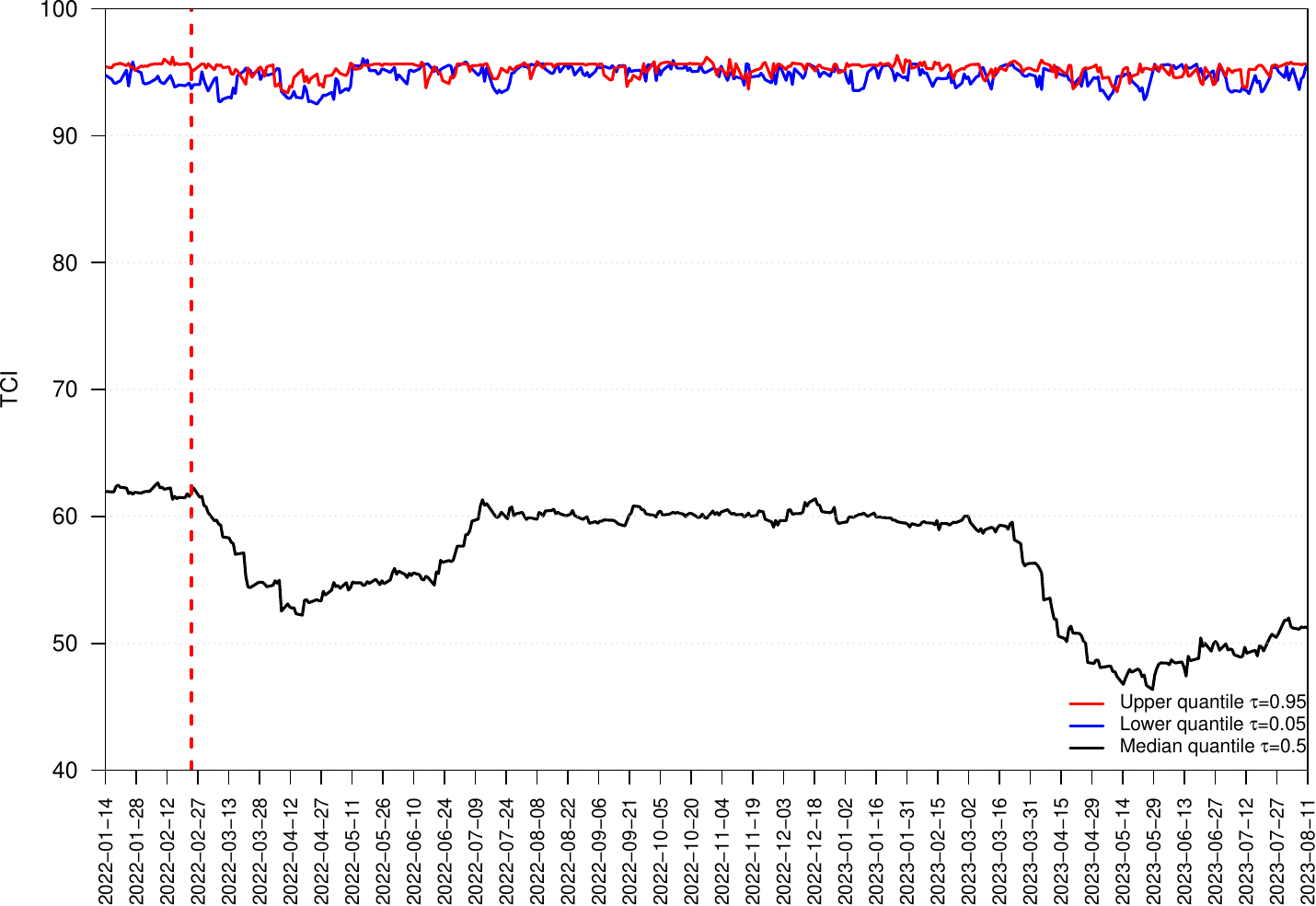}
	\includegraphics[width=8cm]{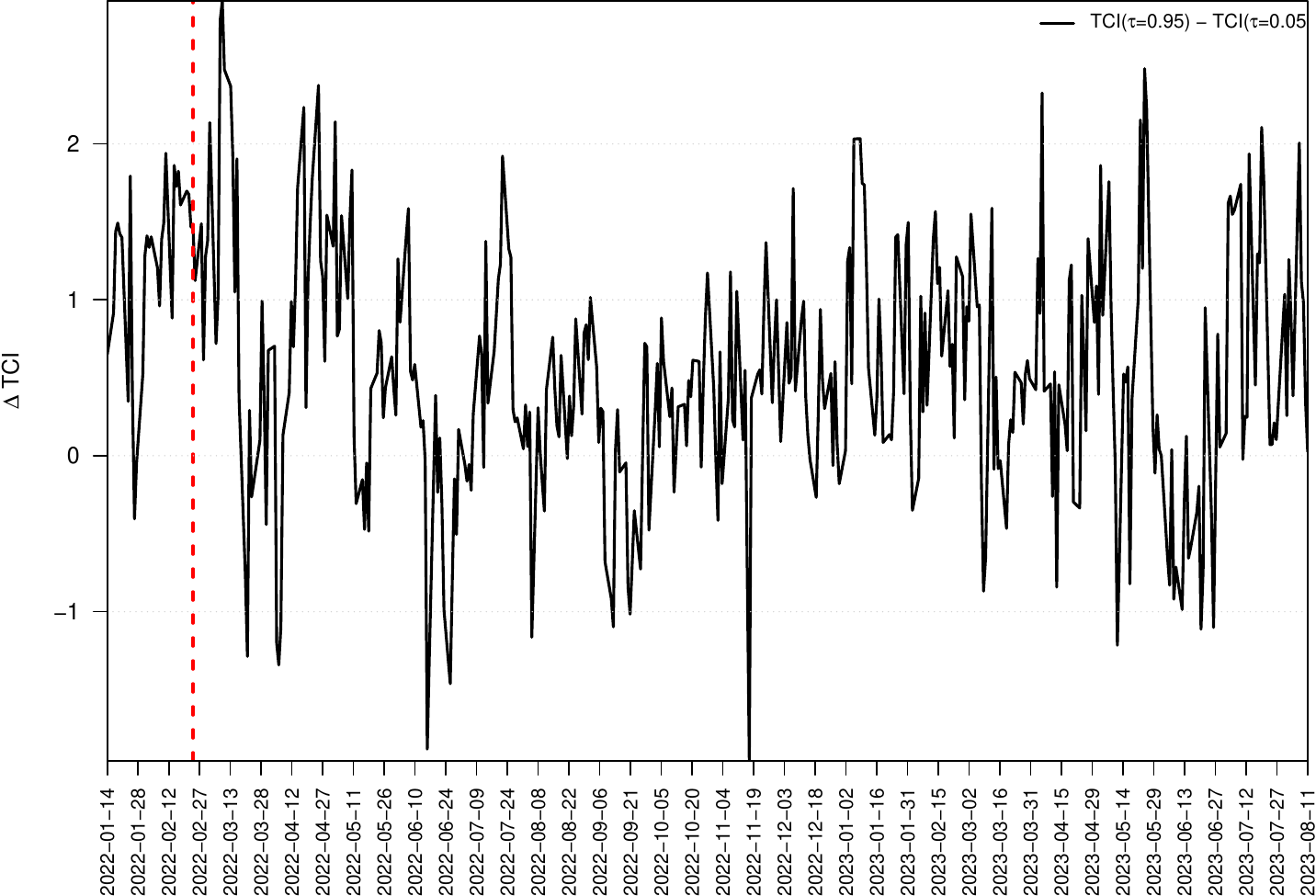}
	\caption{The left plot shows the time-varying total connectedness indices at different quantiles ($\tau=0.05, 0.50, 0.95$) over time. The right plot shows the relative tail dependence, representing the difference between the TCI at the 95th quantile and the 5th quantile. The vertical dashed line marks the start of the Russia-Ukraine conflict on February 24, 2022.}
	\label{Fig:TCI:Difference:Time:domain}
\end{figure}

We now examine time-varying spillover effects among grain futures. By conducting a rolling-window analysis at the median, lower-tail, and upper-tail quantiles, we report results in Fig.~\ref{Fig:TCI:Difference:Time:domain}. The left panel shows the TCI at the conditional median, which ranges from about 46\% to 63\%, indicating substantial time variation. The TCI is elevated early in the sample, then drops sharply at the outset of the Russia-Ukraine conflict. It rises from late April 2022 and reaches a second peak just below 60\% by mid-July 2022, remaining elevated until March 18, 2023. After that date it declines to a trough by the end of May 2023 and then recovers gradually. The first decline coincides with the conflict outbreak. The second decline around March 18, 2023 coincides with Russia's decision to shorten the BSGI renewal to 60 days, which likely increased uncertainty about future food supplies, especially for wheat and corn. The initial rise aligns with United Nations discussions with Turkey, Russia, and Ukraine on April 25 to 27, 2022 that laid the groundwork for the BSGI, and the subsequent rise is consistent with the May 17 agreement to extend the BSGI from May 18 to July 18.

The dynamic TCI in the left and right tails fluctuates around 95\%, well above the conditional median, indicating high sensitivity to both downside and upside conditions. Shortly after the onset of the Russia-Ukraine conflict, the TCI exhibits pronounced swings, consistent with a persistent and sizable impact of extreme events. The right panel of Fig.~\ref{Fig:TCI:Difference:Time:domain} illustrates the relative tail dependence of spillovers among grains, showing that tail spillovers are asymmetric and heterogeneous. Spillovers at $\tau=0.95$ are generally stronger than at $\tau=0.05$, and their trajectories differ over time. This result is consistent with \citet{ren2024dynamic}.

\subsubsection{Net total directional connectedness measures in the time domain}

\begin{figure}[!ht]
	\centering
	\includegraphics[width=13cm]{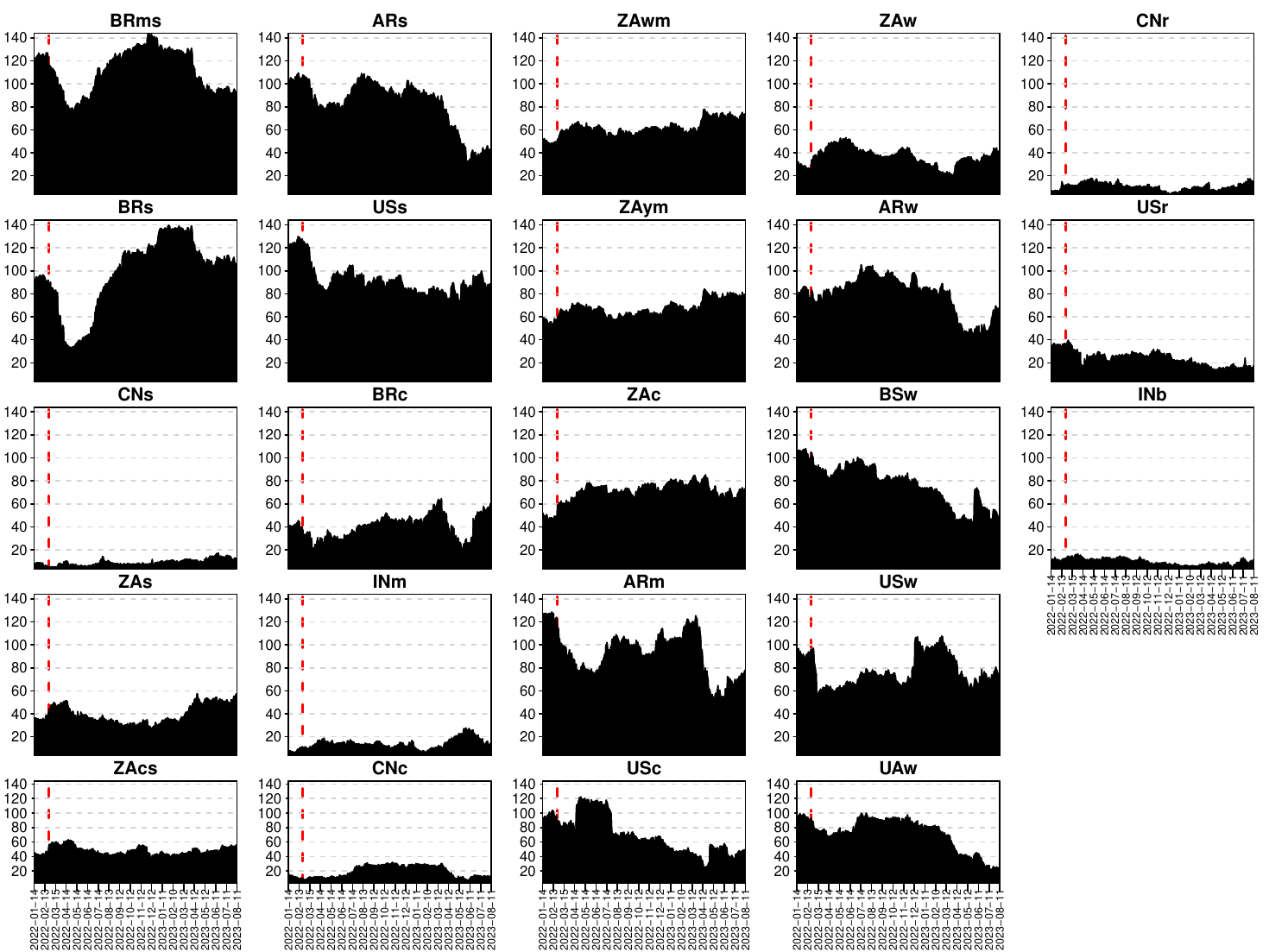}
	\caption{TO connectedness at the 0.5 quantile in the time domain.}
	\label{Fig:TO:median:Time:domain}
\end{figure}

\begin{figure}[!ht] 
	
	\centering
	\includegraphics[width=13cm]{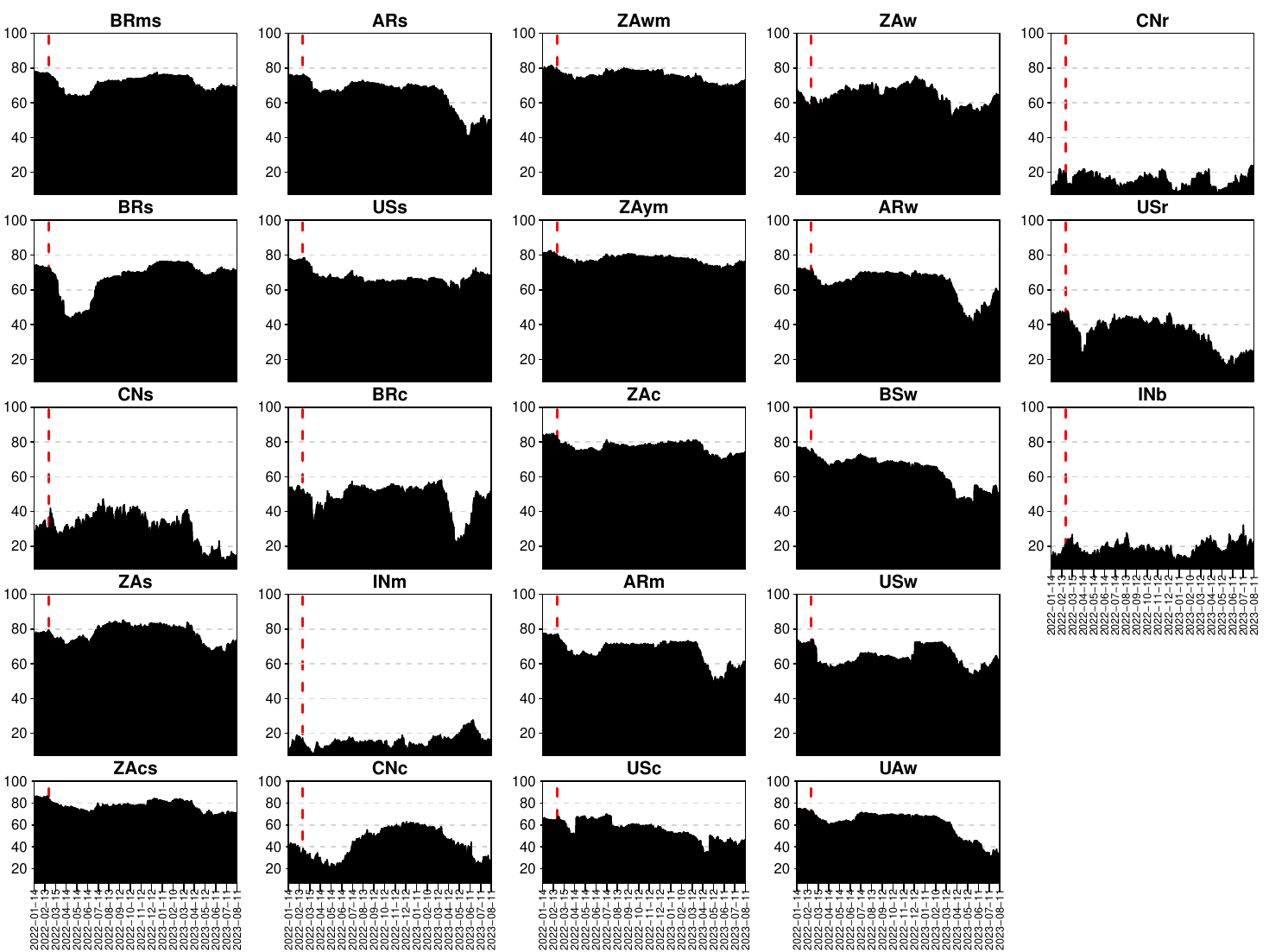}
	\caption{FROM connectedness at the 0.5 quantile in the time domain.}
	\label{Fig:FROM:median:Time:domain}
\end{figure}

\begin{figure}[!ht]
	\centering
	\includegraphics[width=13cm]{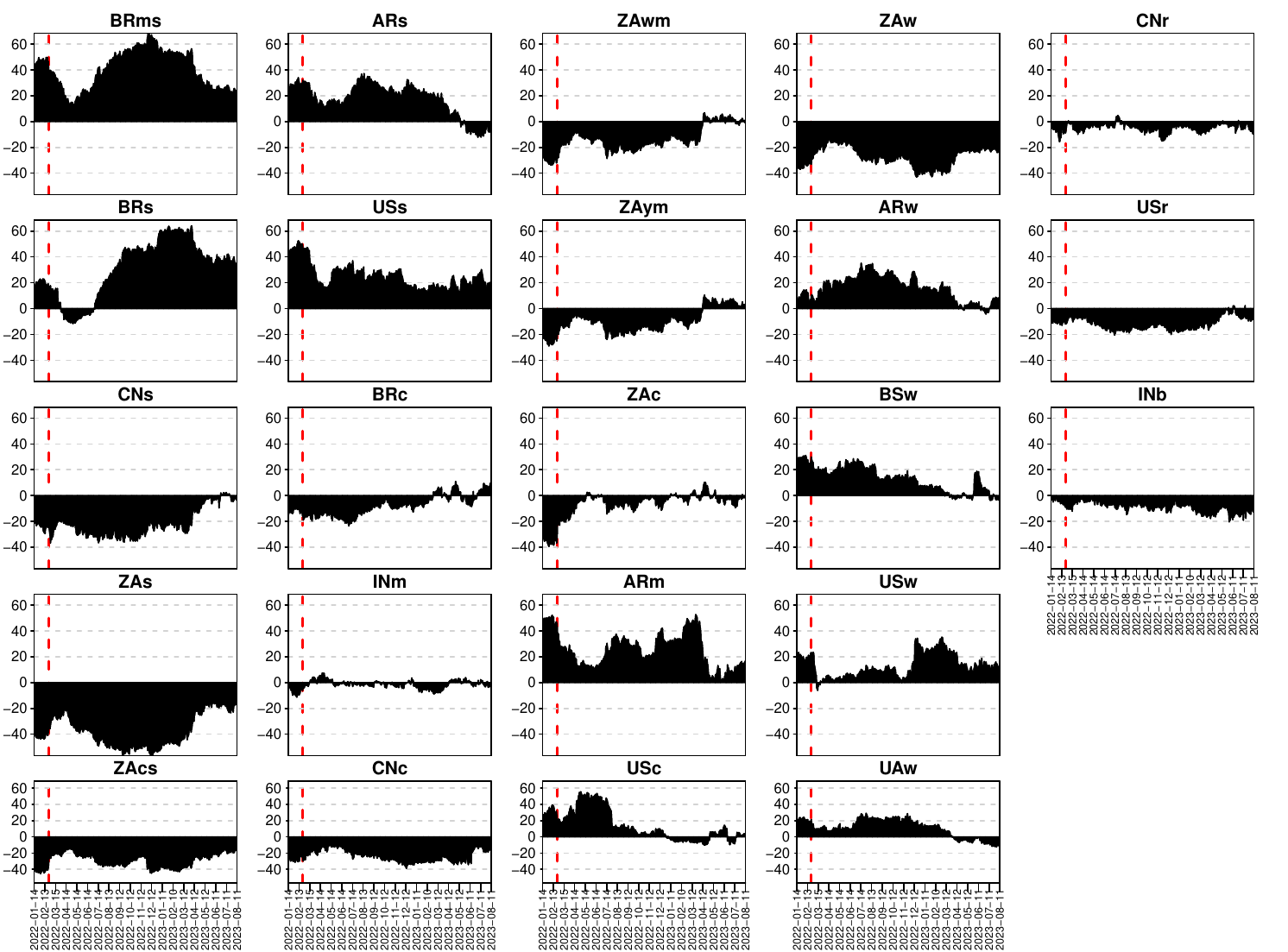}
	\caption{NET connectedness at the 0.5 quantile in the time domain.}
	\label{Fig:NET:median:Time:domain}
\end{figure}

Next, Figs.~\ref{Fig:TO:median:Time:domain}, \ref{Fig:FROM:median:Time:domain}, and \ref{Fig:NET:median:Time:domain} present time-domain TO, FROM, and NET connectedness at the median quantile, highlighting how the Russia-Ukraine conflict affects BRICS and international grain markets.
Overall, Fig.~\ref{Fig:TO:median:Time:domain} shows that BRms, ARs, BRs, USs, ZAwm, ZAym, ARw, ZAc, BSw, ARm, USw, USc, and UAw transmit relatively strong spillovers to other grain futures throughout the sample, with Brazilian soybeans the most influential and peaking near 140\%. By contrast, TO spillovers from USr, CNc, and INm are moderate and do not exceed 40\%, while CNr, CNs, and INb remain small with maxima below 20\%. The conflict's impact on TO connectedness is evident. After the outbreak, the spillovers from BRms, BRs, ARs, USs, ARm, BSw, USw, and UAw to other grain markets decline sharply, whereas TO connectedness for ZAw, ARw, and ZAc rises markedly, underscoring pronounced heterogeneity.

Correspondingly, Fig.~\ref{Fig:FROM:median:Time:domain} shows that, except for CNr, INb, and INm whose FROM connectedness averages below 20\%, the remaining grains receive strong, time-varying spillovers from other grain markets, with South African maize the largest recipient, averaging above 80\% over the sample. Unlike TO connectedness, FROM connectedness for nearly all grains declines in the early stage after the Russia-Ukraine conflict began. U.S. wheat and Brazilian soybeans react the most quickly, with both TO and FROM measures adjusting sharply around the outbreak.

Moving ahead, we present dynamic NET interlinkages among grain futures, representing the combined effect of TO minus FROM. 
We evaluate NET across investment horizons (short, medium, long) and market states (median, lower tail, upper tail), as shown in Figs.\ref{Fig:NET:median:Time:domain}, \ref{Fig:NET:median:fre}, \ref{Fig:TO:FROM:NET:left:right:tails:Time}, and \ref{Fig:TO:FROM:NET:left:right:tails:fre}. We begin with the time-domain results at the median in Fig.\ref{Fig:NET:median:Time:domain}. As defined in Section~\ref{S4:Methodology}, positive values indicate net transmitters of systemwide shocks and negative values indicate net receivers. BRms, USs, ARw, BSw, ARm, and USw consistently exhibit positive NET connectedness over the sample, acting as net transmitters. ZAw, CNr, USr, CNs, INb, ZAs, ZAcs, and CNc are net receivers throughout. ZAwm and ZAym are receivers until late March 2023 and then become transmitters, although they are receivers for most of the sample. UAw shows the opposite pattern, shifting from a transmitter early on to a receiver after late March 2023, a timing that coincides with the shortened BSGI renewal. Over the full sample its transmitter role still predominates. BRs is a transmitter before April 2022, a receiver from April to July 2022, and a transmitter again thereafter, with transmission intensifying following the implementation of the BSGI. BRc, ZAc, and INm tend to be receivers for much of the early sample. A structural shift in NET series is visible around the outbreak of the Russia-Ukraine conflict, and noticeable changes in dynamic TO, FROM, and NET follow the BSGI renewal in March 2023, as seen in Figs.~\ref{Fig:TO:median:Time:domain}, \ref{Fig:FROM:median:Time:domain}, and \ref{Fig:NET:median:Time:domain}.

The preceding discussion of TO, FROM, and NET connectedness pertains to normal market conditions. We next analyze time-varying spillovers under extreme bearish and bullish environments. Fig.~\ref{Fig:TO:FROM:NET:left:right:tails:Time} plots TO, FROM, and NET connectedness at the left ($\tau=0.05$) and right ($\tau=0.95$) tails. TO and FROM remain elevated in both tails, in contrast with the median quantile where several grains show weaker connectedness. Throughout the sample period, TO connectedness at both tails generally exceeds 100\%. In the left tail, some grains reach TO values as high as 500\%, indicating a strong spillover effect. In the right tail, certain grains show TO spillover well beyond 500\%, with Black Sea wheat the most prominent at about 900\% by late August 2022, which signals an extremely strong spillover. The impact of the Russia-Ukraine conflict is also evident. For instance, the right-tail TO of Ukrainian wheat falls from roughly 780\% to about 100\% shortly after the outbreak. FROM connectedness in both tails hovers around 95\% on average, and the right tail is generally higher than the left tail, consistent with the static connectedness results. NET connectedness in both tails changes sign over time, and no grain maintains a persistent role as a net transmitter or receiver. Even so, net roles differ across grains, with some transmitting risk more often than receiving it, while others show the opposite pattern. The asymmetry between the left and right tails is evident in the dynamics of TO, FROM, and NET connectedness.

\subsubsection{Net pairwise directional connectedness measures in the time domain}

\begin{figure}[!ht]
	\centering
	\includegraphics[width=0.98\linewidth]{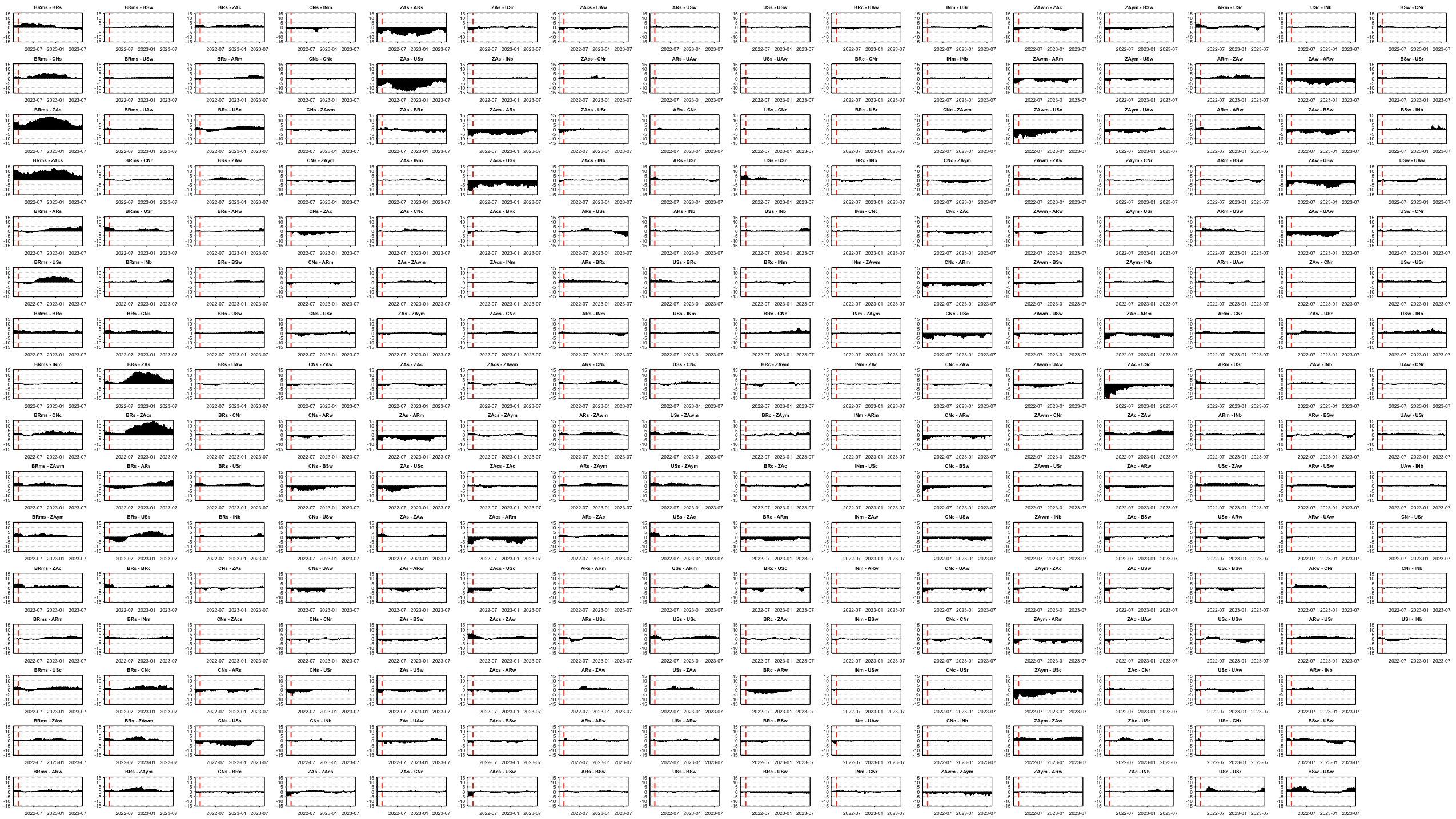}
	\caption{Dynamic net pairwise directional connectedness at the 0.5 quantile in the time domain. }
	\label{Fig:NPDC:median:time:domain}
\end{figure}

Fig.~\ref{Fig:NPDC:median:time:domain} depicts the dynamic net pairwise directional connectedness of returns, highlighting bilateral net spillovers and how each grain futures market influences the others. Brazilian soybean futures are persistently dominant transmitters to most other markets over the sample, especially to soybean contracts in other BRICS countries, which receive large spillovers from Brazil. Spillovers decline immediately after the onset of the Russia-Ukraine conflict, then rise following the BSGI announcement, peak, and subsequently ease. Specifically, before the BSGI, Brazilian soybeans are net receivers from the U.S. soybean market, and after the initiative they become consistent net transmitters.

In the net pairwise spillovers between Chinese soybean futures and other grain futures, most contracts are net transmitters to Chinese soybeans, with USs, BSw, and UAw exhibiting relatively strong spillovers for a period after the conflict began. Over the sample, South African soybean futures consistently receive strong net spillovers from Argentine soybean, U.S. soybean, and Argentine wheat. South African maize futures are persistent net receivers from U.S. corn, and South African wheat futures are consistently net receivers from Argentine, Black Sea, U.S., and Ukrainian wheat.

Before the BSGI was announced, Black Sea wheat was a net transmitter to Ukrainian wheat. After the initiative, Ukrainian wheat became the net transmitter. More broadly, U.S. grain futures, as benchmarks for international prices, generally hold a dominant position in net pairwise spillovers, influencing BRICS grain futures for most of the sample. Under normal conditions, net pairwise directional connectedness is strongest among contracts on the same grain futures, which is economically intuitive.

\subsubsection{Dynamic total connectedness measures in the frequency domain}

\begin{figure}[!ht]
	\centering
	\includegraphics[width=5.4cm]{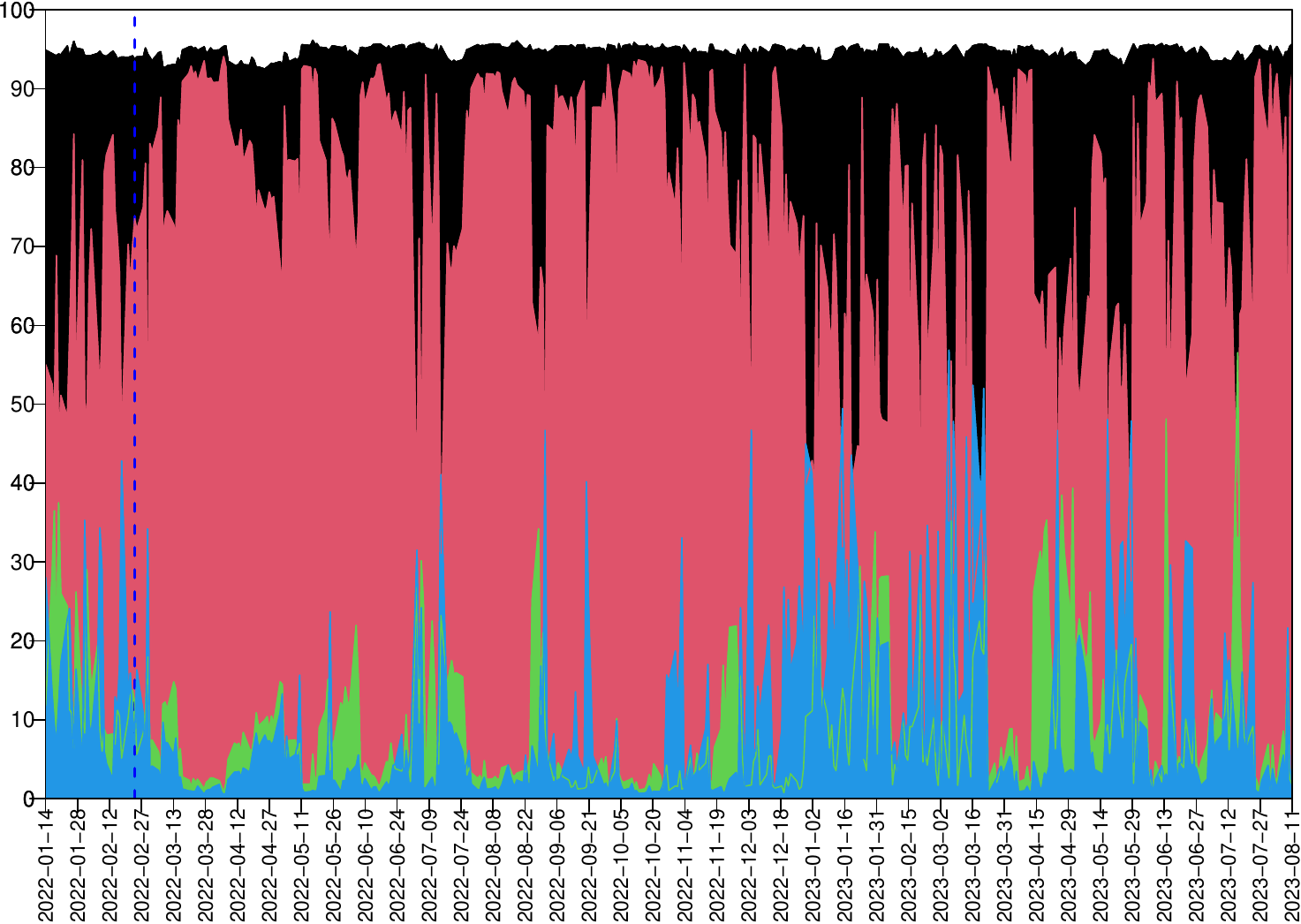}
	\includegraphics[width=5.4cm]{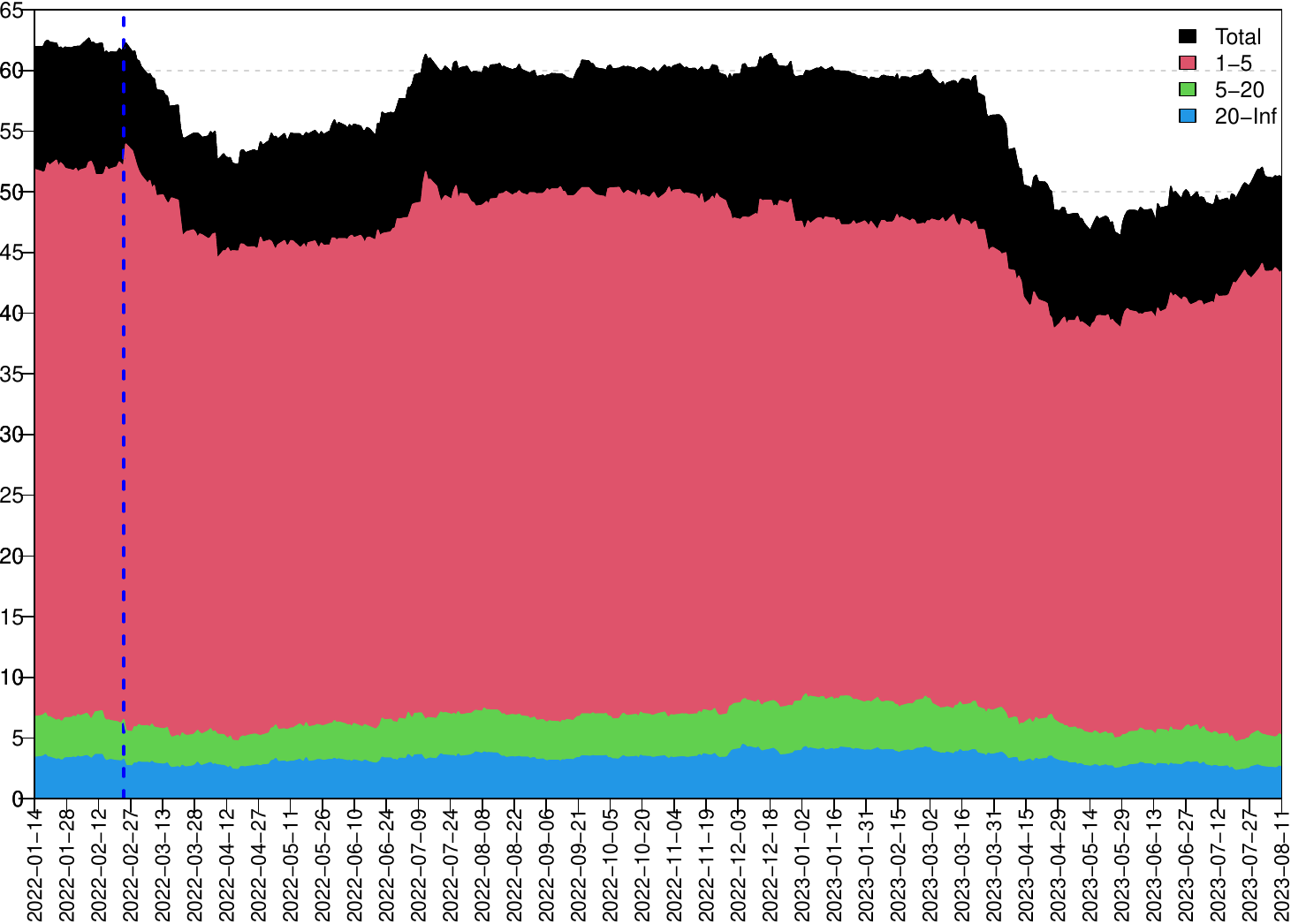}
	\includegraphics[width=5.4cm]{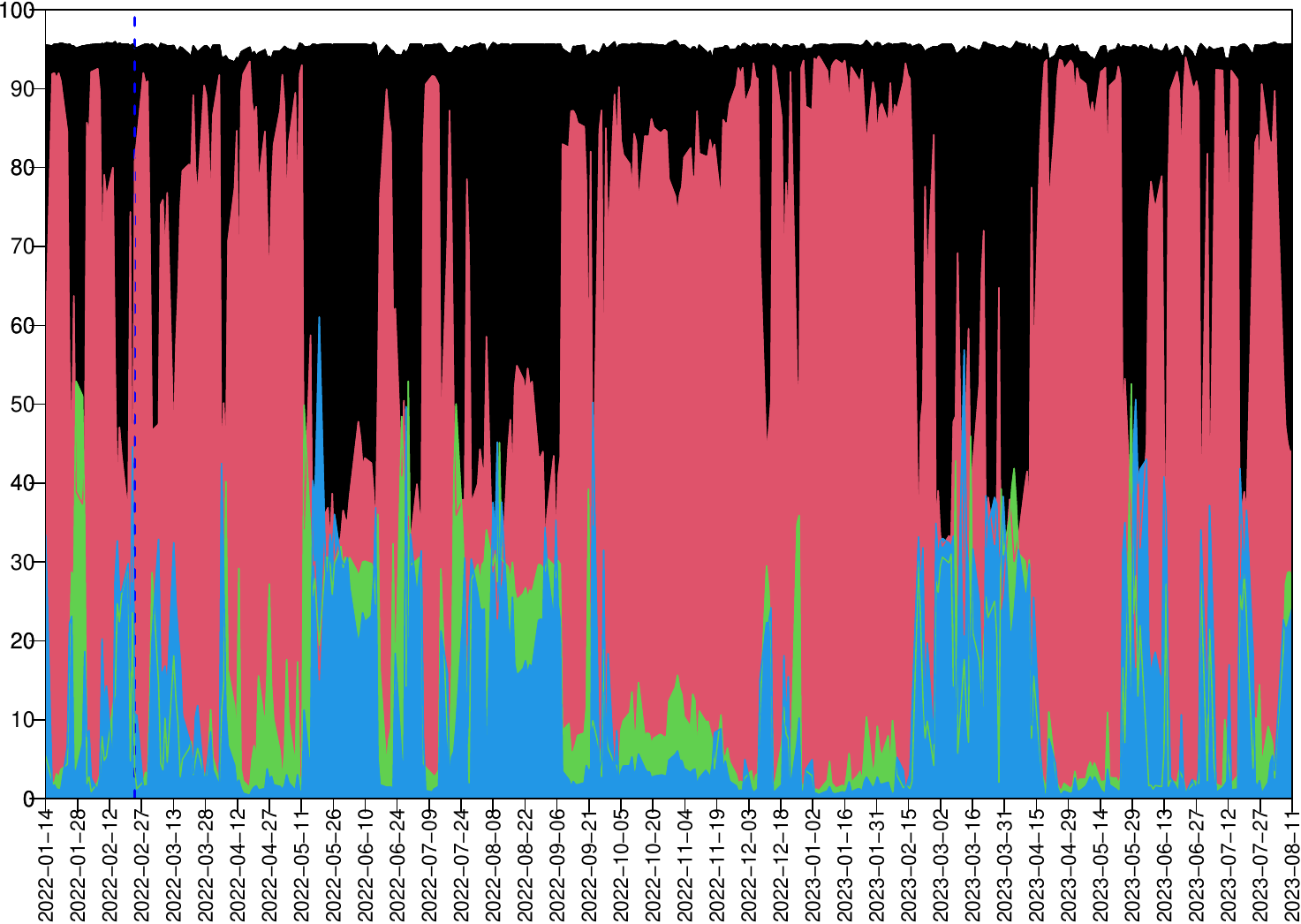}
	\caption{Short-term, medium-term, long-term, and overall dynamic total connectedness in the frequency domain. The left, middle, and right subfigures correspond to the extreme lower ($\tau = 0.05$), conditional median ($\tau = 0.5$), and extreme upper ($\tau = 0.95$) quantiles, respectively.}
	\label{Fig:TCI:fre}
\end{figure}

Fig.~\ref{Fig:TCI:fre} illustrates time-varying total connectedness at short-, medium-, and long-term frequencies. Under normal conditions, total connectedness is driven by the short-term component, and the short-term series closely mirrors the overall path. Two sharp declines in short-term connectedness coincide with the onset of the Russia-Ukraine conflict and subsequent changes to the BSGI. By contrast, medium- and long-term components contribute only modestly over the sample, with the medium term ranging from 5\% to 10\% and the long term generally below 5\%.

Under extreme market conditions, total connectedness, as at the median, is predominantly driven by short-term shocks for most of the sample. Medium- and long-term components also contribute meaningfully in specific intervals. Fig.~\ref{Fig:TCI:fre} further highlights asymmetry between the left and right tails and shows that, in the tails, medium-term connectedness does not consistently exceed long-term connectedness.

\subsubsection{Net total directional connectedness measures in the frequency domain}

\begin{figure}[!ht]
	\centering
	\includegraphics[width=0.98\textwidth]{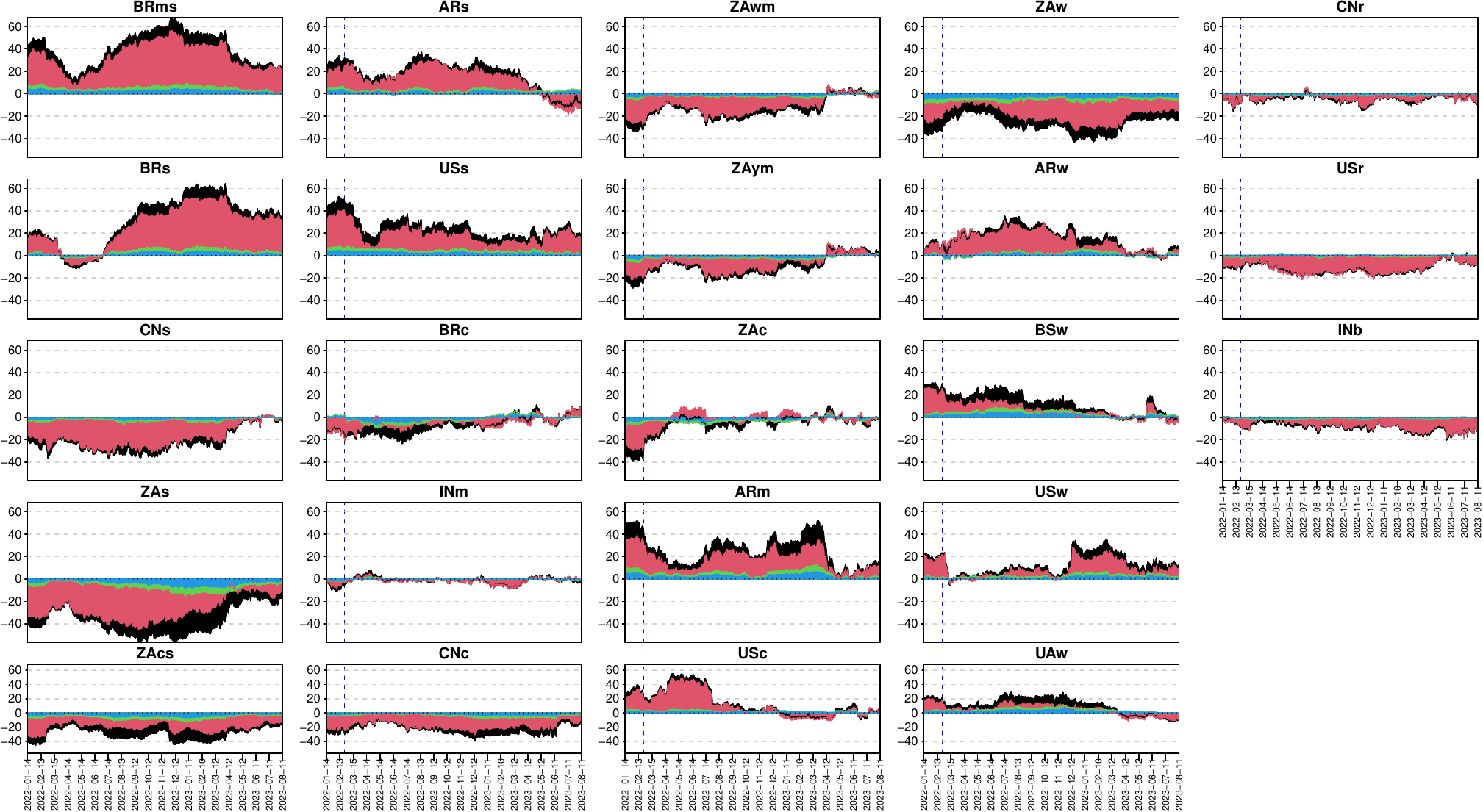}
	\caption{Time-varying NET connectedness at the conditional median ($\tau=0.50$) quantile in the frequency domain.}
	\label{Fig:NET:median:fre}
\end{figure}

Fig.~\ref{Fig:NET:median:fre} plots time-varying NET spillover indices for all grain futures at the median quantile in the frequency domain. Some grains act as persistent net transmitters across short, medium, and long horizons, such as BRms and USs, whereas others are persistent net receivers, such as USr and INb. In Fig.~\ref{Fig:NET:median:fre}, short-horizon components dominate the net effects, followed by the medium horizon, with the long horizon weakest. This implies that the net imbalance between shocks transmitted to and received from the system is smallest at long horizons. The time paths in Fig.~\ref{Fig:NET:median:fre} are influenced by the Russia-Ukraine conflict and the major policy announcements discussed earlier, and they closely track the evolution seen in Fig.~\ref{Fig:NET:median:Time:domain}.

\subsection{Sensitivity to quantiles}

\begin{figure}[!ht]
	\centering
	\includegraphics[width=12cm]{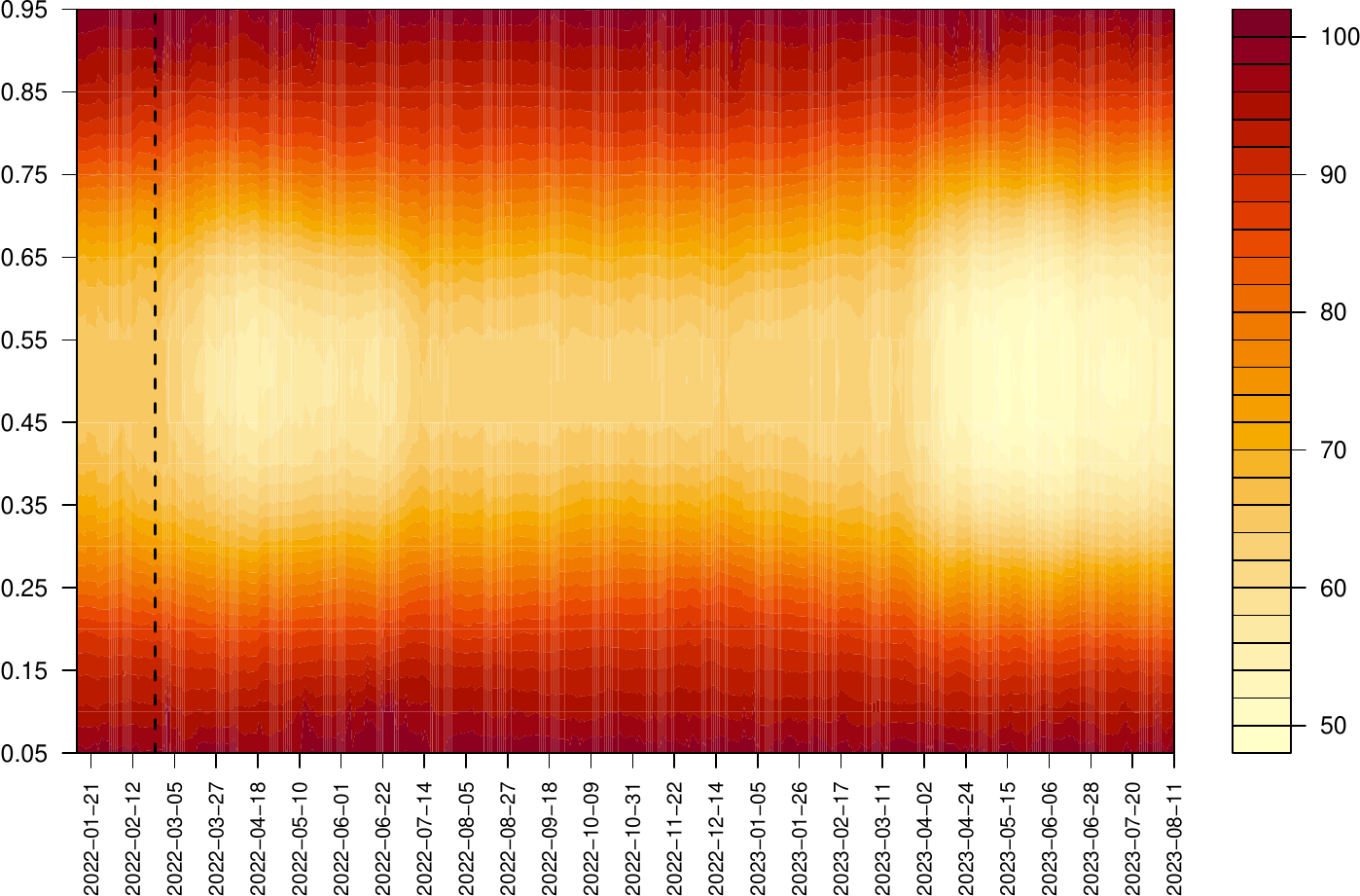}
	\caption{Dynamic total connectedness across time and quantiles in the time domain.}
	\label{Fig:TCI:quantile:time:domain}
\end{figure}

To further characterize fluctuations in total spillovers, we visualize the sensitivity of connectedness to quantiles in both the time and frequency domains using heatmaps. Fig.~\ref{Fig:TCI:quantile:time:domain} displays dynamic total connectedness over time and quantiles in the time domain, where warmer colors indicate higher connectedness. Fig.~\ref{Fig:TCI:time:quantiles:fre} depicts short-, medium-, and long-term dynamic total connectedness across time and quantiles.

In Fig.~\ref{Fig:TCI:quantile:time:domain}, total connectedness is elevated in the tails, with warmer bands below about the 25th quantile and above about the 75th quantile. Between roughly $\tau=0.30$ and $\tau=0.70$, connectedness declines after the onset of the Russia-Ukraine conflict and the subsequent shortening of the BSGI. Visually, the time-domain surface appears approximately symmetric across quantiles.

In Fig.~\ref{Fig:TCI:time:quantiles:fre}, the short-term component closely tracks the overall surface and dominates across quantiles, the medium-term component is weak at all quantiles, and the long-term component is weaker still. Unlike the time-domain heatmap, the frequency-domain panels reveal asymmetry across quantiles, particularly in the tails. Taken together, these results corroborate Fig.~\ref{Fig:TCI:fre} and indicate that, across quantiles, short-term connectedness is typically the largest.

\begin{figure}[!ht]
	\centering
	\includegraphics[width=0.98\linewidth]{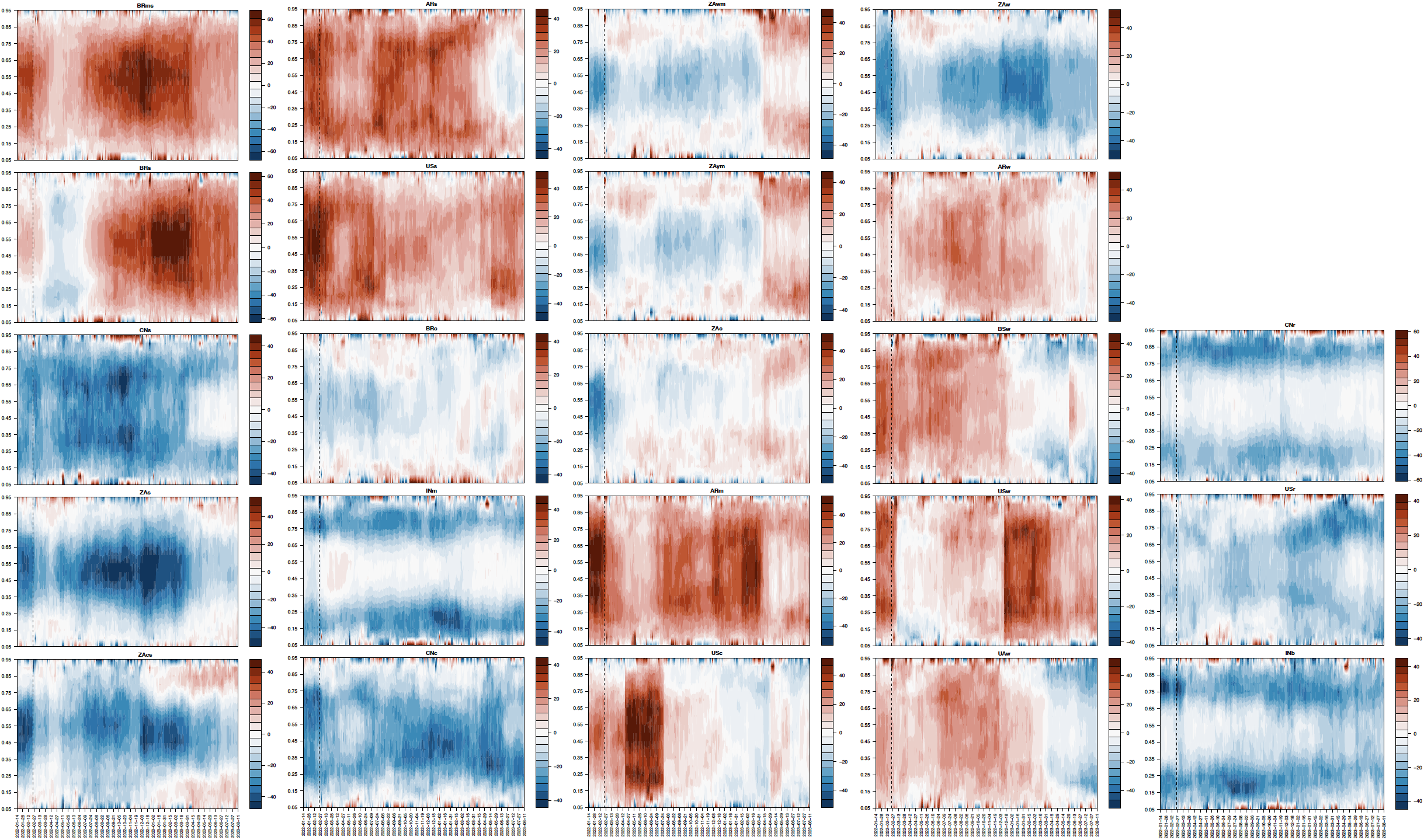}
	\caption{Net total directional connectedness across time and quantiles in the time domain.}
	\label{Fig:NET:quantile:time:domain}
\end{figure}

We also analyze the quantile sensitivity of net total directional connectedness for each grain futures market. Fig.~\ref{Fig:NET:quantile:time:domain} displays a time-domain heatmap of net directional connectedness. Fig.~\ref{Fig:NET:time:quantiles:fre} presents heatmaps of net total directional connectedness at short-, medium-, and long-term horizons. In each panel, warmer colors (red) indicate net transmitters and cooler colors (blue) indicate net receivers. Darker shades denote stronger net transmission or reception of systemwide shocks.

Referring to Fig.~\ref{Fig:NET:quantile:time:domain}, net directional connectedness is heterogeneous across quantiles and evolves over time. The visualization also highlights persistent net transmitters of systemwide shocks, including BRms, BRs, ARs, USs, ARw, ARm, and USw. After the onset of the Russia-Ukraine conflict, noticeable changes in color intensity indicate a regime shift. Many markets alternate between net-receiver and net-transmitter roles across time and quantiles, notably ZAwm, ZAym, ZAc, USc, BSw, and UAw. The March 2023 shortening of the BSGI coincides with divergent movements. ZAwm, ZAym, and ZAc deepen toward red, whereas UAw shifts toward blue. USc and BSw also show weakening net-transmitter roles following the BSGI announcement, alternating between net transmission and reception.

Correspondingly, ZAw, ZAs, CNs, INm, CNc, CNr, USr, and INb are net receivers for most of the sample, but in the tails ($\tau=0.95$ and $\tau=0.05$) several of them alternate between receiver and transmitter. Notably, the heatmap varies across quantiles even for markets that are usually net receivers. As an illustration, INm and CNr exhibit weaker net reception around the median quantiles, whereas ZAs and CNc show stronger net reception there. Overall, Fig.~\ref{Fig:NET:quantile:time:domain} indicates that net directional connectedness is highly state dependent, especially in bullish and bearish conditions.

When we examine quantile sensitivity in the frequency domain as shown in Fig.~\ref{Fig:NET:time:quantiles:fre}, the short-term panel closely mirrors the overall time-domain pattern, whereas the medium- and long-term panels diverge more visibly. For instance, at long horizons USr does not remain a net receiver across quantiles and primarily acts as a net transmitter.

\subsection{Structural breaks and sub-sample analysis}

We run a Chow test with 24 February 2022 as the break date to detect a structural change in the TCI \citep{Naifar-2025-JCommodMark}. The data are split into pre-conflict and post-conflict periods so we can see whether links among grain futures markets shift after the Russia-Ukraine conflict. Table~\ref{Tb:Structural:break} shows that the Chow test rejects the null hypothesis at the 5th, 50th, and 95th conditional quantiles, and the Wilcoxon rank-sum test confirms a significant location shift at the same quantiles. Together these findings indicate that the conflict brought a clear change in the dependence structure and risk transmission of grain futures markets, with the adjustment most evident around the median market conditions.

\begin{table}[!ht]
	\centering
	\setlength{\abovecaptionskip}{0pt}
	\setlength{\belowcaptionskip}{10pt}
	\caption{Structural breaks and sub-sample analyses for the 5th, 50th, and 95th conditional quantiles.}
	\label{Tb:Structural:break}
	\resizebox{\textwidth}{!}{
		\begin{tabular*}{\textwidth}{@{\extracolsep{\fill}} l
				d{2.3} @{\hspace{-1em}}d{1.3}  
				d{2.3} @{\hspace{-1em}} d{1.3} @{\hspace{2.5em}}   
				d{2.3} @{\hspace{-1em}} d{1.3} @{\hspace{2em}}                    
				@{}}
			\toprule
			& \multicolumn{2}{c}{$\tau=0.05$}
			& \multicolumn{2}{c}{$\tau=0.50$}
			& \multicolumn{2}{c}{$\tau=0.95$} \\
			\cmidrule(lr){2-3}\cmidrule(lr){4-5}\cmidrule(lr){6-7}
			& \multicolumn{1}{c}{F/M-statistic}
			& \multicolumn{1}{c}{$p$-value}
			& \multicolumn{1}{c}{F/M-statistic}
			& \multicolumn{1}{c}{$p$-value}
			& \multicolumn{1}{c}{F/M-statistic}
			& \multicolumn{1}{c}{$p$-value} \\
			\midrule
			Chow Test & 6.317 & 0.012 & 48.644 &1.241 \times 10^{-11}  & 13.659 & 2.491\times10^{-4} \\
			Wilcoxon Rank-Sum Test & 3527.000 & 0.001 & 11035.000 & 2.200 \times 10^{-16}  & 7949.000 & 9.347\times10^{-5} \\
			\bottomrule
	\end{tabular*}}
	\begin{flushleft}
		\footnotesize
		\justifying Note: The Chow test sets February 24, 2022 as the exogenous break date. The Wilcoxon rank-sum test compares the pre-conflict and post-conflict samples; the alternative hypothesis is $H_1$: the true location shift $\neq 0$.
	\end{flushleft} 
\end{table}%

\section{Portfolio implications}
\label{S6:Portofolio}

Having documented how connectedness among grain futures evolves over time, we now ask what this means for investors. To that end, we examine two classic allocation rules, the minimum variance portfolio (MVP) of \cite{Markowitz-1952-JFinance} and the minimum correlation portfolio (MCP) of \cite{Christoffersen-Errunza-Jacobs-Jin-2014-IntJForecast}, and a newer strategy that targets network spillovers, the minimum connectedness portfolio (MCoP). \cite{Broadstock-Chatziantoniou-Gabauer-2022-ApplEnergyFinance} show that the MCoP reduces risk and outperforms the other two methods, underscoring the value of incorporating connectedness information into portfolio design.

\begin{table}[htp]
	\centering
	\setlength{\abovecaptionskip}{0pt}
	\setlength{\belowcaptionskip}{10pt}
	\captionsetup{width=0.68\textwidth, justification=raggedright,singlelinecheck=false}
	\caption{Multivariate portfolio estimates at the conditional median ($\tau = 0.50$).}
	\label{Tb:portfolio}
	\resizebox{0.68\textwidth}{!}{
		\begin{tabular*}{\textwidth}{@{\extracolsep{\fill}} c l *{4}{d{1.3}} d{2.3} d{1.3} @{}}
			\toprule		
			&	& \multicolumn{1}{c}{Mean} & \multicolumn{1}{c}{Std.Dev.} & \multicolumn{1}{c}{5\%} & \multicolumn{1}{c}{95\%} & \multicolumn{1}{c}{HE} & \multicolumn{1}{c}{$p$-value} \\
			\midrule
			MVP&BRms & 0.017 & 0.014 & 0.000 & 0.052 & 0.923 & 0.000 \\
			&BRs & 0.020 & 0.021 & 0.000 & 0.064 & 0.921 & 0.000 \\
			&CNs & 0.067 & 0.039 & 0.018 & 0.122 & 0.843 & 0.000\\
			&ZAs & 0.036 & 0.020 & 0.000 & 0.069 & 0.919 & 0.000\\
			&ZAcs & 0.004 & 0.008 & 0.000 & 0.024 & 0.941 & 0.000\\
			&ARs & 0.026 & 0.012 & 0.004 & 0.045 & 0.904 & 0.000\\
			&USs & 0.002 & 0.006 & 0.000 & 0.017 & 0.950 & 0.000 \\
			&BRc & 0.036 & 0.022 & 0.005 & 0.075 & 0.908 & 0.000 \\
			&INm & 0.025 & 0.015 & 0.007 & 0.053 & 0.943 & 0.000\\
			&CNc & 0.103 & 0.024 & 0.060 & 0.136 & 0.766 & 0.000 \\
			&ZAwm & 0.000 & 0.001 & 0.000 & 0.000 & 0.933 & 0.000 \\
			&ZAym & 0.062 & 0.026 & 0.005 & 0.102 & 0.920 & 0.000 \\
			&ZAc & 0.000 & 0.000 & 0.000 & 0.000 & 0.960 & 0.000 \\
			&ARm & 0.004 & 0.007 & 0.000 & 0.021 & 0.946 & 0.000 \\
			&USc & 0.009 & 0.008 & 0.000 & 0.020 & 0.977 & 0.000 \\
			&ZAw & 0.096 & 0.028 & 0.066 & 0.168 & 0.833 & 0.000 \\
			&ARw & 0.004 & 0.007 & 0.000 & 0.019 & 0.965 & 0.000 \\
			&BSw & 0.054 & 0.052 & 0.000 & 0.171 & 0.888 & 0.000 \\
			&USw & 0.001 & 0.003 & 0.000 & 0.012 & 0.984 & 0.000 \\
			&UAw & 0.007 & 0.012 & 0.000 & 0.036 & 0.914 & 0.000 \\
			&CNr & 0.350 & 0.065 & 0.255 & 0.457 & 0.323 & 0.000 \\
			&USr & 0.040 & 0.014 & 0.022 & 0.065 & 0.941 & 0.000 \\
			&INb & 0.036 & 0.019 & 0.015 & 0.073 & 0.933 & 0.000 \\
			\midrule
			MCP&BRms & 0.015 & 0.020 & 0.000 & 0.056 & 0.846 & 0.000 \\
			&BRs & 0.038 & 0.034 & 0.000 & 0.086 & 0.843 & 0.000 \\
			&CNs & 0.082 & 0.028 & 0.038 & 0.124 & 0.687 & 0.000 \\
			&ZAs & 0.059 & 0.023 & 0.013 & 0.092 & 0.839 & 0.000 \\
			&ZAcs & 0.013 & 0.019 & 0.000 & 0.052 & 0.882 & 0.000 \\
			&ARs & 0.035 & 0.019 & 0.003 & 0.062 & 0.809 & 0.000 \\
			&USs & 0.025 & 0.020 & 0.000 & 0.054 & 0.901 & 0.000 \\
			&BRc & 0.038 & 0.022 & 0.005 & 0.073 & 0.817 & 0.000 \\
			&INm & 0.097 & 0.025 & 0.062 & 0.132 & 0.886 & 0.000 \\
			&CNc & 0.066 & 0.027 & 0.020 & 0.100 & 0.534 & 0.000 \\
			&ZAwm & 0.011 & 0.018 & 0.000 & 0.047 & 0.866 & 0.000 \\
			&ZAym & 0.035 & 0.031 & 0.000 & 0.091 & 0.841 & 0.000 \\
			&ZAc & 0.007 & 0.013 & 0.000 & 0.037 & 0.921 & 0.000 \\
			&ARm & 0.008 & 0.015 & 0.000 & 0.043 & 0.893 & 0.000 \\
			&USc & 0.047 & 0.031 & 0.000 & 0.089 & 0.954 & 0.000 \\
			&ZAw & 0.050 & 0.017 & 0.013 & 0.072 & 0.668 & 0.000 \\
			&ARw & 0.025 & 0.025 & 0.000 & 0.067 & 0.929 & 0.000 \\
			&BSw & 0.020 & 0.021 & 0.000 & 0.054 & 0.777 & 0.000 \\
			&USw & 0.029 & 0.026 & 0.000 & 0.067 & 0.968 & 0.000 \\
			&UAw & 0.017 & 0.026 & 0.000 & 0.079 & 0.829 & 0.000 \\
			&CNr & 0.109 & 0.011 & 0.091 & 0.130 & -0.349 & 0.003 \\
			&USr & 0.068 & 0.022 & 0.034 & 0.101 & 0.882 & 0.000 \\
			&INb & 0.107 & 0.021 & 0.078 & 0.141 & 0.868 & 0.000 \\
			\midrule
			MCoP&BRms & 0.022 & 0.024 & 0.000 & 0.068 & 0.835 & 0.000 \\
			&BRs & 0.036 & 0.033 & 0.000 & 0.082 & 0.833 & 0.000 \\
			&CNs & 0.078 & 0.008 & 0.061 & 0.089 & 0.666 & 0.000 \\
			&ZAs & 0.031 & 0.011 & 0.014 & 0.048 & 0.828 & 0.000 \\
			&ZAcs & 0.045 & 0.016 & 0.017 & 0.072 & 0.874 & 0.000 \\
			&ARs & 0.043 & 0.014 & 0.019 & 0.063 & 0.797 & 0.000 \\
			&USs & 0.033 & 0.029 & 0.000 & 0.079 & 0.895 & 0.000 \\
			&BRc & 0.062 & 0.009 & 0.044 & 0.078 & 0.805 & 0.000 \\
			&INm & 0.081 & 0.008 & 0.069 & 0.094 & 0.878 & 0.000 \\
			&CNc & 0.069 & 0.011 & 0.053 & 0.087 & 0.502 & 0.000 \\
			&ZAwm & 0.043 & 0.024 & 0.005 & 0.086 & 0.858 & 0.000 \\
			&ZAym & 0.006 & 0.011 & 0.000 & 0.033 & 0.830 & 0.000 \\
			&ZAc & 0.004 & 0.007 & 0.000 & 0.021 & 0.916 & 0.000 \\
			&ARm & 0.013 & 0.010 & 0.000 & 0.029 & 0.886 & 0.000 \\
			&USc & 0.039 & 0.020 & 0.000 & 0.061 & 0.951 & 0.000 \\
			&ZAw & 0.046 & 0.010 & 0.028 & 0.062 & 0.646 & 0.000 \\
			&ARw & 0.026 & 0.021 & 0.000 & 0.055 & 0.925 & 0.000 \\
			&BSw & 0.028 & 0.023 & 0.000 & 0.059 & 0.762 & 0.000 \\
			&USw & 0.020 & 0.018 & 0.000 & 0.048 & 0.966 & 0.000 \\
			&UAw & 0.040 & 0.031 & 0.000 & 0.092 & 0.817 & 0.000 \\
			&CNr & 0.084 & 0.008 & 0.072 & 0.100 & -0.439 & 0.000 \\
			&USr & 0.072 & 0.005 & 0.064 & 0.080 & 0.874 & 0.000 \\
			&INb & 0.082 & 0.005 & 0.074 & 0.092 & 0.859 & 0.000 \\
			\bottomrule
	\end{tabular*}}
\end{table}

 The MCoP builds a portfolio by minimising the pairwise connectedness of asset returns. At each time $t$, the weight vector is computed as
\begin{equation}
	\boldsymbol{\omega}_{C,t}= \frac{PCI_t^{-1}\mathbf{1}}{\mathbf{1}^{\prime}PCI_t^{-1}\mathbf{1}},
\end{equation}
where $PCI_t$ is the pairwise connectedness index matrix that summarises spillovers among the contracts, $\boldsymbol{\omega}_{C,t}$ is the vector of portfolio weights, and $\mathbf{1}$ is a conformable vector of ones.

In addition, to provide a comprehensive assessment of the portfolio's risk and return profile, we employ two metrics, the Sharpe ratio and Hedging Effectiveness (HE). The Sharpe ratio normalises the portfolio excess return by its risk. A higher value signals superior performance for a given risk level. For robustness we also compute risk-adjusted returns using Value at Risk (\text{VaR}) and Conditional VaR (\text{CVaR}) instead of volatility. HE compares the variance of the portfolio with that of a single asset $i$ to measure the percentage reduction in risk. A higher value indicates greater risk reduction relative to asset $i$.

Most studies examine how asset connectedness affects portfolio performance and compare the results with the MVP and MCP. These studies typically measure connectedness only at the conditional mean. Because investors may react differently across the return distribution, portfolios constructed with connectedness estimated at varying conditional quantiles are likely to show distinct performance patterns \citep{Shi-Chen-2025-GlobFinJ}. 

Table~\ref{Tb:portfolio} summarizes how the three multivariate strategies allocate capital and the extent of risk reduction they achieve under normal market conditions. For the MVP portfolio, HE mostly falls between 0.80 and 0.95 and is statistically significant, indicating sizable risk reduction relative to the underlying assets. 
CNr stands out because it receives a large average weight of 35\% while its HE is 32.3\%, implying modest risk reduction from this holding. This result is consistent with the weak spillover profile documented earlier for CNr and likely limits its contribution to portfolio risk sharing.

The MCP and MCoP portfolios show greater dispersion in HE, and CNr records a negative HE, indicating that diversification benefits are not uniform under these constructions. Even so, risk reduction for each commodity generally exceeds 50 percent. Weights are broadly diversified, with many small or near zero allocations. Among the three strategies, MCoP has the lowest HE, indicating the weakest overall risk reduction. According to the Sharpe ratios in Table~\ref{Tb:portfolio:performance} under normal market conditions, MCoP is the highest and positive, MCP ranks second and is negative, and MVP is the lowest and is also negative.
Under the lower tail market scenario ($\tau=0.05$), the Sharpe ratios follow the median pattern. MCoP posts the highest and positive value, MCP is also positive, and MVP is negative. By contrast, in the upper-tail scenario ($\tau=0.95$), MVP and MCP remain positive, whereas MCoP turns negative.

\begin{table}[htp]
	\centering
	\setlength{\abovecaptionskip}{0pt}
	\setlength{\belowcaptionskip}{10pt}
	\caption{Performance of the MVP, MCP, and MCoP portfolios at the 5th, 50th, and 95th conditional quantiles.}
	\label{Tb:portfolio:performance}
	\resizebox{\textwidth}{!}{
		\begin{tabular*}{\textwidth}
			{@{\extracolsep{\fill}} l *{9}{d{2.3}} @{}}
			\toprule
			& \multicolumn{3}{c}{$\tau=0.05$}
			& \multicolumn{3}{c}{$\tau=0.50$}
			& \multicolumn{3}{c}{$\tau=0.95$} \\
			\cmidrule(lr){2-4}\cmidrule(lr){5-7}\cmidrule(lr){8-10}
			& \multicolumn{1}{c}{\MVP}
			& \multicolumn{1}{c}{\MCP}
			& \multicolumn{1}{c}{\MCoP}
			& \multicolumn{1}{c}{\MVP}
			& \multicolumn{1}{c}{\MCP}
			& \multicolumn{1}{c}{\MCoP}
			& \multicolumn{1}{c}{\MVP}
			& \multicolumn{1}{c}{\MCP}
			& \multicolumn{1}{c}{\MCoP} \\
			\midrule
			\multicolumn{10}{@{}l}{\textit{Panel A: Full sample}}\\
			Return & -0.019 & 0.006 & 0.008 & -0.003 & -0.002 & 0.015 & 0.024 & 0.028 & -0.015\\
			StdDev & 0.064 & 0.111 & 0.126 & 0.059 & 0.084 & 0.087 & 0.071 & 0.104 & 0.121\\
			Sharpe Ratio (StdDev) & -0.291 & 0.054 & 0.066 & -0.048 & -0.029 & 0.168 & 0.337 & 0.271 & -0.124\\
			Sharpe Ratio (VaR) & -3.084 & 0.553 & 0.676 & -0.513 & -0.297 & 1.724 & 3.565 & 2.694 & -1.227\\
			Sharpe Ratio (CVaR) & -2.449 & 0.422 & 0.437 & -0.408 & -0.235 & 1.333 & 2.729 & 1.918 & -0.879 \\\\
			
			\multicolumn{10}{@{}l}{\textit{Panel B: Post‑conflict sample}}\\
			Return & -0.043 & -0.042 & -0.051 & -0.024 & -0.042 & -0.029 & -0.007 & -0.020 & -0.066\\
			StdDev & 0.063 & 0.112 & 0.129 & 0.060 & 0.085 & 0.088 & 0.072 & 0.105 & 0.122\\
			Sharpe Ratio (StdDev) & -0.680 & -0.372 & -0.399 & -0.391 & -0.491 & -0.327 & -0.097 & -0.193 & -0.545\\
			Sharpe Ratio (VaR) & -6.829 & -3.787 & -4.045 & -4.130 & -5.002 & -3.318 & -0.997 & -1.898 & -5.311\\
			Sharpe Ratio (CVaR) & -5.255 & -2.936 & -2.681 & -3.304 & -3.989 & -2.600 & -0.761 & -1.368 & -3.833\\
			\bottomrule
	\end{tabular*}}
	\begin{flushleft}
		\footnotesize
		\justifying Note: The pre‑conflict window provides fewer than 30 observations for calculating the portfolio metrics, so the Sharpe ratios, particularly the VaR‑ and CVaR‑based measures, are statistically unreliable and are therefore not reported.
	\end{flushleft} 
\end{table}

Moreover, Table~\ref{Tb:portfolio:performance} indicates that in the post-conflict period, the average returns of all three portfolios turn negative, as do their StdDev-, VaR-, and CVaR-Sharpe ratios. Compared with the full-sample results, these risk-adjusted measures deteriorate further, indicating that the geopolitical shock lowered returns without a commensurate reduction in risk and thus markedly eroded the portfolios' risk compensation.

\begin{figure}[!ht]
	\centering
	\includegraphics[width=0.95\linewidth]{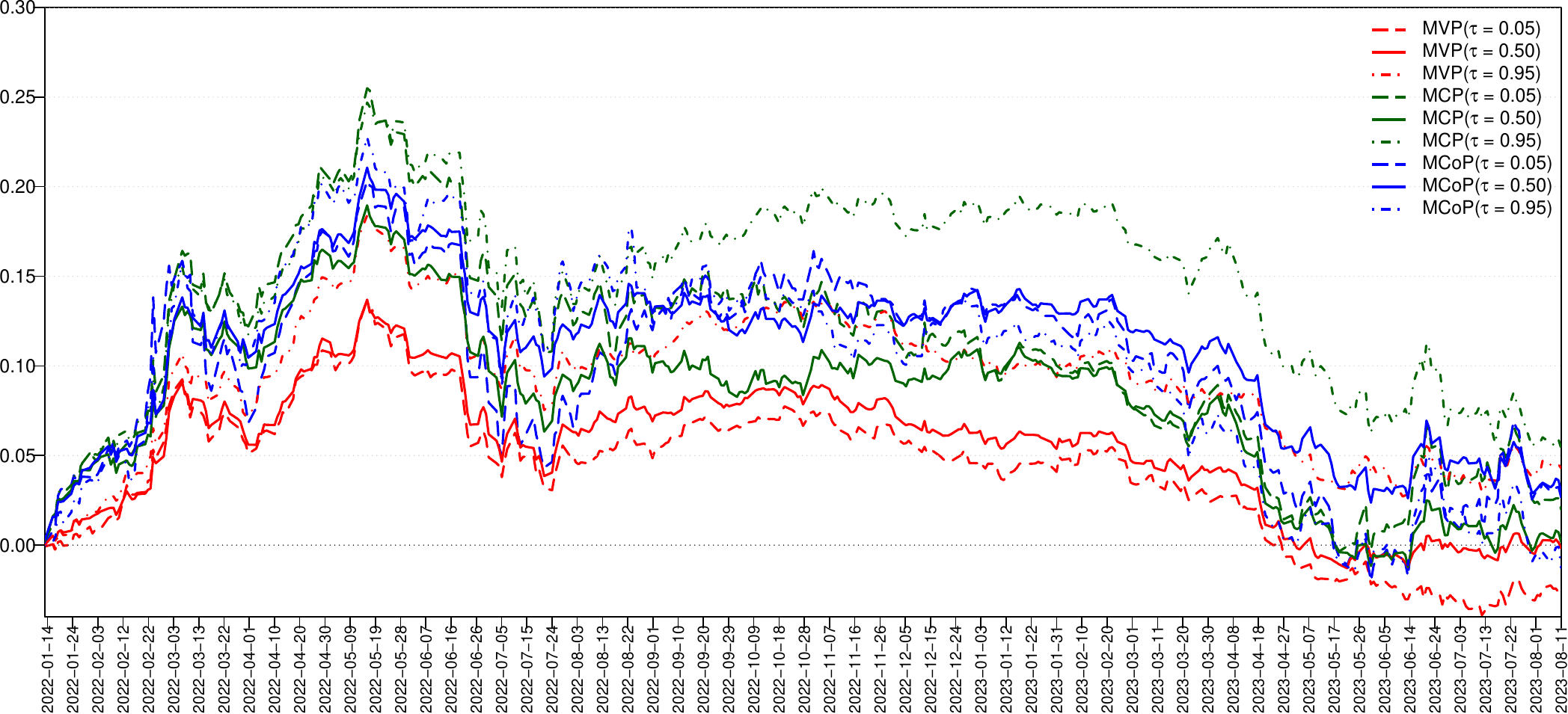}
	\caption{Cumulative returns of different portfolios.}
	\label{Fig:cumsumR}
\end{figure}

To further profile portfolio performance, Fig.~\ref{Fig:cumsumR} plots cumulative returns for the three strategies at $\tau=0.05$, $0.50$, and $0.95$. The nine paths move broadly together over the sample. Around key geopolitical dates the trends diverge. Cumulative returns decline after the start of the Russia-Ukraine war in February 2022, rise following the signing of the BSGI in July~2022, and decline again after the March 2023 extension that shortened the term to sixty days. Overall, MCP delivers the strongest performance at $\tau=0.95$, while MCoP remains comparatively stable across all three quantiles.

Therefore, the evidence indicates that portfolios constructed to minimize cross-asset connectedness strengthen systemic stability in grain markets and yield a more stable market portfolio, particularly under the $\tau=0.05$ and $\tau=0.50$ market conditions.

\section{Conclusion}
\label{S7:conclusion}

Global food security is critical, especially in periods of geopolitical instability, because it shapes welfare, economic stability, and financial markets \citep{behnassi2022implications}. Grain futures function as benchmarks for price discovery and risk management, aggregating information and enabling hedging that supports market stability and reliable supplies. Existing research mostly examines spillovers between grain futures and other asset classes, while spillovers within grain futures are less explored and evidence on how they vary across conditional quantiles and across major producers such as the BRICS remains limited. Addressing this gap matters for investors and policymakers who manage grain related risks.

We analyze quantile connectedness between the BRICS and international grain futures with particular attention to the Russia-Ukraine conflict. Methodologically, we combine the dynamic connectedness framework based on the quantile vector autoregression of \cite{ando2022quantile} with the frequency domain approach of \cite{barunik2018measuring}. This design provides evidence on spillovers across time and frequency, from both static and dynamic perspectives, under normal and tail market conditions.

Results show that connectedness is heterogeneous across quantiles. In the time domain, static measures indicate that under normal and tail states each grain primarily explains its own forecast error variance, while systemwide spillovers vary by regime. The TCI is generally higher in the pre-outbreak period across quantiles, pointing to weaker interdependence after the conflict. Average dynamic connectedness further indicates that grain type and regional proximity strengthen pairwise links. Moreover, throughout the entire sample period, South African grain futures tend to be net receivers, whereas Argentine grain futures together with U.S. soybean and Ukrainian wheat typically act as key transmitters.

Under extreme bearish conditions, U.S. corn, wheat, and rice, Black Sea wheat, Chinese soybean, and South African maize are persistent net receivers across the pre- and post-outbreak periods, which suggests heightened vulnerability to external shocks. Brazilian soybean, Argentine maize, Brazilian corn, Argentine soybean, U.S. soybean, South African wheat, and Ukrainian wheat act as key transmitters. The Indian market shifts from a net receiver to a transmitter, Argentine wheat becomes a net receiver, Brazilian soybean and corn strengthen their transmitter roles, and Ukrainian wheat’s transmission weakens.

Under extreme bullish conditions, Brazilian soybean futures are net receivers before the outbreak. U.S., Argentine, and Black Sea wheat futures switch from net receivers to net transmitters after the outbreak, with Black Sea wheat moving from -0.49\% to 13.95\%. South African soybean and Chinese corn futures move from net transmitters to net receivers, and Ukrainian wheat shifts from a dominant net transmitter at 49.92\% to a net receiver at -0.67\%. At the median quantile, the transmission mechanism differs from the tails. South African grain futures are the main absorbers, Argentine grain futures are dominant transmitters, and transmission strength declines for most transmitters from pre- to post-outbreak, including Black Sea wheat from 28.32\% to 4.24\% and Ukrainian wheat from 20.86\% to 2.74\%, making them the weakest transmitters after the conflict.

In the frequency domain, static results show that total spillovers decrease as the horizon lengthens, with short-term spillovers dominating, consistent with evidence from other markets \citep{liu2023spillover,le2023price,lei2024nexus}. Under extreme conditions, short-term TCIs are higher post-outbreak, whereas medium- and long-term TCIs are higher pre-outbreak. Under normal conditions, the short-, medium-, and long-term patterns reverse.

From a time-varying perspective, the dynamic TCI at the left and right tails fluctuates around 95\%, notably above the conditional median. Additionally, the systemic risk within the grain futures system is heavily influenced by the onset of the Russia-Ukraine conflict, as well as the signing, modification, and termination of the BSGI. Frequency-based dynamics are largely driven by short-term shocks over most of the sample. Dynamic NET spillovers show Brazilian soybeans, U.S. soybeans, Argentine wheat, Black Sea wheat, Argentine maize, and U.S. wheat as persistent transmitters, while South African wheat, Chinese rice, U.S. rice, Chinese soybeans, Indian barley, South African soybeans, South African corn, and Chinese corn are persistent receivers. Notably, South African white and yellow maize move from receivers to transmitters around late March 2023, whereas Ukrainian wheat moves in the opposite direction, which is consistent with the shortened extension of the BSGI. Overall, U.S. grain futures, serving as a benchmark for international grain prices, consistently held a dominant position in net pairwise spillovers, exerting influence on the grain futures markets of BRICS countries for most of the period. This finding is consistent with \cite{zhu2024uncovering}'s research on risk spillovers between Sino-US agricultural futures.

Finally, the analysis of quantile sensitivity reveals that spillover effects are generally symmetric across quantiles, with a significant decline in connectedness among grain futures observed between the 0.3 and 0.7 quantiles following the Russia-Ukraine conflict and the shortening of the BSGI. Notably, while different quantile levels do not significantly change the overall roles of certain grain futures as dominant transmitters or receivers, net directional connectedness remains highly sensitive to market conditions, especially during bullish and bearish periods.

From a portfolio perspective, the MVP, MCP, and MCoP strategies deliver state dependent payoffs. Under normal and lower tail conditions, MCoP posts the highest and positive Sharpe ratio, whereas in the upper tail MVP and MCP lead. After the conflict, average returns and Sharpe ratios based on volatility, VaR, and CVaR are negative, indicating lower returns without commensurate risk declines. These patterns support minimizing connectedness across assets in tranquil or bearish markets and rotating toward rules based on variance or correlation in strong rallies. Investors should carefully hedge contracts that are persistent net receivers, use persistent net transmitters as strategic hedging instruments, and rebalance more frequently when roles switch across regimes.
For policymakers, the results highlight the need for measures to stabilize markets and risk monitoring in real time that is sensitive to tail conditions and to short-, medium-, and long-term horizons. The benchmark role of U.S. grain futures and their transmission to BRICS markets calls for coordinated international responses to limit cascade effects. Transparent communication around major policy changes, including those related to the BSGI, can sustain confidence, while stronger international cooperation and robust regulatory frameworks help reduce systemic risk in global grain futures.

Several limitations could be addressed in future research. First, the analysis is constrained by the availability of data for Black Sea and Ukrainian wheat futures through August 11, 2023. Additional observations may yield further insights into spillovers among grain futures. Second, while this study focuses on connectedness across quantiles and frequencies, future work could examine decomposed and partial connectedness to deepen understanding of spillovers \citep{Naifar-2025-JCommodMark}. Finally, extending the analysis to other emerging markets or to different geopolitical events could provide a more comprehensive view of market dynamics.

\appendix
\newpage
\section{TO, FROM, and NET connectedness at the left and right tails}

\setcounter{figure}{0}
\setcounter{table}{0}

	\begin{figure}[!htb]
	\centering
	\begin{tikzpicture}
		\node[anchor=south west,inner sep=0] (imageA) at (0,0) {\includegraphics[width=8.1cm]{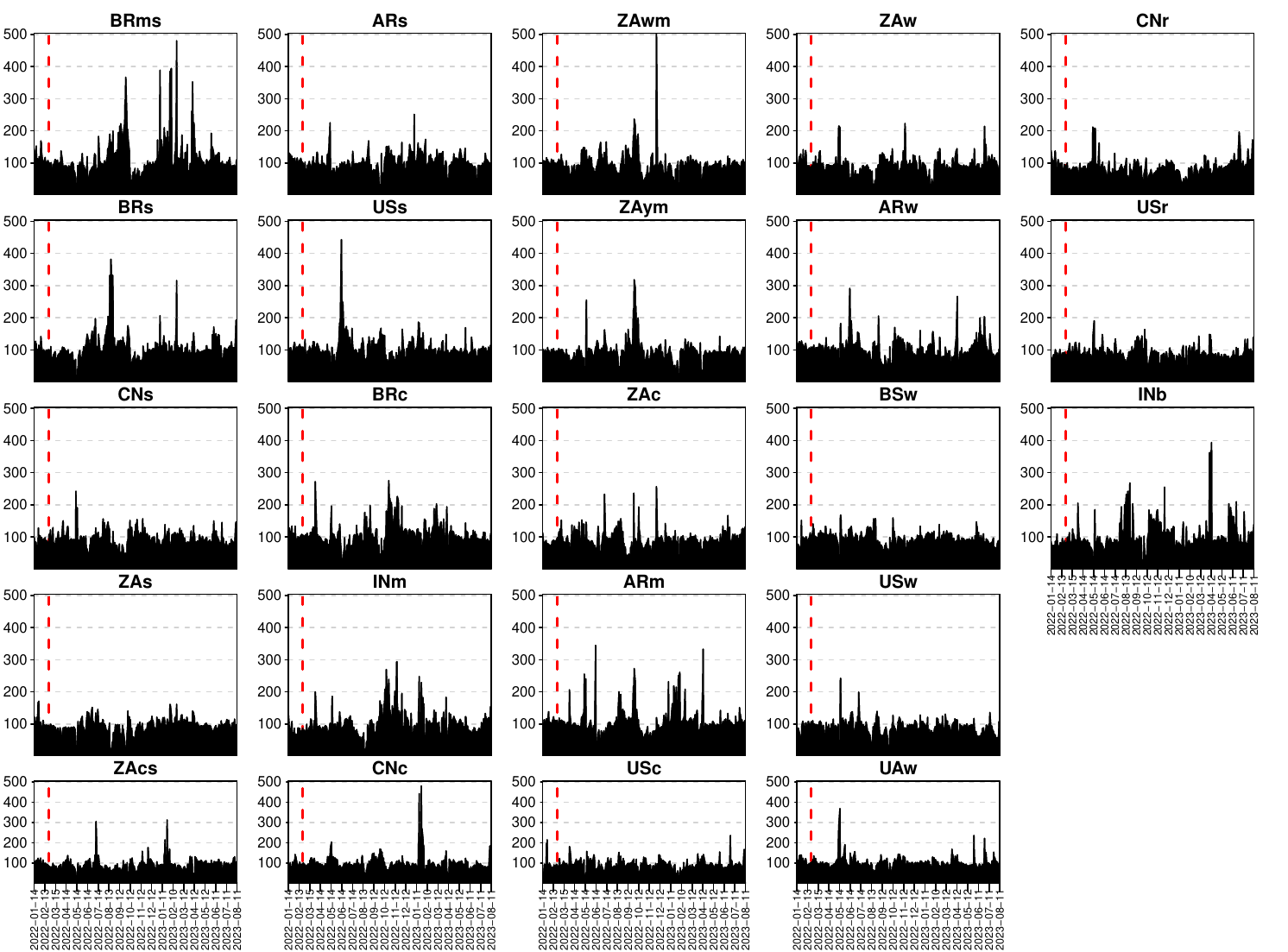}};
		\node[anchor=south east, xshift=-5mm, yshift=5mm] at (imageA.south east) {(A)};
	\end{tikzpicture}
	\begin{tikzpicture}
		\node[anchor=south west,inner sep=0] (imageB) at (0,0) {\includegraphics[width=8.1cm]{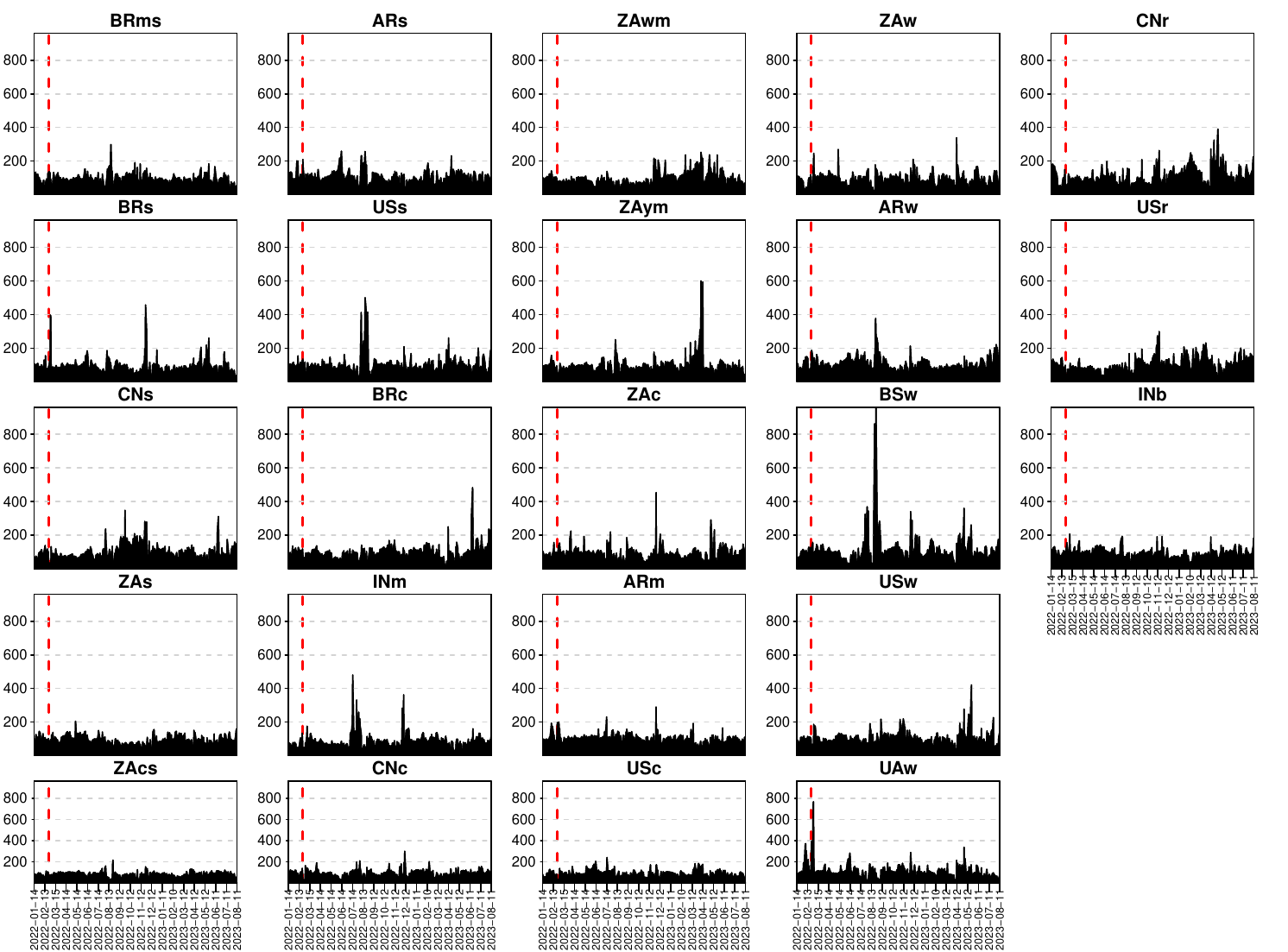}};
		\node[anchor=south east, xshift=-5mm, yshift=5mm] at (imageB.south east) {(B)};
	\end{tikzpicture}
	\\
	\begin{tikzpicture}
		\node[anchor=south west,inner sep=0] (imageC) at (0,0) {\includegraphics[width=8.1cm]{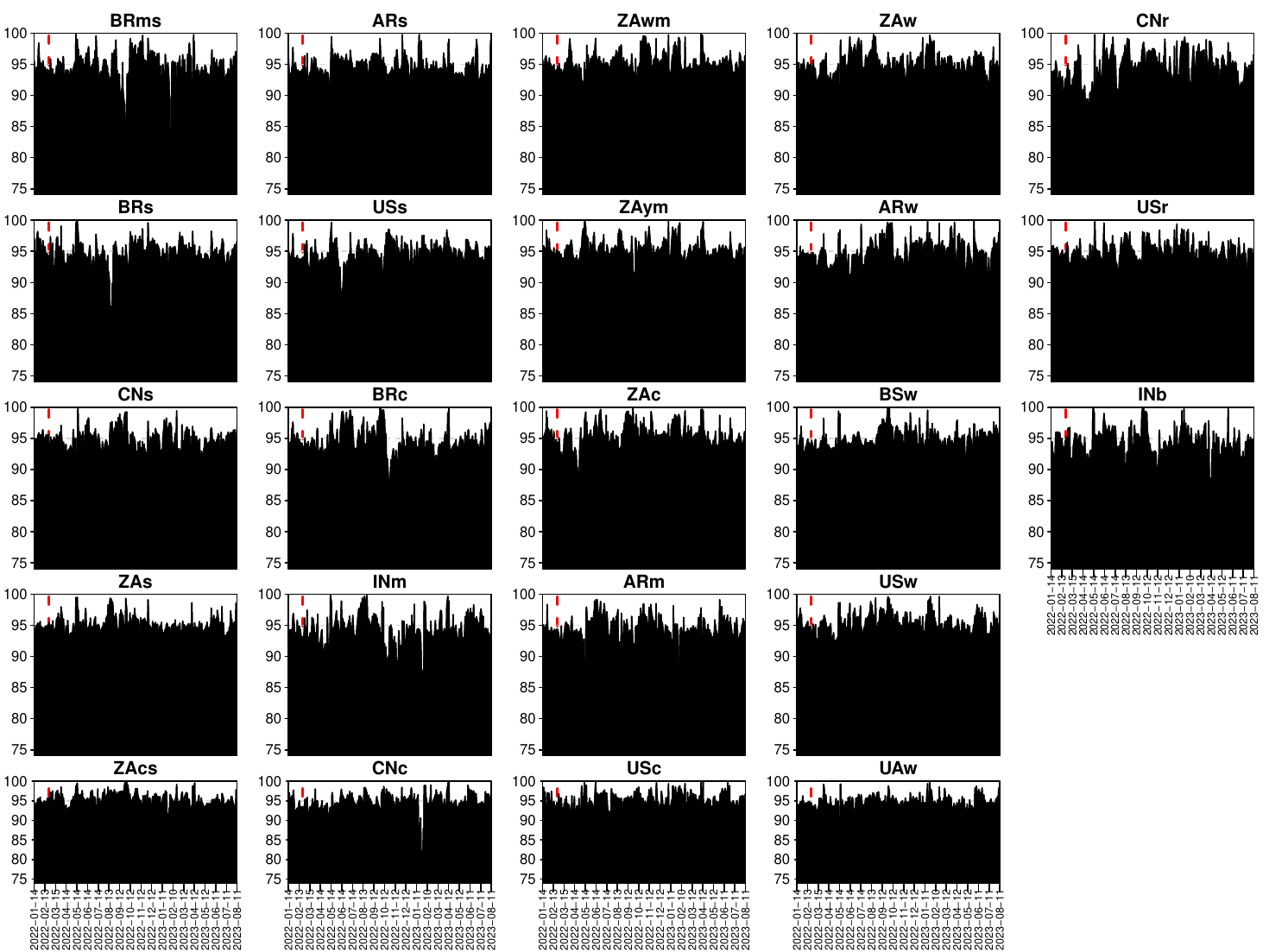}};
		\node[anchor=south east, xshift=-5mm, yshift=5mm] at (imageC.south east) {(C)};
	\end{tikzpicture}
	\begin{tikzpicture}
		\node[anchor=south west,inner sep=0] (imageD) at (0,0) {\includegraphics[width=8.1cm]{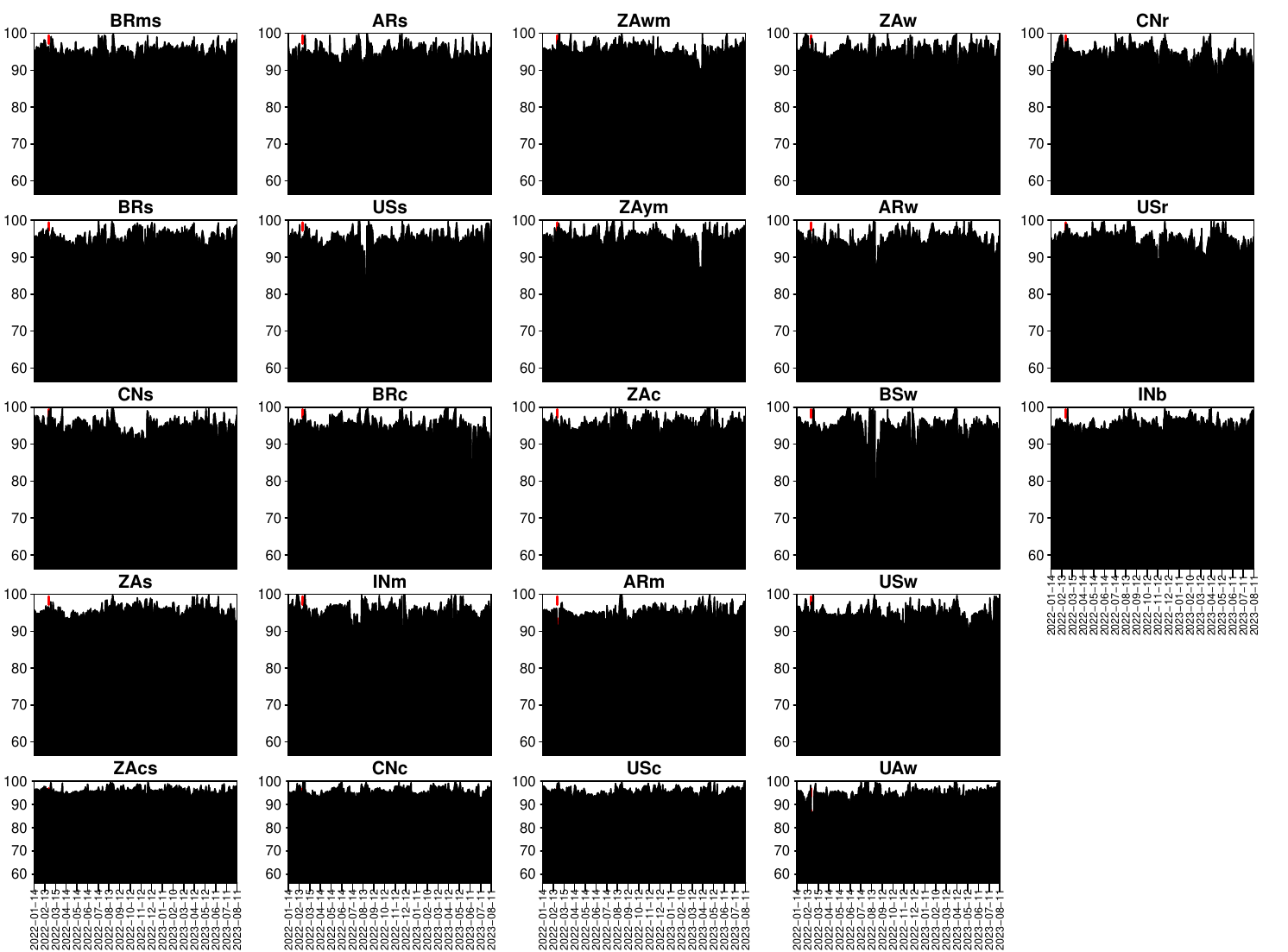}};
		\node[anchor=south east, xshift=-5mm, yshift=5mm] at (imageD.south east) {(D)};
	\end{tikzpicture}
	\\
	\begin{tikzpicture}
	\node[anchor=south west,inner sep=0] (imageA) at (0,0) {\includegraphics[width=8.1cm]{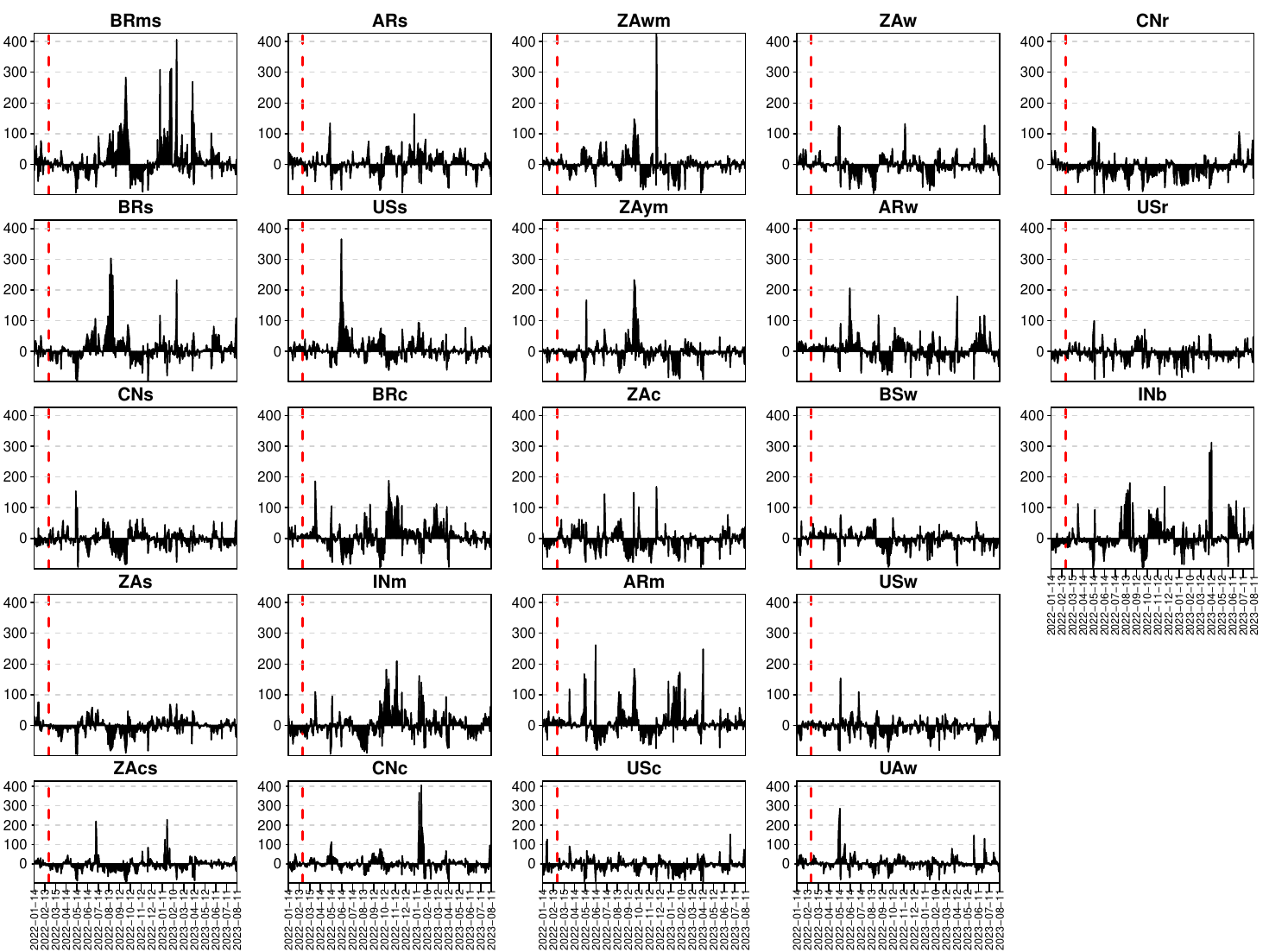}};
	\node[anchor=south east, xshift=-5mm, yshift=5mm] at (imageA.south east) {(E)};
\end{tikzpicture}
\begin{tikzpicture}
	\node[anchor=south west,inner sep=0] (imageB) at (0,0) {\includegraphics[width=8.1cm]{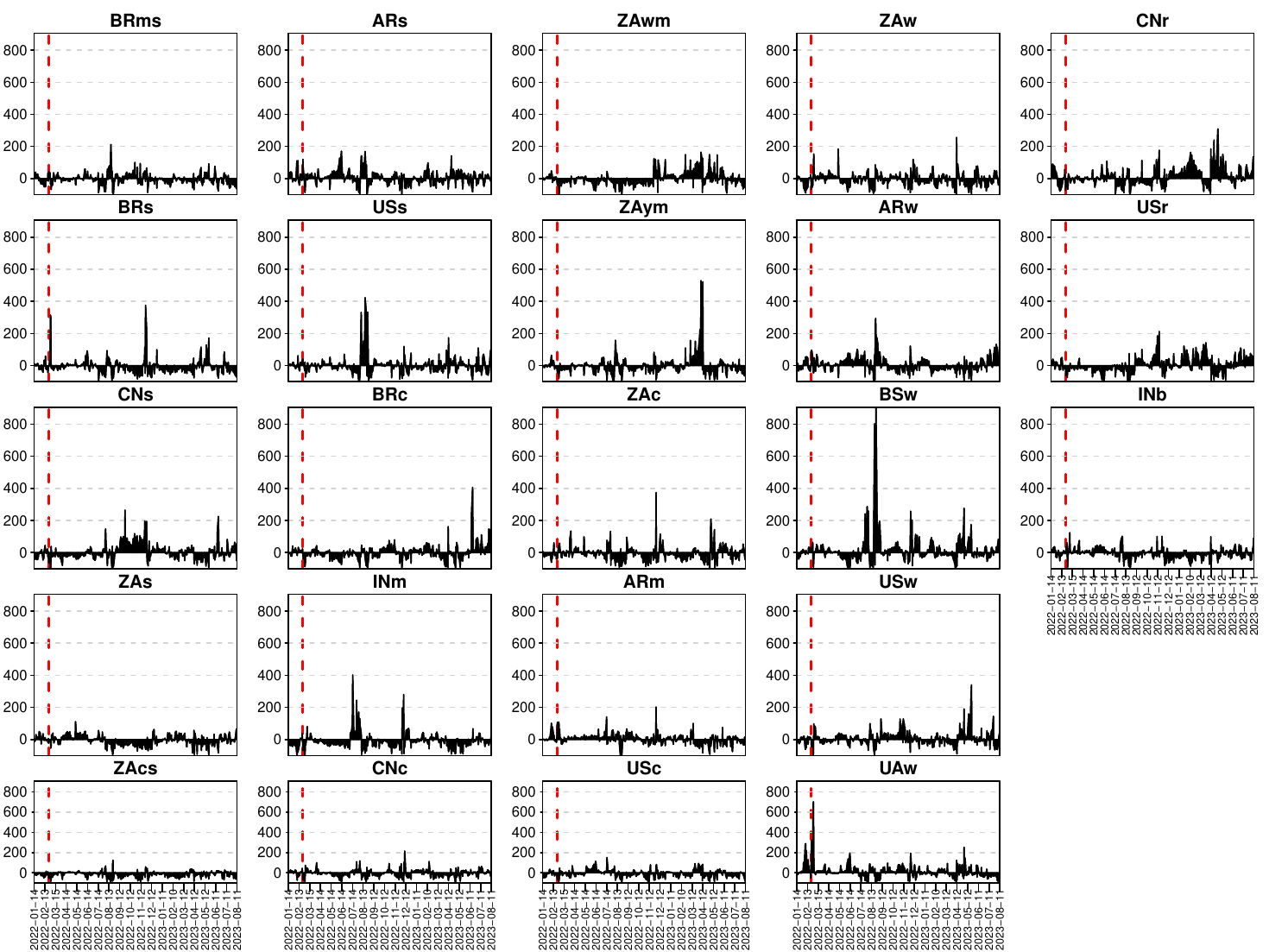}};
	\node[anchor=south east, xshift=-5mm, yshift=5mm] at (imageB.south east) {(F)};
\end{tikzpicture}
\caption{(A) and (B) show TO connectedness, (C) and (D) display FROM connectedness, and (E) and (F) represent NET connectedness at the left ($\tau=0.05$) and right tails ($\tau=0.95$) in the time domain, respectively.}
  \label{Fig:TO:FROM:NET:left:right:tails:Time}
\end{figure}

\begin{figure}[!htb]
	\centering
	\begin{tikzpicture}
		\node[anchor=south west,inner sep=0] (imageA) at (0,0) {\includegraphics[width=0.95\textwidth]{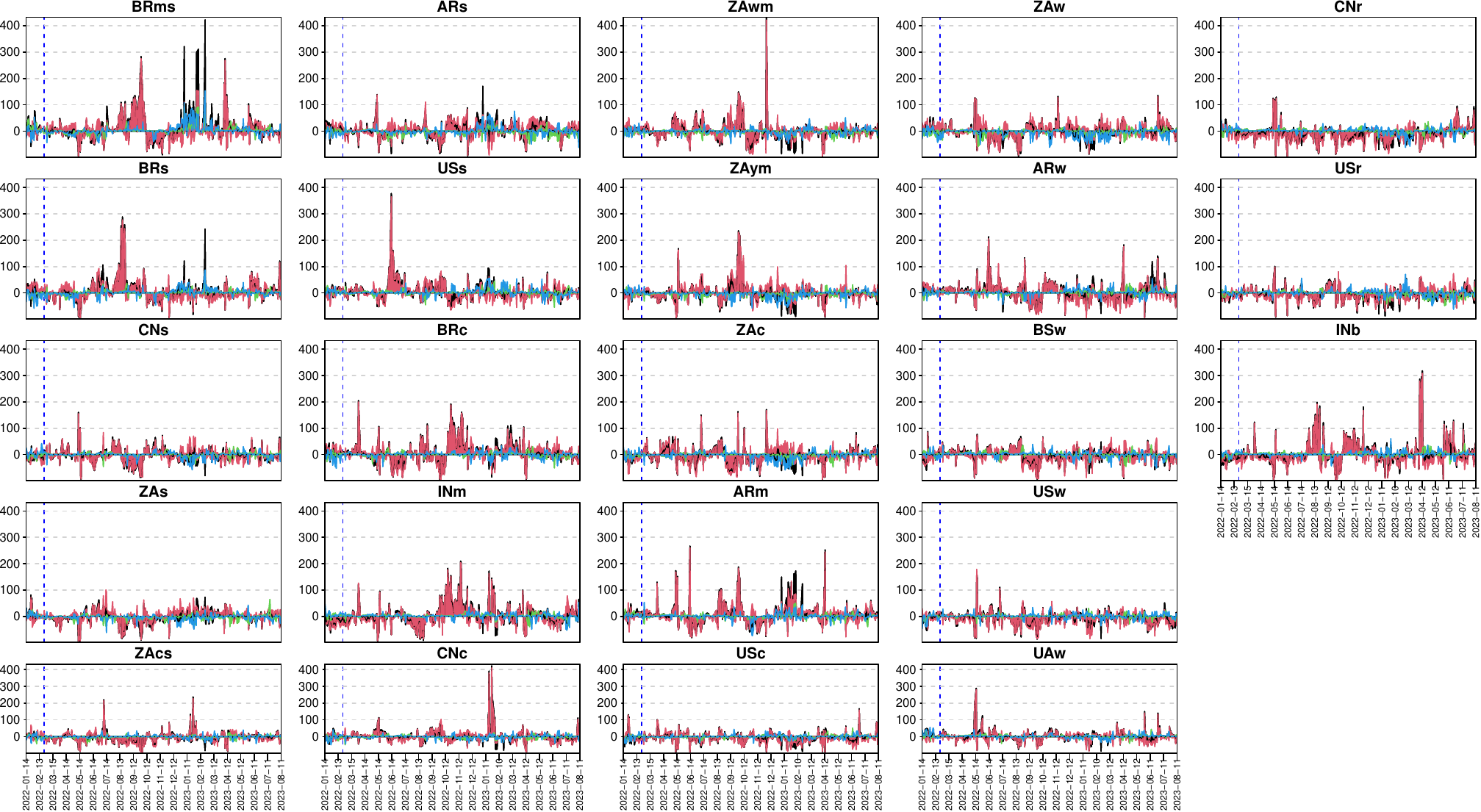}};
		\node[anchor=south east, xshift=-5mm, yshift=5mm] at (imageA.south east) {(A)};
	\end{tikzpicture}
	\\
	\begin{tikzpicture}
		\node[anchor=south west,inner sep=0] (imageB) at (0,0) {\includegraphics[width=0.95\textwidth]{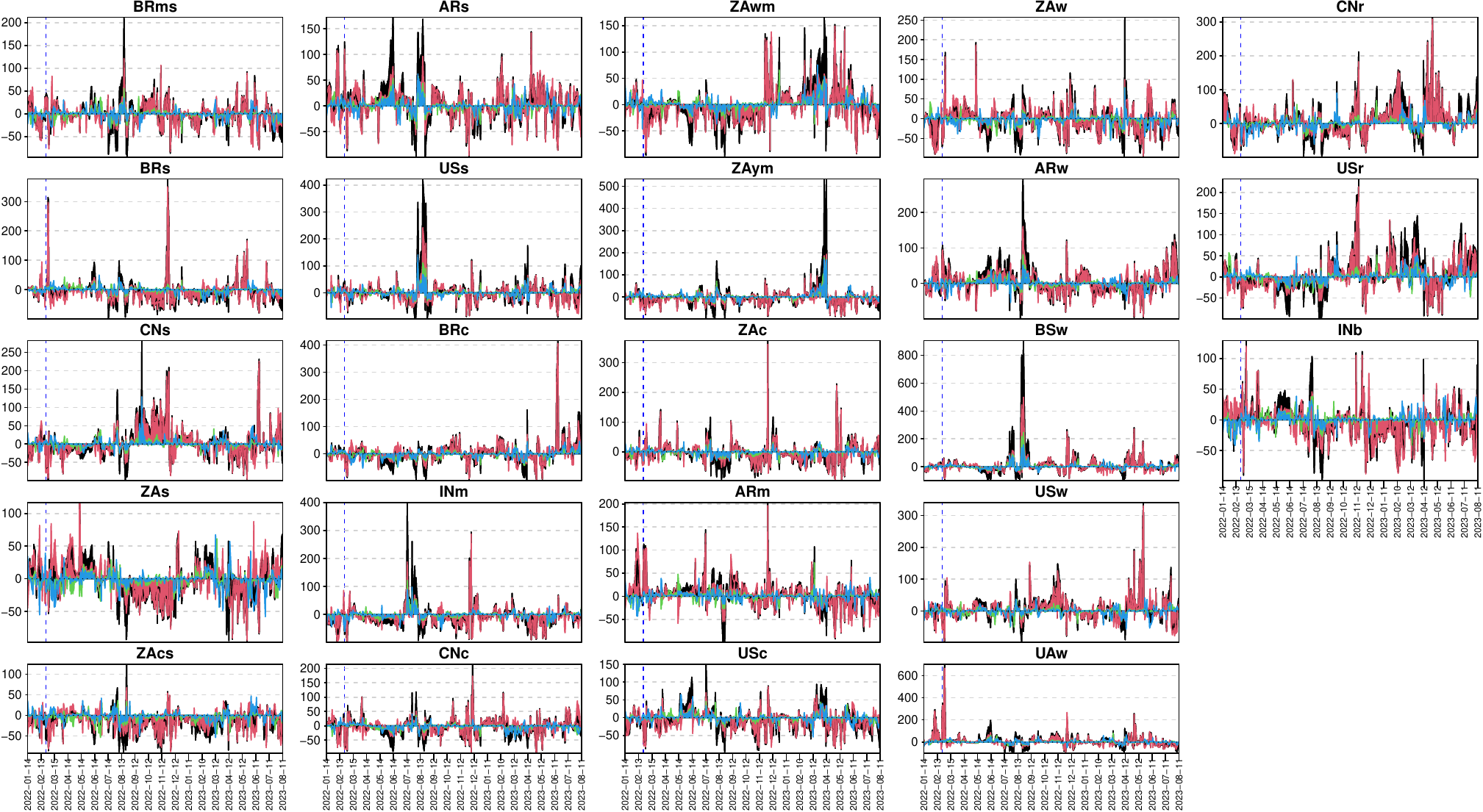}};
		\node[anchor=south east, xshift=-5mm, yshift=5mm] at (imageB.south east) {(B)};
	\end{tikzpicture}
    \caption{(A) and (B) show time-varying NET connectedness at the left ($\tau=0.05$) and right tails ($\tau=0.95$) in the frequency domain, respectively.}
	\label{Fig:TO:FROM:NET:left:right:tails:fre}
\end{figure}

\newpage

\section{Average connectedness for the pre-outbreak and post-outbreak periods}

\begin{table}[!htb]
	\centering
	\setlength{\abovecaptionskip}{0pt}
	\setlength{\belowcaptionskip}{10pt}
	   \captionsetup{width=0.89\textwidth, justification=raggedright,singlelinecheck=false}
	\caption{Average dynamic connectedness in the time domain at the 0.05, 0.50 and 0.95 quantiles during the pre-outbreak and post-outbreak periods.}
	\label{Tb:Average:Dynamic:Connectedness:Time:pre:post}
	\resizebox{0.89\textwidth}{!}{
		\begin{tabular*}{\textwidth}
			{@{\extracolsep{\fill}} l *{9}{d{3.2}} @{}}
			\toprule
			& \multicolumn{3}{c}{FROM}
			& \multicolumn{3}{c}{TO}
			& \multicolumn{3}{c}{NET} \\
			\cmidrule(lr){2-4}\cmidrule(lr){5-7}\cmidrule(lr){8-10}
			& \multicolumn{1}{c}{$\tau=0.05$}
			& \multicolumn{1}{c}{$\tau=0.50$}
			& \multicolumn{1}{c}{$\tau=0.95$}
			& \multicolumn{1}{c}{$\tau=0.05$}
			& \multicolumn{1}{c}{$\tau=0.50$}
			& \multicolumn{1}{c}{$\tau=0.95$}
			& \multicolumn{1}{c}{$\tau=0.05$}
			& \multicolumn{1}{c}{$\tau=0.50$}
			& \multicolumn{1}{c}{$\tau=0.95$}\\
			\midrule
			\multicolumn{10}{@{}l}{\textit{Panel A: Pre‑conflict sample}}\\
			BRms & 94.54 & 77.22 & 96.05 & 105.24 & 124.14 & 89.22 & 10.70 & 46.91 & -6.82 \\
			BRs  & 95.04 & 73.26 & 96.08 &  98.86 &  93.35 & 89.74 &  3.83 & 20.09 & -6.34 \\
			CNs  & 95.16 & 30.76 & 96.10 &  84.96 &   7.24 & 85.86 & -10.20 & -23.52 & -10.23 \\
			ZAs  & 94.54 & 77.81 & 95.26 & 103.72 &  35.89 &102.09 &  9.18 & -41.92 &  6.84 \\
			ZAcs & 94.52 & 85.60 & 96.36 &  99.95 &  43.44 & 78.16 &  5.44 & -42.16 & -18.20 \\
			ARs  & 94.05 & 75.54 & 94.15 & 101.30 & 104.52 &110.52 &  7.25 &  28.98 &  16.37 \\
			USs  & 94.03 & 77.18 & 95.53 & 102.03 & 124.85 & 99.98 &  8.00 &  47.67 &   4.45 \\
			BRc  & 94.26 & 52.67 & 95.32 & 101.94 &  40.70 &102.73 &  7.68 & -11.97 &   7.41 \\
			INm  & 94.17 & 15.17 & 97.11 &  75.28 &   8.10 & 66.70 & -18.89 &  -7.07 & -30.41 \\
			CNc  & 93.59 & 39.28 & 95.67 &  91.53 &  11.71 & 96.92 &  -2.06 & -27.57 &   1.25 \\
			ZAwm & 94.71 & 79.99 & 95.33 & 100.27 &  49.16 &101.67 &   5.56 & -30.84 &   6.34 \\
			ZAym & 95.19 & 81.32 & 95.45 &  91.21 &  55.97 & 99.97 &  -3.99 & -25.34 &   4.52 \\
			ZAc  & 95.54 & 84.00 & 96.15 &  78.99 &  48.15 & 83.27 & -16.54 & -35.85 & -12.88 \\
			ARm  & 94.27 & 76.73 & 95.33 & 103.91 & 125.49 &108.06 &   9.64 &  48.76 &  12.73 \\
			USc  & 94.18 & 65.00 & 96.15 &  89.23 &  96.79 & 83.23 &  -4.95 &  31.79 & -12.92 \\
			ZAw  & 94.30 & 62.42 & 96.72 & 100.61 &  28.05 & 74.64 &   6.31 & -34.37 & -22.08 \\
			ARw  & 94.14 & 71.49 & 95.65 & 107.30 &  81.48 & 94.45 &  13.17 &   9.99 &  -1.20 \\
			BSw  & 93.81 & 76.03 & 95.62 &  88.57 & 104.35 & 95.13 &  -5.23 &  28.32 &  -0.49 \\
			USw  & 95.10 & 72.14 & 95.76 &  86.76 &  92.20 & 92.14 &  -8.34 &  20.06 &  -3.62 \\
			UAw  & 94.07 & 73.85 & 93.14 & 103.84 &  94.71 &143.06 &   9.78 &  20.86 &  49.92 \\
			CNr  & 92.75 & 15.83 & 95.23 &  96.25 &   8.35 &103.83 &   3.50 &  -7.48 &   8.61 \\
			USr  & 94.80 & 46.13 & 95.00 &  83.44 &  35.24 &101.75 & -11.36 & -10.89 &   6.76 \\
			INb  & 94.07 & 15.71 & 95.51 &  75.61 &  11.26 & 95.51 & -18.46 &  -4.45 &   0.00 \\\\
			
			\multicolumn{10}{@{}l}{\textit{Panel B: Post‑conflict sample}}\\
			BRms & 93.74 & 72.25 & 95.80 & 118.38 & 114.29 &  86.08 &  24.64 &  42.04 &  -9.72 \\
			BRs  & 94.58 & 72.94 & 95.93 & 100.98 & 121.01 &  82.89 &   6.41 &  48.07 & -13.05 \\
			CNs  & 94.50 & 24.80 & 95.32 &  89.38 &  11.03 &  92.55 &  -5.12 & -13.77 &  -2.77 \\
			ZAs  & 94.77 & 76.33 & 95.65 &  97.59 &  43.07 &  90.65 &   2.82 & -33.26 &  -5.00 \\
			ZAcs & 94.84 & 76.71 & 96.02 &  94.25 &  46.20 &  79.24 &  -0.59 & -30.51 & -16.77 \\
			ARs  & 94.23 & 59.20 & 95.15 & 101.19 &  68.82 &  95.41 &   6.96 &   9.62 &   0.26 \\
			USs  & 94.82 & 65.98 & 95.28 &  99.02 &  84.30 &  95.61 &   4.19 &  18.32 &   0.33 \\
			BRc  & 94.24 & 46.07 & 94.72 & 103.66 &  44.77 & 101.79 &   9.41 &  -1.29 &   7.07 \\
			INm  & 94.09 & 17.19 & 95.39 &  98.29 &  15.13 &  86.19 &   4.20 &  -2.07 &  -9.19 \\
			CNc  & 94.40 & 46.59 & 95.63 &  94.83 &  19.78 &  90.79 &   0.43 & -26.81 &  -4.84 \\
			ZAwm & 95.17 & 73.22 & 94.47 &  89.17 &  66.52 & 112.68 &  -5.99 &  -6.70 &  18.21 \\
			ZAym & 95.26 & 76.13 & 94.80 &  86.51 &  72.73 & 102.27 &  -8.75 &  -3.40 &   7.47 \\
			ZAc  & 95.30 & 76.11 & 95.68 &  90.65 &  74.25 &  89.36 &  -4.65 &  -1.87 &  -6.32 \\
			ARm  & 94.00 & 63.98 & 95.47 & 107.67 &  88.31 &  91.00 &  13.67 &  24.33 &  -4.47 \\
			USc  & 95.20 & 47.97 & 95.84 &  83.28 &  47.27 &  84.74 & -11.92 &  -0.71 & -11.10 \\
			ZAw  & 94.94 & 62.16 & 95.44 &  91.03 &  32.07 &  86.63 &  -3.92 & -30.09 &  -8.81 \\
			ARw  & 94.99 & 58.75 & 95.28 &  94.64 &  68.16 &  95.70 &  -0.36 &   9.41 &   0.42 \\
			BSw  & 95.06 & 56.62 & 94.43 &  87.72 &  60.86 & 108.38 &  -7.33 &   4.24 &  13.95 \\
			USw  & 95.13 & 64.21 & 94.93 &  85.29 &  83.35 &  99.50 &  -9.84 &  19.15 &   4.57 \\
			UAw  & 94.87 & 53.35 & 95.36 &  95.05 &  56.09 &  94.69 &   0.18 &   2.74 &  -0.67 \\
			CNr  & 94.84 & 14.32 & 93.56 &  83.49 &   9.24 & 128.91 & -11.34 &  -5.08 &  35.36 \\
			USr  & 94.90 & 29.48 & 94.31 &  83.54 &  18.62 & 112.91 & -11.36 & -10.85 &  18.61 \\
			INb  & 93.77 & 18.97 & 95.91 & 102.03 &   7.45 &  82.38 &   8.26 & -11.52 & -13.53 \\
			\bottomrule
	\end{tabular*}}
\end{table}

\begin{table}[htp]
	\centering
	\setlength{\abovecaptionskip}{0pt}
	\setlength{\belowcaptionskip}{10pt}
	\captionsetup{width=0.66\textwidth, justification=raggedright,singlelinecheck=false}
	\caption{Average dynamic connectedness in the frequency domain at three quantile levels before and after the outbreak, with values in parentheses indicating medium- and long-term measures and all other values indicating short-term.}
	\label{Tb:Average:Dynamic:Connectedness:frequency:pre:post}
	{\footnotesize
		\setlength{\tabcolsep}{4pt} 
		\resizebox{0.66\textwidth}{!}{
			\begin{tabular}{l*{9}{r}}
				\toprule
				& \multicolumn{3}{c}{FROM} & \multicolumn{3}{c}{TO} & \multicolumn{3}{c}{NET}\\
				\cmidrule(lr){2-4}\cmidrule(lr){5-7}\cmidrule(lr){8-10}
				& $\tau=0.05$ & $\tau=0.50$ & $\tau=0.95$
				& $\tau=0.05$ & $\tau=0.50$ & $\tau=0.95$
				& $\tau=0.05$ & $\tau=0.50$ & $\tau=0.95$\\
				\midrule			
				\multicolumn{10}{l}{\textit{Panel A: Pre‑conflict sample}}\\
				BRms & 66.63 & 67.17 & 64.64 & 65.71 & 103.08 & 64.59 & -0.92 & 35.91 & -0.05 \\
				& (12.84, 15.07) & (6.75, 3.30) & (16.74, 14.66) & (23.37,  17.15) & (13.97,  7.08) & (12.83,  11.66) & (10.53,   2.09) & (7.22,  3.78) & (-3.90,  -3.00) \\
				BRs & 47.59 & 62.74 & 66.61 & 67.25 & 79.45 & 65.26 & 19.66 & 16.70 & -1.34 \\
				& (21.70, 25.73) & (7.05, 3.46) & (14.33, 15.11) & (18.00,  14.54) & (9.23,  4.68) & (12.73,  12.16) & (-3.70, -11.19) & (2.17,  1.22) & (-1.60,  -2.95) \\
				CNs & 71.66 & 24.23 & 67.02 & 59.62 & 6.02 & 57.63 & -12.03 & -18.21 & -9.39 \\
				& (16.78,  6.74) & (4.35, 2.19) & (15.33, 13.69) & (12.45,  12.17) & (0.81,  0.41) & (14.22,  14.51) & (-4.33,   5.43) & (-3.53, -1.78) & (-1.11,   0.82) \\
				ZAs & 65.67 & 62.49 & 61.55 & 67.66 & 29.02 & 72.46 & 1.99 & -33.48 & 10.91 \\
				& (12.55, 16.33) & (10.24, 5.08) & (16.78, 16.98) & (16.61,  19.79) & (4.46,  2.41) & (15.57,  14.42) & (4.06,   3.46) & (-5.78, -2.67) & (-1.22,  -2.56) \\
				ZAcs & 50.69 & 67.58 & 65.32 & 63.30 & 35.26 & 50.71 & 12.61 & -32.32 & -14.61 \\
				& (20.78, 23.00) & (11.98, 6.04) & (17.27, 13.85) & (17.87,  19.22) & (5.43,  2.74) & (13.66,  12.88) & (-2.90,  -3.78) & (-6.54, -3.30) & (-3.61,  -0.97) \\
				ARs & 44.40 & 65.93 & 77.76 & 62.36 & 86.35 & 86.07 & 17.95 & 20.42 & 8.32 \\
				& (20.85, 28.73) & (6.47, 3.14) & (8.05,  8.62) & (20.05,  19.92) & (11.99,  6.19) & (12.86,  11.92) & (-0.80,  -8.82) & (5.51,  3.04) & (4.82,   3.30) \\
				USs & 67.46 & 66.38 & 68.51 & 65.39 & 103.16 & 69.88 & -2.07 & 36.78 & 1.36 \\
				& (11.10, 15.52) & (7.26, 3.54) & (14.15, 12.76) & (17.58,  19.04) & (14.34,  7.35) & (15.63,  14.19) & (6.48,   3.51) & (7.08,  3.81) & (1.48,   1.43) \\
				BRc & 44.06 & 44.69 & 69.68 & 65.26 & 30.92 & 67.41 & 21.21 & -13.76 & -2.27 \\
				& (20.80, 29.38) & (5.31, 2.67) & (14.46, 11.12) & (19.49,  18.04) & (6.44,  3.34) & (17.99,  17.13) & (-1.31, -11.34) & (1.12,  0.67) & (3.53,   6.01) \\
				INm & 80.45 & 11.92 & 63.42 & 50.02 & 6.94 & 42.31 & -30.43 & -4.98 & -21.11 \\
				& (8.27,  5.54) & (2.15, 1.10) & (15.24, 18.26) & (12.34,  12.70) & (0.78,  0.38) & (11.66,  11.63) & (4.07,   7.16) & (-1.38, -0.71) & (-3.59,  -6.63) \\
				CNc & 67.70 & 31.57 & 68.21 & 68.09 & 9.88 & 67.32 & 0.39 & -21.68 & -0.89 \\
				& (13.88, 12.09) & (5.07, 2.64) & (15.31, 12.03) & (12.34,  10.68) & (1.21,  0.62) & (14.80,  14.63) & (-1.54,  -1.40) & (-3.86, -2.02) & (-0.51,   2.61) \\
				ZAwm & 57.09 & 65.86 & 65.85 & 68.39 & 41.61 & 70.72 & 11.30 & -24.25 & 4.87 \\
				& (21.69, 15.99) & (9.40, 4.73) & (16.25, 13.18) & (16.15,  15.85) & (5.02,  2.53) & (15.79,  15.10) & (-5.54,  -0.14) & (-4.38, -2.21) & (-0.45,   1.91) \\
				ZAym & 66.12 & 65.24 & 77.07 & 58.33 & 47.34 & 71.87 & -7.78 & -17.90 & -5.21 \\
				& (16.26, 12.90) & (10.63, 5.45) & (11.16,  7.13) & (15.56,  16.90) & (5.77,  2.86) & (14.77,  13.98) & (-0.70,   4.00) & (-4.85, -2.59) & (3.60,   6.85) \\
				ZAc & 66.63 & 67.45 & 72.47 & 51.59 & 40.64 & 55.67 & -15.04 & -26.80 & -16.81 \\
				& (15.12, 13.86) & (10.92, 5.63) & (14.25,  9.30) & (13.10,  13.67) & (5.01,  2.49) & (13.82,  13.71) & (-2.02,  -0.19) & (-5.91, -3.14) & (-0.43,   4.40) \\
				ARm & 70.92 & 69.50 & 53.17 & 70.48 & 104.46 & 80.52 & -0.45 & 34.96 & 27.35 \\
				& (9.79, 13.63) & (4.96, 2.28) & (20.70, 21.43) & (17.23,  15.79) & (14.00,  7.04) & (14.57,  13.12) & (7.44,   2.16) & (9.04,  4.76) & (-6.13,  -8.31) \\
				USc & 64.38 & 56.53 & 65.70 & 67.04 & 82.91 & 57.72 & 2.66 & 26.39 & -7.98 \\
				& (14.42, 15.53) & (5.56, 2.92) & (15.41, 15.00) & (11.40,  10.64) & (9.41,  4.46) & (13.19,  12.40) & (-3.02,  -4.89) & (3.85,  1.55) & (-2.23,  -2.60) \\
				ZAw & 43.22 & 46.69 & 78.12 & 62.82 & 23.94 & 47.85 & 19.59 & -22.75 & -30.28 \\
				& (23.44, 27.70) & (10.33, 5.39) & (8.09, 10.48) & (16.75,  21.74) & (2.83,  1.28) & (13.31,  13.05) & (-6.69,  -5.97) & (-7.51, -4.11) & (5.21,   2.57) \\
				ARw & 52.68 & 61.97 & 67.10 & 67.79 & 70.11 & 71.22 & 15.11 & 8.15 & 4.11 \\
				& (20.07, 21.38) & (6.37, 3.15) & (14.30, 14.36) & (18.18,  21.14) & (7.65,  3.71) & (12.06,  11.26) & (-1.89,  -0.24) & (1.28,  0.56) & (-2.25,  -3.10) \\
				BSw & 75.67 & 66.99 & 62.19 & 66.59 & 90.66 & 60.43 & -9.07 & 23.67 & -1.75 \\
				& (7.95, 10.14) & (6.09, 2.95) & (16.47, 16.99) & (11.23,  11.13) & (9.25,  4.44) & (17.04,  16.83) & (3.28,   0.99) & (3.16,  1.49) & (0.58,  -0.16) \\
				USw & 61.54 & 61.10 & 65.32 & 59.20 & 78.84 & 59.24 & -2.34 & 17.74 & -6.08 \\
				& (15.73, 17.93) & (7.46, 3.58) & (16.04, 14.41) & (13.34,  13.65) & (8.97,  4.39) & (16.34,  15.90) & (-2.38,  -4.28) & (1.51,  0.81) & (0.30,   1.49) \\
				UAw & 73.18 & 64.61 & 65.52 & 66.31 & 81.04 & 111.15 & -6.87 & 16.42 & 45.63 \\
				& (12.52,  8.44) & (6.18, 3.05) & (13.87, 13.90) & (15.97,  21.60) & (9.25,  4.42) & (16.58,  15.96) & (3.45,  13.15) & (3.07,  1.36) & (2.70,   2.06) \\
				CNr & 74.79 & 13.67 & 72.05 & 66.68 & 6.29 & 70.97 & -8.11 & -7.37 & -1.08 \\
				& (14.04,  3.97) & (1.43, 0.73) & (12.59, 10.60) & (15.66,  13.82) & (1.35,  0.71) & (17.15,  15.94) & (1.62,   9.85) & (-0.09, -0.02) & (4.56,   5.34) \\
				USr & 64.20 & 37.81 & 75.87 & 55.39 & 28.35 & 75.82 & -8.81 & -9.46 & -0.05 \\
				& (13.96, 16.74) & (5.56, 2.76) & (8.19, 10.80) & (11.95,  15.80) & (4.56,  2.33) & (14.27,  12.69) & (-2.00,  -0.94) & (-1.00, -0.43) & (6.08,   1.89) \\
				INb & 72.95 & 12.95 & 52.09 & 54.41 & 8.77 & 68.42 & -18.55 & -4.18 & 16.34 \\
				& (11.97,  9.29) & (1.83, 0.94) & (19.39, 24.04) & (9.86,  10.66) & (1.63,  0.85) & (13.55,  13.66) & (-2.12,   1.38) & (-0.19, -0.08) & (-5.84, -10.39) \\\\
				
				\multicolumn{10}{l}{\textit{Panel B: Post‑conflict sample}}\\
				BRms & 71.44 & 59.46 & 72.37 & 76.93 & 94.20 & 66.85 & 5.49 & 34.74 & -5.52 \\
				& (8.54, 13.75) & (8.49, 4.30) & (11.45, 12.01) & (16.69,  25.30) & (13.43,  6.66) & (8.83,  10.22) & (8.15, 11.56) & (4.94,  2.36) & (-2.63, -1.79) \\
				BRs & 70.54 & 58.97 & 70.33 & 69.06 & 100.14 & 64.50 & -1.48 & 41.17 & -5.84 \\
				& (10.33, 13.72) & (9.26, 4.71) & (12.05, 13.60) & (12.86,  19.17) & (13.98,  6.89) & (8.40,   9.44) & (2.53,  5.45) & (4.72,  2.18) & (-3.65, -4.16) \\
				CNs & 70.91 & 20.39 & 72.12 & 69.78 & 9.38 & 68.74 & -1.13 & -11.02 & -3.38 \\
				& (9.63, 14.01) & (2.94, 1.47) & (11.58, 11.66) & (9.27,  10.75) & (1.11,  0.54) & (10.96,  13.12) & (-0.36, -3.26) & (-1.83, -0.92) & (-0.62,  1.46) \\
				ZAs & 58.10 & 55.64 & 66.19 & 68.22 & 36.51 & 64.96 & 10.12 & -19.13 & -1.23 \\
				& (14.75, 21.95) & (13.59, 7.10) & (12.73, 16.71) & (12.17,  17.08) & (4.38,  2.18) & (11.83,  13.54) & (-2.59, -4.87) & (-9.22, -4.92) & (-0.90, -3.17) \\
				ZAcs & 72.90 & 60.11 & 73.16 & 73.47 & 39.17 & 60.82 & 0.57 & -20.95 & -12.34 \\
				& (9.50, 12.50) & (10.98, 5.61) & (11.57, 11.32) & (9.73,  11.35) & (4.71,  2.32) & (8.42,   9.49) & (0.24, -1.15) & (-6.27, -3.29) & (-3.15, -1.83) \\
				ARs & 57.99 & 49.35 & 73.88 & 65.31 & 54.43 & 77.00 & 7.32 & 5.08 & 3.12 \\
				& (18.32, 17.99) & (6.59, 3.26) & (10.17, 11.14) & (14.47,  21.08) & (9.58,  4.80) & (8.58,   9.75) & (-3.85,  3.08) & (3.00,  1.54) & (-1.59, -1.39) \\
				USs & 70.47 & 55.83 & 75.37 & 67.74 & 69.41 & 70.40 & -2.72 & 13.57 & -4.98 \\
				& (10.36, 13.99) & (6.78, 3.37) & (10.23,  9.61) & (12.47,  18.68) & (9.94,  4.95) & (12.54,  12.45) & (2.11,  4.69) & (3.16,  1.59) & (2.30,  2.84) \\
				BRc & 63.05 & 37.37 & 71.49 & 72.34 & 34.78 & 79.49 & 9.29 & -2.59 & 7.99 \\
				& (12.98, 18.26) & (5.80, 2.90) & (11.36, 11.88) & (12.88,  18.17) & (6.68,  3.32) & (10.53,  12.42) & (-0.11, -0.09) & (0.88,  0.42) & (-0.84,  0.54) \\
				INm & 76.65 & 15.39 & 72.84 & 79.41 & 13.19 & 63.25 & 2.76 & -2.20 & -9.59 \\
				& (8.49,  8.99) & (1.21, 0.59) & (10.41, 12.17) & (9.50,  10.19) & (1.30,  0.63) & (10.68,  12.22) & (1.00,  1.20) & (0.09,  0.05) & (0.28,  0.05) \\
				CNc & 71.26 & 34.42 & 68.83 & 74.06 & 15.89 & 73.13 & 2.80 & -18.52 & 4.30 \\
				& (11.44, 11.69) & (8.08, 4.09) & (12.59, 14.25) & (9.82,  11.65) & (2.59,  1.30) & (8.14,   9.50) & (-1.61, -0.04) & (-5.50, -2.78) & (-4.45, -4.75) \\
				ZAwm & 69.21 & 59.35 & 71.22 & 67.98 & 54.29 & 80.96 & -1.22 & -5.06 & 9.74 \\
				& (13.24, 12.78) & (9.22, 4.65) & (10.56, 12.72) & (9.73,  11.12) & (8.11,  4.12) & (14.63,  17.08) & (-3.50, -1.66) & (-1.12, -0.53) & (4.07,  4.35) \\
				ZAym & 64.23 & 60.97 & 71.59 & 65.18 & 59.69 & 68.74 & 0.96 & -1.27 & -2.85 \\
				& (12.99, 18.10) & (10.09, 5.07) & (11.13, 12.19) & (9.77,  11.16) & (8.67,  4.37) & (15.27,  17.45) & (-3.21, -6.94) & (-1.42, -0.70) & (4.14,  5.27) \\
				ZAc & 64.93 & 60.27 & 70.11 & 68.08 & 60.34 & 66.35 & 3.15 & 0.07 & -3.76 \\
				& (11.86, 18.54) & (10.53, 5.32) & (11.72, 13.86) & (9.96,  12.52) & (9.26,  4.65) & (10.48,  12.42) & (-1.90, -6.02) & (-1.27, -0.67) & (-1.25, -1.44) \\
				ARm & 62.74 & 53.25 & 74.68 & 73.74 & 69.67 & 70.36 & 11.00 & 16.41 & -4.32 \\
				& (12.95, 18.32) & (7.26, 3.47) & (10.19, 10.66) & (14.20,  19.84) & (12.42,  6.22) & (10.02,  10.21) & (1.25,  1.52) & (5.17,  2.75) & (-0.17, -0.46) \\
				USc & 79.00 & 40.70 & 71.39 & 63.08 & 38.04 & 62.94 & -15.93 & -2.66 & -8.45 \\
				& (7.04,  9.20) & (4.92, 2.35) & (11.03, 13.44) & (9.32,  11.02) & (6.14,  3.08) & (10.00,  11.57) & (2.28,  1.82) & (1.22,  0.74) & (-1.03, -1.86) \\
				ZAw & 62.90 & 48.50 & 69.06 & 70.02 & 26.49 & 64.25 & 7.13 & -22.01 & -4.80 \\
				& (12.03, 20.05) & (9.07, 4.58) & (11.74, 14.60) & (9.62,  11.25) & (3.70,  1.88) & (10.32,  11.82) & (-2.40, -8.79) & (-5.38, -2.70) & (-1.43, -2.77) \\
				ARw & 76.95 & 47.25 & 72.62 & 67.95 & 53.79 & 75.13 & -9.00 & 6.54 & 2.51 \\
				& (7.61, 10.45) & (7.68, 3.82) & (10.51, 12.20) & (11.88,  14.96) & (9.58,  4.79) & (9.71,  10.77) & (4.27,  4.50) & (1.90,  0.97) & (-0.80, -1.43) \\
				BSw & 65.20 & 47.35 & 74.97 & 61.40 & 48.98 & 84.65 & -3.80 & 1.63 & 9.68 \\
				& (12.29, 17.63) & (6.19, 3.09) & (9.13, 10.33) & (11.09,  15.01) & (7.88,  4.00) & (11.61,  12.92) & (-1.20, -2.62) & (1.70,  0.92) & (2.48,  2.59) \\
				USw & 73.99 & 53.57 & 71.21 & 62.24 & 66.72 & 77.62 & -11.75 & 13.14 & 6.41 \\
				& (9.57, 11.65) & (7.12, 3.51) & (10.46, 13.29) & (10.11,  12.51) & (11.07,  5.56) & (10.29,  11.56) & (0.53,  0.86) & (3.95,  2.06) & (-0.17, -1.73) \\
				UAw & 72.57 & 46.67 & 74.12 & 71.35 & 45.13 & 71.54 & -1.22 & -1.53 & -2.58 \\
				& (9.29, 13.03) & (4.51, 2.18) & (10.29, 10.95) & (10.24,  13.27) & (7.28,  3.68) & (11.18,  12.71) & (0.96,  0.24) & (2.77,  1.50) & (0.89,  1.77) \\
				CNr & 73.50 & 12.27 & 69.55 & 63.75 & 7.80 & 98.64 & -9.75 & -4.46 & 29.09 \\
				& (9.56, 11.82) & (1.37, 0.69) & (10.85, 13.18) & (9.07,  10.62) & (0.96,  0.48) & (14.67,  16.95) & (-0.49, -1.20) & (-0.41, -0.21) & (3.81,  3.78) \\
				USr & 64.54 & 25.30 & 65.87 & 61.07 & 14.89 & 78.39 & -3.47 & -10.41 & 12.52 \\
				& (15.42, 15.00) & (2.90, 1.28) & (12.03, 16.29) & (9.68,  12.57) & (2.48,  1.25) & (15.89,  19.29) & (-5.74, -2.42) & (-0.41, -0.03) & (3.86,  3.01) \\
				INb & 78.82 & 16.98 & 75.26 & 79.70 & 6.43 & 59.54 & 0.88 & -10.54 & -15.72 \\
				& (6.81,  8.16) & (1.33, 0.66) & (9.44, 11.25) & (10.46,  12.29) & (0.68,  0.34) & (10.27,  12.40) & (3.64,  4.13) & (-0.65, -0.32) & (0.83,  1.15) \\
				\bottomrule
	\end{tabular}}}
\end{table}

\newpage

\section{Network connectedness in the frequency domain}

\begin{figure}[!b]
	\centering
	\begin{tikzpicture}
		\node[anchor=south west,inner sep=0] (imageA) at (0,0) {\includegraphics[width=0.495\textwidth]{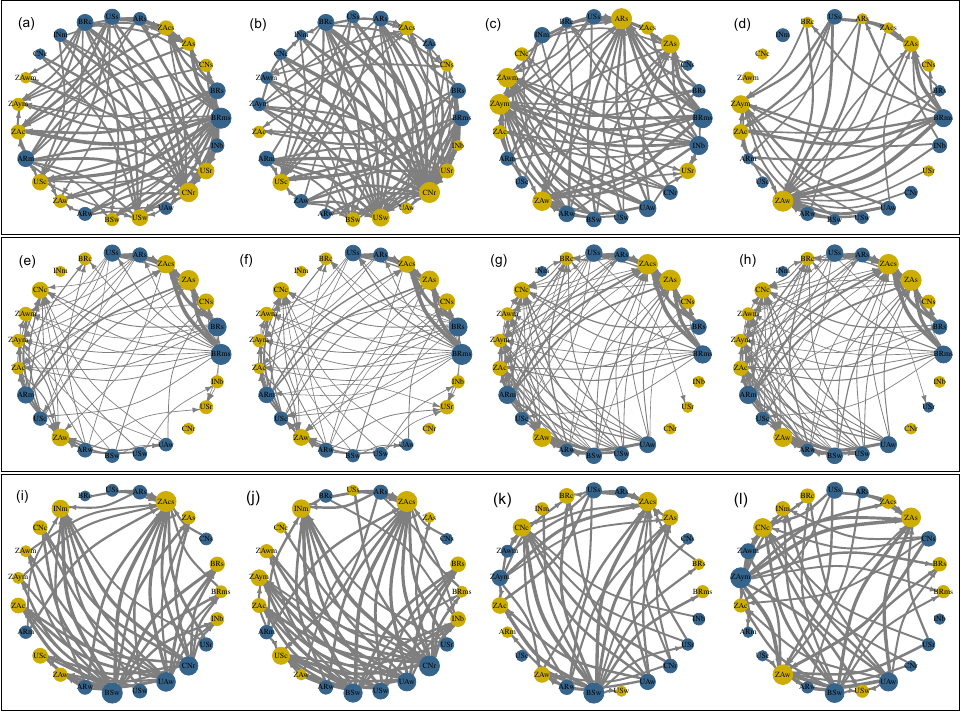}};
		\node[anchor=south east, xshift=-36mm, yshift=60mm] at (imageA.south east) {(A)};
	\end{tikzpicture}
	\begin{tikzpicture}
		\node[anchor=south west,inner sep=0] (imageB) at (0,0) {\includegraphics[width=0.499\textwidth]{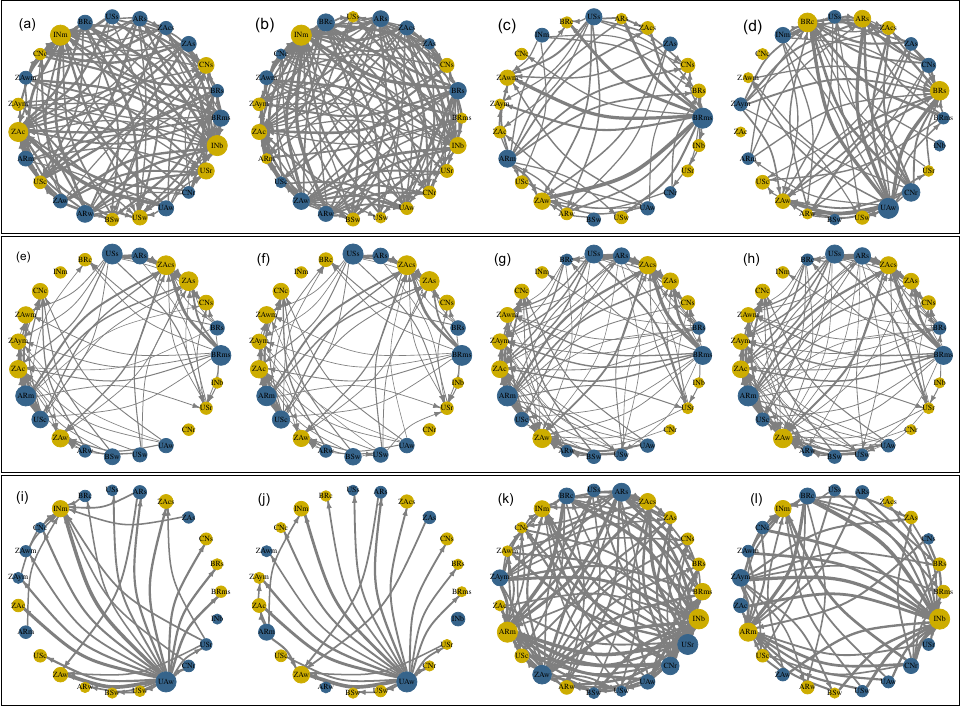}};
		\node[anchor=south east, xshift=-36mm, yshift=60mm] at (imageB.south east) {(B)};
	\end{tikzpicture}
	\\
	\begin{tikzpicture}
		\node[anchor=south west,inner sep=0] (imageB) at (0,0) {\includegraphics[width=0.495\textwidth]{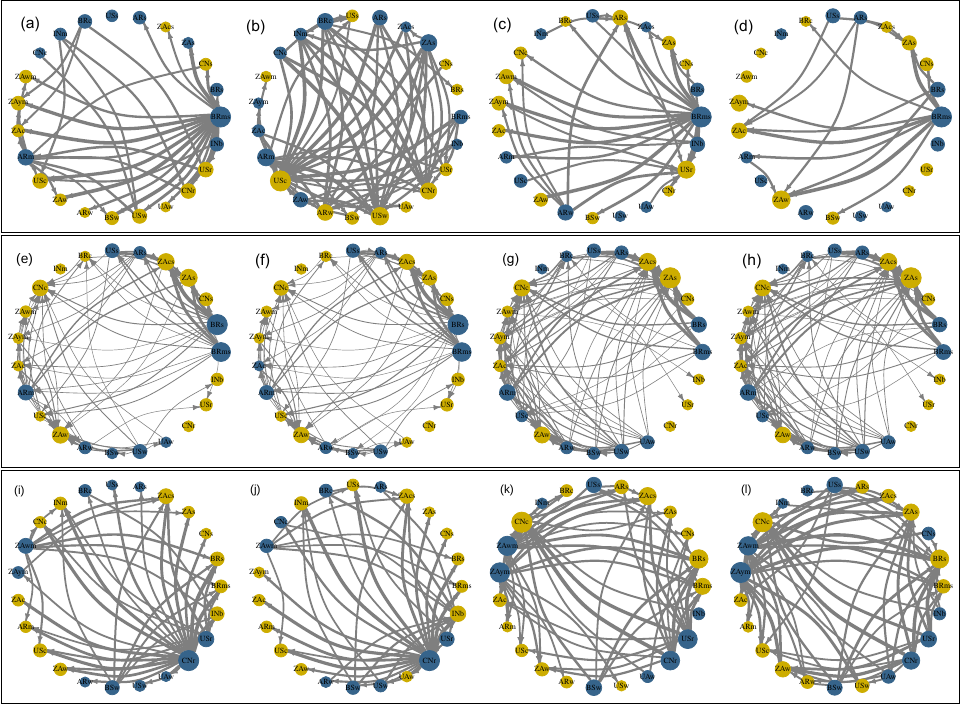}};
		\node[anchor=south east, xshift=7mm, yshift=30mm] at (imageB.south east) {(C)};
	\end{tikzpicture}
	\caption{Subfigures (A)–(C) display frequency-domain network connectedness for the full sample, the pre-outbreak period, and the post-outbreak period, respectively. Within each subfigure, the upper, middle, and lower rows correspond to the extreme lower ($\tau = 0.05$), conditional median ($\tau = 0.50$), and extreme upper ($\tau = 0.95$) quantiles, while the first through fourth columns report overall, short-term, medium-term, and long-term connectedness, respectively.}
	\label{Fig:Network:fre}
\end{figure}

\newpage
\section{Dynamic total and net connectedness across time and quantiles in the frequency domain}

\begin{figure}[!b]
	\centering
	\begin{tikzpicture}
		\node[anchor=south west,inner sep=0] (imageA) at (0,0) {\includegraphics[width=0.495\textwidth]{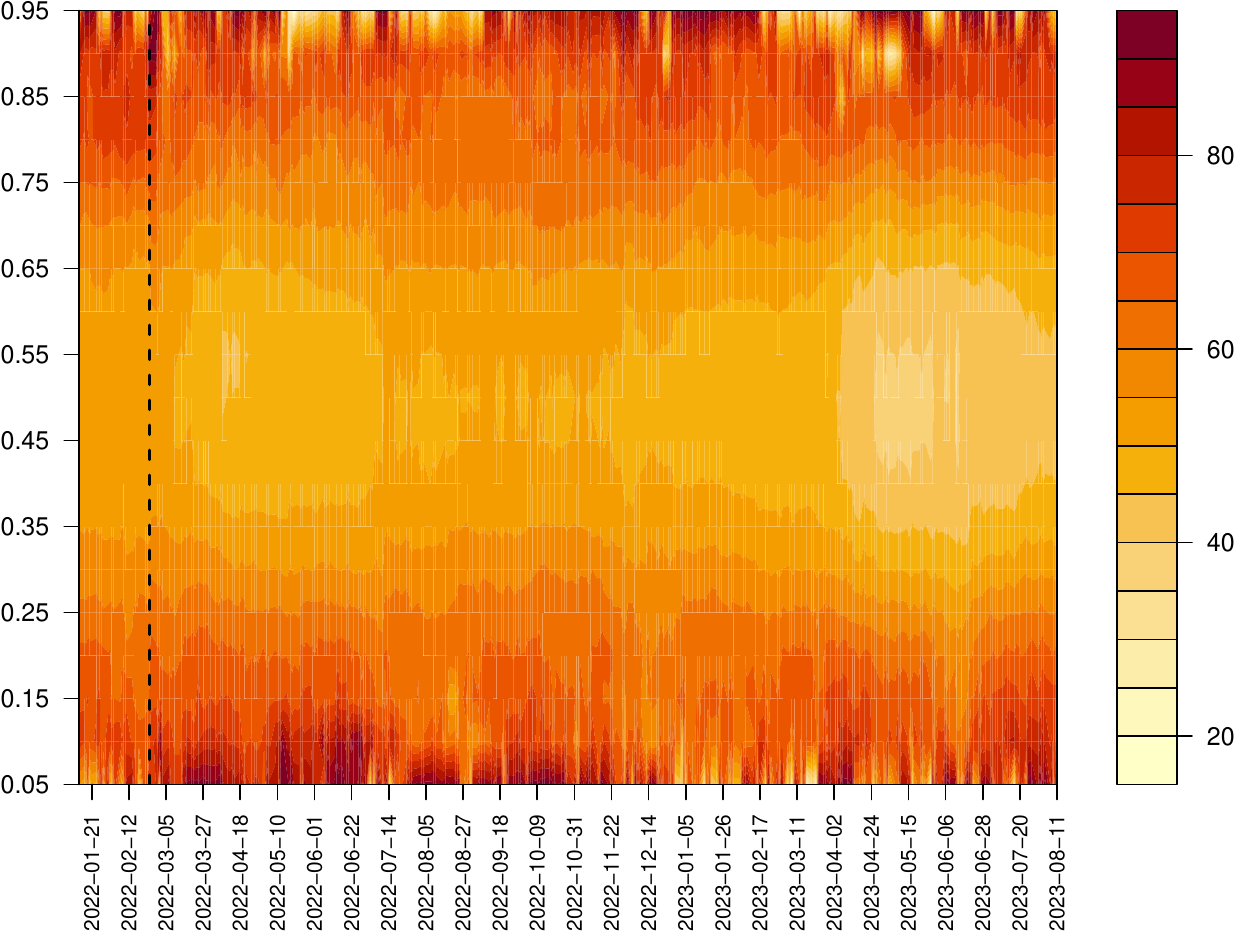}};
		\node[anchor=south east, xshift=-2mm, yshift=2mm] at (imageA.south east) {(A)};
	\end{tikzpicture}
	\begin{tikzpicture}
		\node[anchor=south west,inner sep=0] (imageB) at (0,0) {\includegraphics[width=0.495\textwidth]{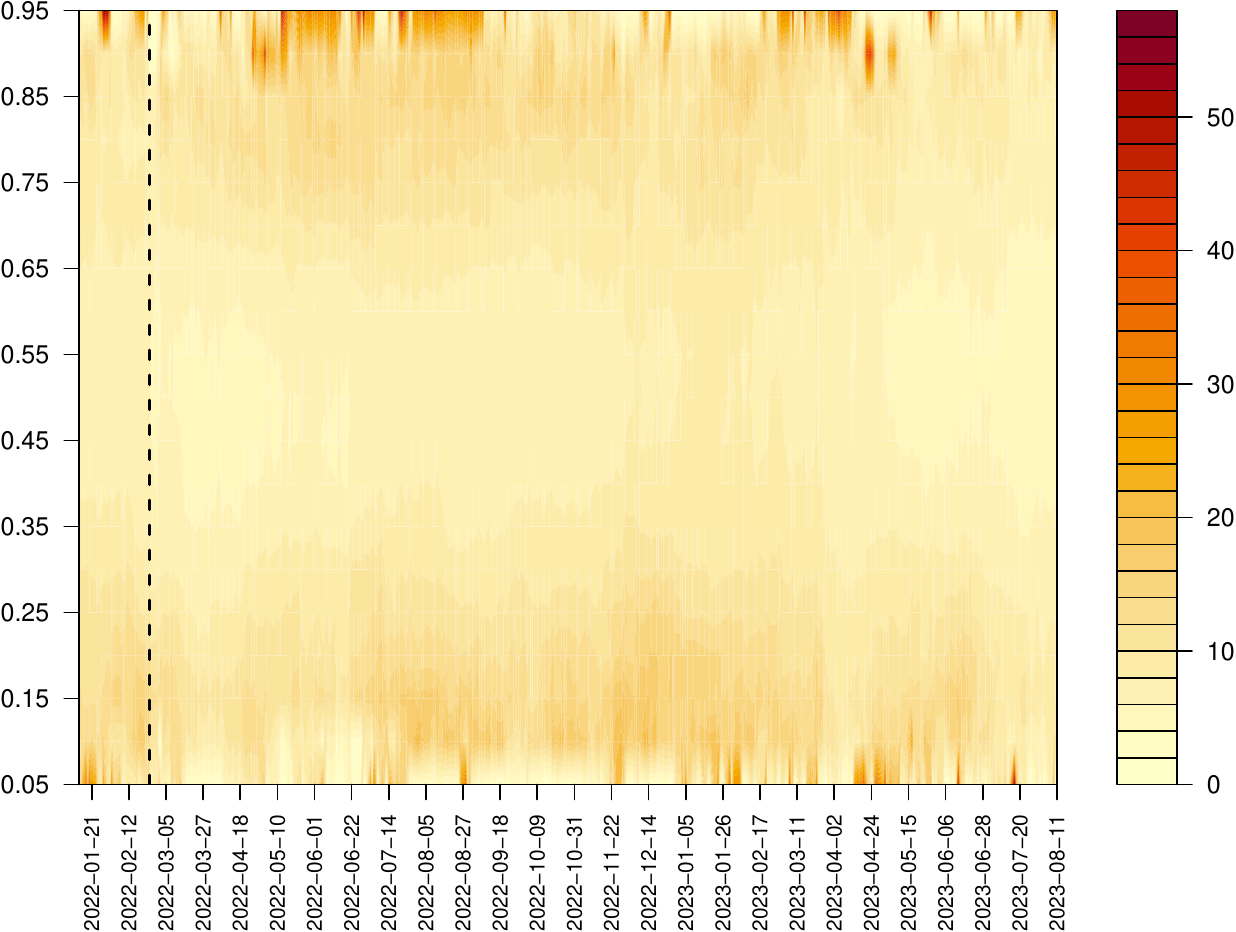}};
		\node[anchor=south east, xshift=-2mm, yshift=2mm] at (imageB.south east) {(B)};
	\end{tikzpicture}
		\\
	\begin{tikzpicture}
		\node[anchor=south west,inner sep=0] (imageB) at (0,0) {\includegraphics[width=0.495\textwidth]{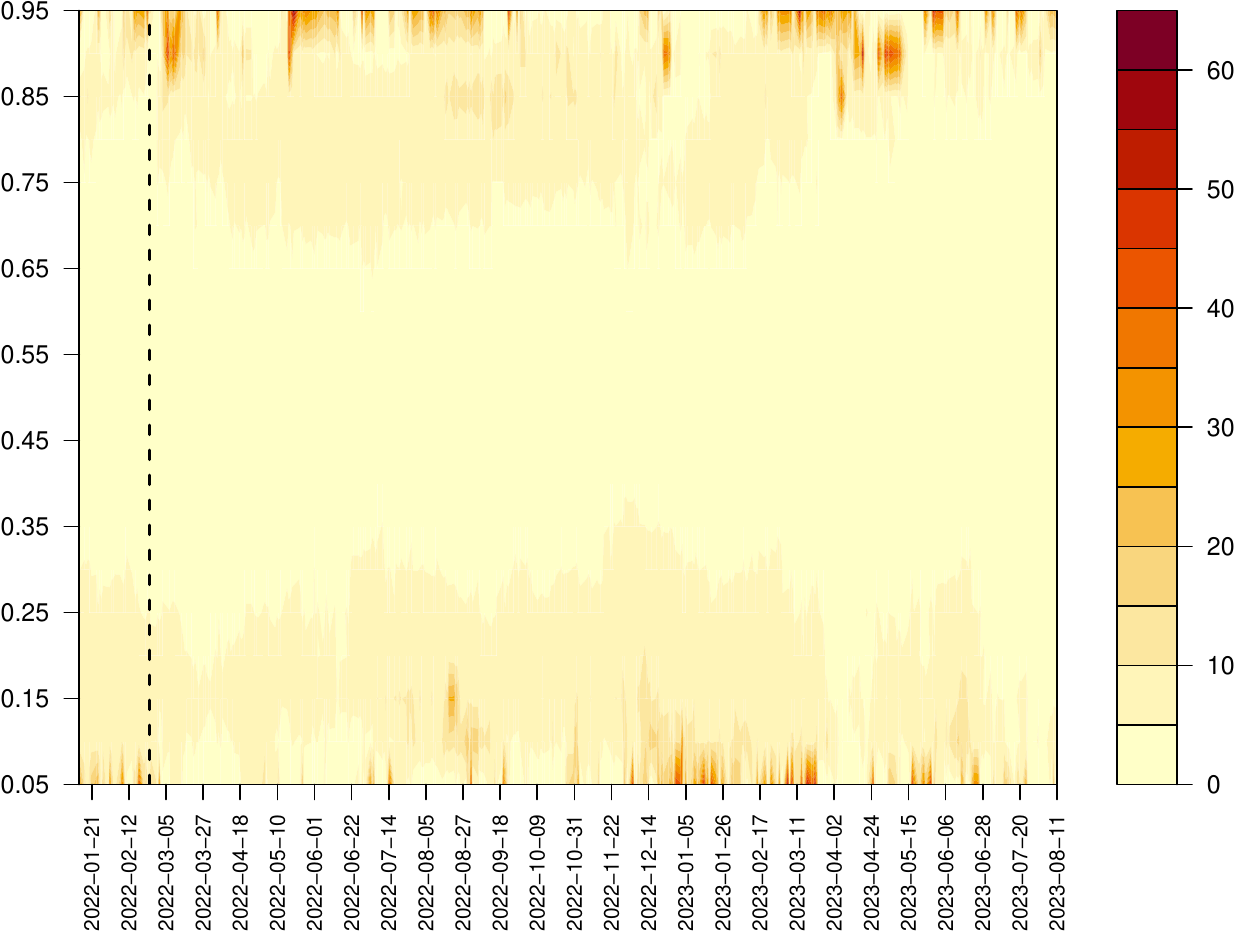}};
		\node[anchor=south east, xshift=-2mm, yshift=2mm] at (imageB.south east) {(C)};
	\end{tikzpicture}
	\caption{Panels (A)–(C) depict the short-, medium-, and long-term dynamic total connectedness across time and quantiles, respectively.}
	\label{Fig:TCI:time:quantiles:fre}
\end{figure}

\begin{figure}[!htp]
	\centering
	\begin{tikzpicture}
		\node[anchor=south west,inner sep=0] (imageA) at (0,0) {\includegraphics[width=0.71\linewidth]{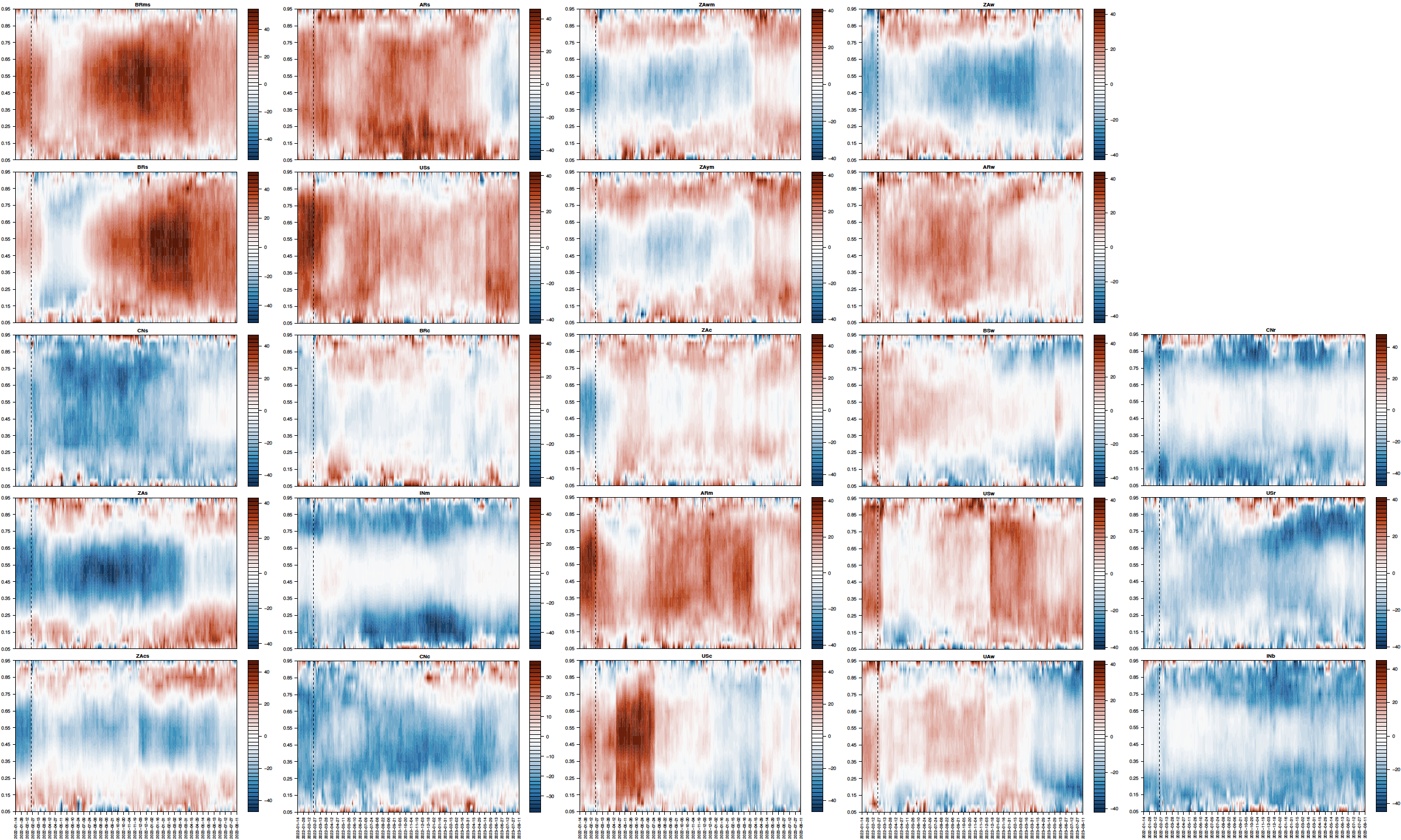}};
		\node[anchor=south east, xshift=-8mm, yshift=52mm] at (imageA.south east) {(A)};
	\end{tikzpicture}
	\begin{tikzpicture}
		\node[anchor=south west,inner sep=0] (imageB) at (0,0) {\includegraphics[width=0.71\linewidth]{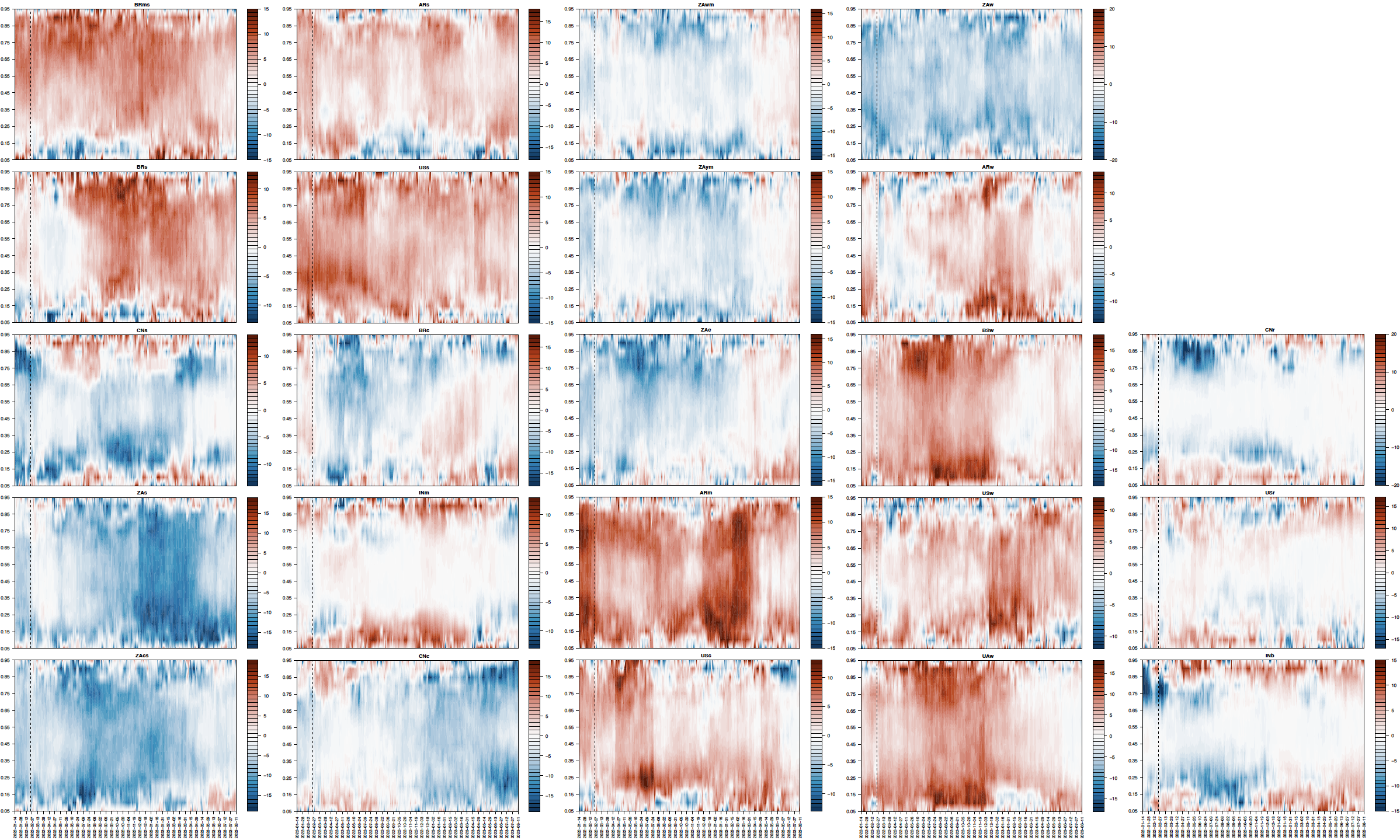}};
		\node[anchor=south east, xshift=-8mm, yshift=52mm] at (imageB.south east) {(B)};
	\end{tikzpicture}
	\\
	\begin{tikzpicture}
		\node[anchor=south west,inner sep=0] (imageB) at (0,0) {\includegraphics[width=0.71\linewidth]{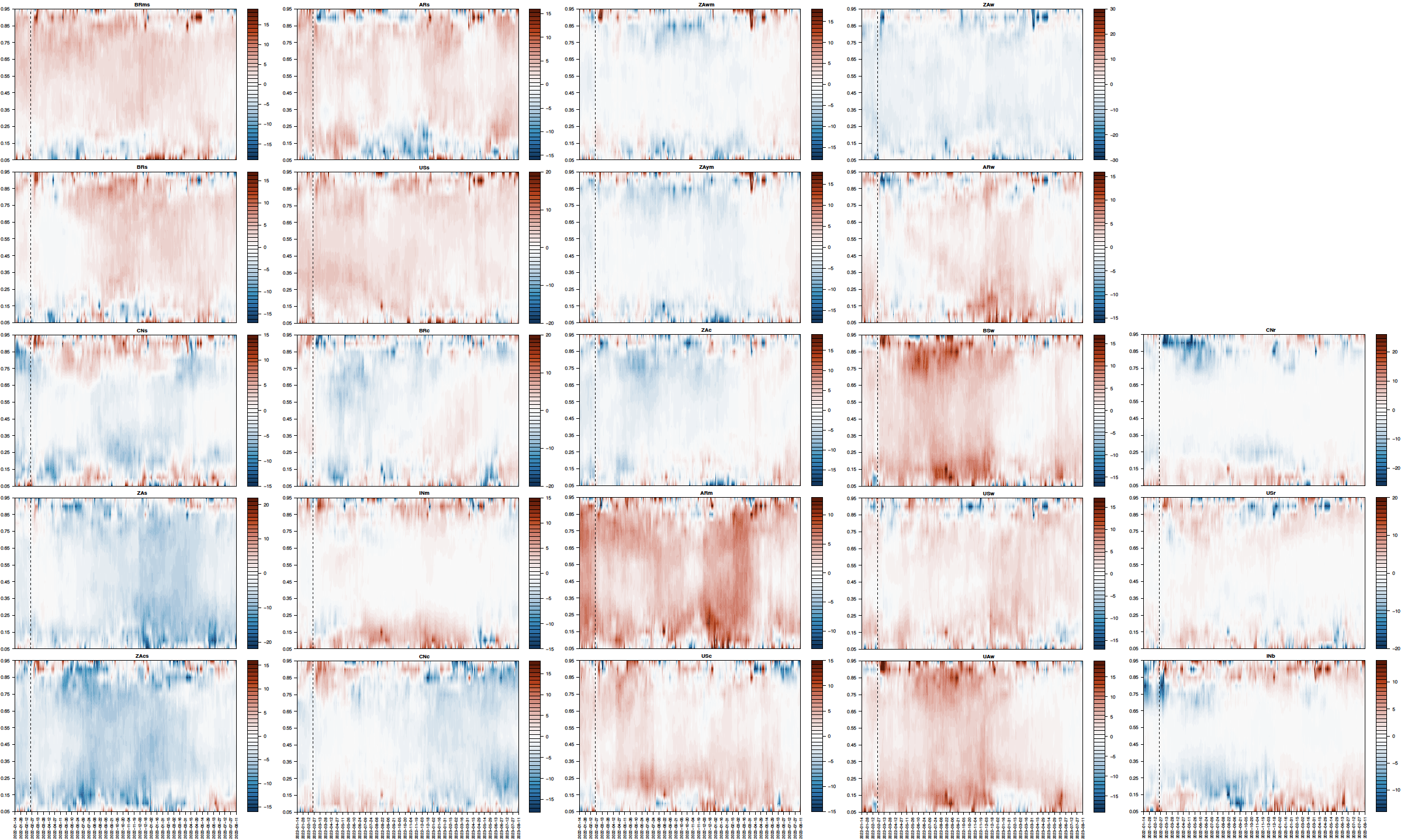}};
		\node[anchor=south east, xshift=-8mm, yshift=52mm] at (imageB.south east) {(C)};
	\end{tikzpicture}
		\caption{Subfigures (A)–(C) show the net total directional connectedness across time and quantiles in the short-, medium-, and long-term frequency domain, respectively.
		}
		\label{Fig:NET:time:quantiles:fre}
	\end{figure}

\end{document}